%% file: revision.tex
\documentclass[sigplan,10pt]{acmart}
\AtBeginDocument{%
  }

\acmYear{2026}\copyrightyear{2026}
\acmConference[EUROSYS '26]{European Conference on Computer Systems}{April 27--30, 2026}{Edinburgh, Scotland Uk}
\acmBooktitle{European Conference on Computer Systems (EUROSYS '26), April 27--30, 2026, Edinburgh, Scotland Uk}
\acmDOI{10.1145/3767295.3769376}
\acmISBN{979-8-4007-2212-7/26/04}

\usepackage{tikz}

\usepackage{filecontents}
\usepackage[normalem]{ulem}
\usepackage{geometry}
\usepackage{booktabs}
\usepackage{siunitx}
\usepackage{graphicx}
\usepackage{subcaption}
\usepackage{listings}
\usepackage{balance}

\usepackage{amsmath}
\usepackage{blindtext}
\usepackage[utf8]{inputenc}
\usepackage[T1]{fontenc}

\usepackage{caption}
\usepackage{xcolor}
\usepackage{alltt}
\usepackage{comment}
\usepackage{multirow}
\usepackage{array}
\usepackage{algpseudocode}
\usepackage[abbreviations]{foreign}
\usepackage{wrapfig}
\usepackage[many]{tcolorbox}
\usepackage{tikz,pgfplots}
\usepackage{pgfplotstable}
\usepackage{nicematrix}
\usepackage{appendix}
\usepackage{listings}
\usepackage{enumitem}
\usepackage{pifont}
\usepackage{makecell}
\usepackage{soul}
\usepackage[linesnumbered,ruled,vlined]{algorithm2e}
\usepackage{xcolor}
\SetKwComment{Comment}{$\triangleright$\ }{}

\usepackage{cleveref}
\crefformat{chapter}{\S#2#1#3}
\crefformat{section}{\S#2#1#3}
\crefformat{subsection}{\S#2#1#3}
\crefformat{subsubsection}{\S#2#1#3}

\input{configs/commands}

\usepackage{caption}
\usepackage{float}
\usepackage{xspace}

\newcommand{\sys}{\textsc{MinatoLoader}\xspace}
\usepackage[linesnumbered,ruled,vlined]{algorithm2e}
\SetKw{Continue}{continue}
\SetKwComment{Comment}{$\triangleright$\ }{}

\begin{document}

\settopmatter{authorsperrow=4}

\title{\sys: Accelerating Machine Learning Training Through Efficient Data Preprocessing}

%%
%% The "author" command and its associated commands are used to define
%% the authors and their affiliations.
%% Of note is the shared affiliation of the first two authors, and the
%% "authornote" and "authornotemark" commands
%% used to denote shared contribution to the research.

\author{Rahma Nouaji}
\affiliation{%
  \institution{McGill University}
  \country{Canada}
}

\author{Stella Bitchebe}
\affiliation{%
  \institution{McGill University}
  \country{Canada}
}

\author{Ricardo Macedo}
\affiliation{%
  \institution{INESC TEC \& U. Minho}
  \country{Portugal}
}
  
\author{Oana Balmau}
\affiliation{%
  \institution{McGill University}
  \country{Canada}
}

%%
%% By default, the full list of authors will be used in the page
%% headers. Often, this list is too long, and will overlap
%% other information printed in the page headers. This command allows
%% the author to define a more concise list
%% of authors' names for this purpose.
%\renewcommand{\shortauthors}{Nouaji et al.}

%---------------------------------------------------
%\input{abstract}

\begin{abstract}

Machine learning (ML) frameworks, such as PyTorch and TensorFlow, rely on data loaders to preprocess data before feeding it to accelerators.
When preprocessing is inefficiently pipelined, GPUs can remain idle over long periods of time, leading to substantial training delays.
For example, PyTorch's default data loaders can cause up to 76\% GPU idleness.
A key bottleneck is the variability in preprocessing time across samples within the same dataset. 
Existing data loaders are oblivious to this variability, training all samples uniformly. 
In this case, a single slow sample can stall the entire batch, causing head-of-line blocking.

We present \sys, a general-purpose data loader for PyTorch that accelerates training and improves GPU utilization under single-server, multi-GPU settings.
It continuously prepares data in background and constructs batches by prioritizing fast-to-process samples, while slower samples are processed in parallel.

Experiments conducted over NVIDIA V100 and A100 GPUs show that \sys accelerates training by up to 7.5$\times$ (3.6$\times$ on average) over PyTorch DataLoader and Pecan, and up to 3$\times$ (2.2$\times$ on average) over DALI. 
It also increases average GPU utilization from 46\% with PyTorch to 90\%, while preserving model accuracy and enabling faster convergence.

\end{abstract}
%---------------------------------------------------

%%
%% The code below is generated by the tool at http://dl.acm.org/ccs.cfm.
%% Please copy and paste the code instead of the example below.
%%



%\begin{CCSXML}
%     <ccs2012>
%      <concept>
%       <concept_id>00000000.0000000.0000000</concept_id>
%       <concept_desc>Do Not Use This Code, Generate the Correct Terms for Your Paper</concept_desc>
%       <concept_significance>500</concept_significance>
%      </concept>
%      <concept>
%       <concept_id>00000000.00000000.00000000</concept_id>
%       <concept_desc>Do Not Use This Code, Generate the Correct Terms for Your Paper</concept_desc>
%       <concept_significance>300</concept_significance>
%      </concept>
%      <concept>
%       <concept_id>00000000.00000000.00000000</concept_id>
%       <concept_desc>Do Not Use This Code, Generate the Correct Terms for Your Paper</concept_desc>
%       <concept_significance>100</concept_significance>
%      </concept>
%      <concept>
%       <concept_id>00000000.00000000.00000000</concept_id>
%       <concept_desc>Do Not Use This Code, Generate the Correct Terms for Your Paper</concept_desc>
%       <concept_significance>100</concept_significance>
%      </concept>
%     </ccs2012>
%     \end{CCSXML}
    
%     \ccsdesc[500]{Do Not Use This Code~Generate the Correct Terms for Your Paper}
%     \ccsdesc[300]{Do Not Use This Code~Generate the Correct Terms for Your Paper}
%     \ccsdesc{Do Not Use This Code~Generate the Correct Terms for Your Paper}
%     \ccsdesc[100]{Do Not Use This Code~Generate the Correct Terms for Your Paper}

% %%
% %% Keywords. The author(s) should pick words that accurately describe
% %% the work being presented. Separate the keywords with commas.

% %% A "teaser" image appears between the author and affiliation
% %% information and the body of the document, and typically spans the
% %% page.

%  \received{20 February 2007}
%  \received[revised]{12 March 2009}
% \received[accepted]{5 June 2009}

%%
%% This command processes the author and affiliation and title
%% information and builds the first part of the formatted document.

%----------------------------------------------------
\maketitle

%----------------------------------------------------
%\input{introduction-oana}
\section{Introduction}
\label{sec:intro}

The efficacy of Machine Learning (ML) deployments relies on both high-quality algorithms and high-quality data, the latter being shaped through data preprocessing. 
Although considerable research focused on improving the training algorithms -- leading to advances in techniques~\cite{10.5281/zenodo.1082415, nuts-flow/ml, 9036908}, software libraries~\cite{numpy, scikit, pandas, mattson2020mlperf}, and hardware accelerators (\emph{e.g.,} GPUs, TPUs, DPUs)-- the data preprocessing step has been relatively underexplored~\cite{gao2024empirical,sambasivan2021everyone,nouajispeedyloader}.
% \rgmacedo{Include here SpeedyLoader?}
% 
Yet, data preprocessing plays an important role in training efficiency and model generalization. 
Transformations such as cropping, resizing, shuffling, and random augmentations are commonly applied on the fly during training to increase model robustness and accuracy~\cite{zhao2022understanding, 9744492, mazumder2022dataperf, shorten2019survey}. 
These operations expose the model to diverse inputs, improving generalization and training speed. 

%---------------------------------------------------------
\input{floaters/motivation-pytorch-dataloader-pipeline}
% \begin{figure}[t]
%     \centering
%     % Subfigure 1: PyTorch DataLoader pipeline inefficiency
%     \begin{subfigure}[t]{\columnwidth}
%         \centering
%         \includegraphics[width=1\linewidth]{figures/pytorch_dataloader_pipeline.pdf}
%         \vspace{-15pt}
%         \caption{Data preprocessing and training pipeline of PyTorch DataLoader.}
%         \label{fig:pytorch-pipeline}
%     \end{subfigure}
    
%     % Subfigure 2: CPU–GPU usage over time
%     \begin{subfigure}[t]{\columnwidth}
%         \vspace{5pt}
%         \centering
%         \includegraphics[width=1\linewidth]{figures/cpu_gpu_v100pytorch_3dunet_zoom (1).pdf}
%         \vspace{-15pt}
%         \caption{CPU and GPU usage of PyTorch DataLoader during 3D‑UNet training.}
%         \label{fig:motiv-gpu-idle}
%     \end{subfigure}
%     \vspace{-.7em}
%     \caption{Inefficient PyTorch DataLoader pipeline. Slow data samples delay the batch construction process, resulting in GPU under-utilization and poor training performance. }
    
%     \label{fig:combined-pytorch-analysis}
%     \vspace{-10pt}
% \end{figure}
%-------------------------------------------------------
To pipeline the data preprocessing with training, modern ML frameworks rely on \textit{data loaders}. 
These components act as intermediaries between the dataset and the model, streamlining and optimizing the process of accessing data for preprocessing, training, and inference. 
The data loader plays a critical role in the training pipeline, and inefficiencies in this stage can bottleneck the overall training performance. 

In most modern applications, this functionality is handled by the PyTorch DataLoader~\cite{pytorch_dataloaders_tutorial}, which is widely used due to PyTorch's popularity. 
In the PyTorch DataLoader, data preprocessing runs on the CPU, and model training happens on the GPU, as depicted in Figure~\ref{fig:pytorch-pipeline}. 
Data samples are loaded from storage into main memory, transformed in parallel, assembled into batches, and finally transferred to the GPU for training.
To reduce loading latency, the data loader exposes several tuning knobs that enable parallel data loading and prefetching, allowing multiple workers to load and prepare data samples concurrently and in advance of training.

However, when preprocessing times vary across samples, tuning these parameters alone is not enough to avoid training delays and GPU stalls. 
This is because PyTorch constructs each batch synchronously, by waiting for all individual samples to be ready before sending it to the GPU. 
As a result, a single slow sample can block the entire batch and cause head-of-line blocking, where a single sample delays the batch creation to undergo training despite other samples being ready.
This inefficiency is evident in the PyTorch DataLoader pipeline (Figure~\ref{fig:pytorch-pipeline}), and results in significant GPU underutilization (Figure~\ref{fig:motiv-gpu-idle}). 
For instance, during heavy data transformations, the GPU often remains idle.
Our experiments reveal that data preprocessing times can vary significantly across samples, even within the same dataset (\cref{subsec:motivation-preprocessing-variability}). 
For example, in an image segmentation workload using the 3D-UNet model~\cite{10.1007/978-3-319-46723-8_49}, preprocessing times range from 0.01 to 2.2 seconds per sample, with an average of approximately 0.5 seconds (Figure~\ref{subfig:variability-image-segmentation}). 
Importantly, it is not trivial to predict which samples will take longer to process, as no simple heuristic (\emph{e.g.,} sample size, number of transformations) reliably predicts preprocessing time across workloads (\cref{subsec:motivation-heuristics}). 
This unpredictability makes it difficult to schedule data loading efficiently and exacerbates the impact of head-of-line blocking.

In this paper, we address the following question: \emph{How can the data loader deliver preprocessed data to the GPU as fast as possible to minimize idle periods?}

To achieve this, we propose \sys, a general-purpose, drop-in replacement for the PyTorch DataLoader that improves training time and GPU utilization without requiring prior knowledge of the dataset or preprocessing pipeline. 
% {\color{red}This work extends our previous workshop paper~\cite{anonomous}(citation anonymized for review).}\rgmacedo{remove this}
%
Contrary to prior work~\cite{pytorch_dataloaders_tutorial, pecan, dali}, which treats all data samples uniformly and blocks batch creation on the slowest sample, \sys introduces a sample-aware scheduling strategy that dynamically adapts to per-sample preprocessing variability.
The key idea behind \sys is to prioritize fast-to-process samples and defer slow ones for later use.
This is a simple yet powerful approach that improves training performance, is generalizable across workloads, and maintains virtually the same model accuracy despite sample reordering inside the batches.

In a nutshell, \sys provides a dynamic load balancer that classifies samples on the fly during training. 
All samples are optimistically assumed to be fast at first. 
After a brief profiling phase to gather lightweight heuristics, \sys applies a per-sample timeout -- samples that exceed a given threshold are flagged as slow and processed in the background, while the rest are immediately used to construct training batches. 
Fast and slow samples are managed through separate queues, enabling \sys to bypass slow samples and prevent stalling batch construction.

To adapt to changing workload demands, \sys dynamically adjusts the number of CPU worker threads involved in preprocessing during training. 
It starts with a default number of CPU workers per GPU, selected based on factors such as dataset size, preprocessing complexity, and I/O intensity. 
During execution, \sys continuously monitors queue occupancy and CPU utilization to detect underutilization or bottlenecks. 
Based on these, it adaptively scales the number of workers to sustain high throughput.

We validate the efficiency of \sys through a comprehensive experimental evaluation across multiple testing scenarios on a single-server with multiple GPUs setup, over three widely-used data preprocessing workloads from the MLPerf benchmarking suite~\cite{mattson2020mlperf}: image segmentation (3D-UNet)~\cite{10.1007/978-3-319-46723-8_49}, object detection (Mask R-CNN)~\cite{massa2018mrcnn}, and speech recognition (RNN-T)~\cite{rnnt}. 
We compare \sys against three baselines: the native PyTorch DataLoader~\cite{pytorch_dataloaders_tutorial}; DALI \cite{dali}, a data loader framework from NVIDIA that delegates preprocessing to the GPU; and Pecan~\cite{pecan}, a state-of-the-art data loader for TensorFlow that automates data preprocessing worker placement and transformation reordering decisions.
Across all experiments, \sys significantly outperforms prior work, reducing total training time by up to $7.5\times$ compared to PyTorch DataLoader and Pecan, and up to $3\times$ compared to DALI. 
It also improves GPU utilization, increasing the average usage from $46.4\%$ with PyTorch DataLoader to $90.5\%$ with \sys. 
Furthermore, \sys scales effectively with the number of GPUs, and even when using a single GPU, it achieves comparable or better training performance than the remainder of data loaders configured with 4 GPUs (outperforming them by up to $60.6\%$).
In memory-constrained settings, it maintains high throughput and avoids input stalls, outperforming existing data loaders up to 2$\times$. 
Finally, we show that \sys has almost no impact on model accuracy while substantially accelerating convergence. 

\noindent
In summary, this paper makes the following contributions:
\begin{itemize}[leftmargin=*,nosep]
   
    \item We extensively study the causes of GPU idleness during training stemming from inefficient data preprocessing. Notably, we identify that sample preprocessing time variability is a major bottleneck, causing head-of-line blocking in state-of-the-art data loaders, including PyTorch DataLoader and DALI.

    \item We design a \emph{dynamic, sample-aware load balancer} that classifies data samples as fast or slow at runtime, enabling efficient batch construction that avoids pipeline stalls and GPU idleness.
    
    \item We combine the above-mentioned observations and techniques into \emph{\sys, a general-purpose data loader} that proactively prepares training data without requiring any prior knowledge of the dataset or preprocessing pipeline.
    We open source \sys at \url{https://github.com/Rahm-no/MinatoLoader} and Zenodo~\cite{rahma_nouaji_2025}. 

\end{itemize}

%----------------------------------------------------
%\input{background}
\section{Background}
\label{sec:back}

This section first provides background on popular data loaders used throughout this paper. It then outlines the datasets we use in the paper and their data preprocessing pipelines. 

\subsection{Data Loaders}
\label{sec:background-dataloaders}

Data loaders iterate over datasets, serving data to the model. They fetch data from persistent storage, to system memory (DRAM), to GPU memory, enabling on-the-fly data augmentation and tasks like batching, shuffling, and parallelization.

\vspace{1mm}
\noindent\textbf{PyTorch DataLoader~\cite{pytorch_dataloaders_tutorial}} uses a standard iterator to access the dataset and supports parallel data loading with worker threads, where each worker can prefetch several samples (controlled by \texttt{prefetch\_factor}). 
While these settings help increase throughput, they are limited when preprocessing becomes the bottleneck. 
The issue lies in how PyTorch constructs and processes batches. 
It first defines the order in which data indices are drawn from the dataset, and then groups these into batches. 
However, by default, the data loader is unaware of differences in preprocessing cost. % differences between samples. 
Consequently, a batch may contain both fast and slow samples, leading to head-of-line blocking where an entire batch is delayed by its slowest sample, as illustrated in Figure~\ref{fig:pytorch-pipeline}.

\vspace{1mm}
\noindent\textbf{DALI~\cite{dali}} is a GPU-accelerated data preprocessing library from NVIDIA that allows users to define mixed CPU-GPU pipelines for preparing data during training. Its design leverages pipelining and asynchronous execution to improve data throughput. While these features can reduce preprocessing latency, we found that tuning DALI's execution parameters had little effect on overall training time. In practice, DALI's reliance on the GPU comes with trade-offs: it increases memory usage on the accelerator and limits flexibility, as most preprocessing operations are implemented as low-level CUDA kernels that are hard to extend or debug. 
 Most importantly, GPUs are expensive and are best reserved for training. Using them for preprocessing is generally only feasible in environments with abundant GPU capacity, where dedicated GPUs can be allocated to preprocessing tasks.

\vspace{1mm}
\noindent\textbf{Pecan~\cite{pecan}} is a recent research state-of-the-art solution, built on tf.data~\cite{tfdata} (TensorFlow's data loader), which aims to reduce ML training costs by optimizing data preprocessing.
Pecan automates data preprocessing, worker placement, and transformation reordering decisions through two main policies: \emph{AutoPlacement} and \emph{AutoOrder}. 
The AutoPlacement policy dynamically schedules data preprocessing workers across CPU resources. 
AutoOrder reorders transformations in the input pipeline to increase per-worker preprocessing throughput. 
Pecan categorizes transformations as \emph{inflationary}, if they increase data volume (\emph{e.g.,} image padding, one-hot encoding), or \emph{deflationary} if they reduce data volume (\emph{e.g.,} sampling, filtering, cropping). 
Following this categorization, the AutoOrder policy moves deflationary transformations earlier in the pipeline and postpones inflationary transformations.
To maintain correctness, Pecan divides the input pipeline into sections at specific transformations that act as barriers to ensure that reordering is restricted within these sections and does not cross barrier boundaries.

%----------------------
\input{floaters/preprocessing-pipelines-table}

\subsection{Data Preprocessing Pipelines}
\label{subsec:data-prep-pipeline}

Throughout this paper, we use three representative workloads from the MLPerf Training Benchmark suite~\cite{mattson2020mlperf,mlcommons-github}, selected for their diversity in data modalities (3D image, 2D image, and audio), preprocessing pipelines, and dataset sizes. This variety allows us to test \sys under different levels of preprocessing complexity. Table~\ref{tab:pipeline-steps} depicts the preprocessing transformations applied over each workload.

\paragraph{Image Segmentation}
This workload performs 3D segmentation of kidney tumors using the 29GB KiTS19 dataset~\cite{kits19,heller2019kits19}, and is trained with the 3D-UNet model~\cite{10.1007/978-3-319-46723-8_49}. 
The data pipeline applies five sequential preprocessing steps (Table~\ref{tab:pipeline-steps}). 
Before preprocessing, the size of input samples ranges from 30MB to 375MB, with an average of 136MB. 
After preprocessing, all samples are standardized to a uniform size of 10MB.
We include this workload because it presents heavy and variable preprocessing, primarily due to the complexity and size of volumetric 3D medical images. 

\paragraph{Object Detection}
This workload detects objects in 2D images using the Mask R-CNN model~\cite{massa2018mrcnn} with a ResNet50 backbone. 
It operates on the 58GB COCO dataset~\cite{cocodataset}, and applies three common preprocessing steps (Table~\ref{tab:pipeline-steps}). 
Before preprocessing, sample sizes range from 0.1MB to 1MB, with an average of 0.8MB. After preprocessing, the sample size increases to between 4MB and 12MB, averaging 7MB.
We select this task as a representative, widely used computer vision benchmark with relatively lightweight preprocessing. 
Compared to image segmentation, its processing time is lower, which helps evaluate whether \sys can still yield benefits when preprocessing is less accentuated.

\paragraph{Speech Recognition}
This workload performs speech-to-text transcription using the RNN-T model~\cite{rnnt}, trained on the 228GB LibriSpeech dataset~\cite{librispeech}, which contains approximately 1,000 hours of English speech. 
As summarized in Table~\ref{tab:pipeline-steps}, the preprocessing pipeline transforms raw audio inputs, ranging from 0.06MB to 0.34MB (average 0.2MB), into spectrograms of size 0.4MB to 9MB (average 4MB).

We include this workload to evaluate \sys on a different data modality, namely audio, and to demonstrate that its benefits generalize beyond image-based tasks. 
Additionally, we design this workload as a microbenchmark, in order to showcase \sys's performance under heavy compute: all samples undergo a \emph{LightStep} transformation that takes 0.5 seconds and simulates lightweight preprocessing, such as volume normalization or frame splicing. 
Every fifth sample is additionally subjected to a \emph{HeavyStep} transformation that simulates more compute-intensive steps, taking either 3 seconds (Speech-3s) or 10 seconds (Speech-10s). 
These may include complex filtering or augmentation, such as applying long-context time-stretching or multi-pass spectrogram enhancement.

%----------------------------------------------------
%\input{motivation-ricardo}
\section{Issues in the Data Preprocessing Pipeline}
\label{sec:motiv}

Modern accelerators can ingest data at TB/s rates~\cite{nvidia_a100}, yet existing data loaders often fail to feed data to the accelerators at a matching pace.
In this section, we present a comprehensive experimental study to uncover the bottlenecks introduced by the data preprocessing pipeline during training. 

\paragraph{Hardware and OS configurations}
\label{sec:motiv-testbed}
We run experiments in two hardware configurations: \emph{Config. A} is a server with two 64-core AMD EPYC processors, 512~GB of memory, 4$\times$ A100 40~GB NVIDIA GPUs, connected to a shared Lustre file system via a 200~Gb/s interconnect, using Rocky Linux 8; \emph{Config. B} corresponds to a server with two 40-core Intel Xeon processors, 512GB of memory, 8$\times$ V100 32GB NVIDIA GPUs, and a 7TB NVMe SSD, using Ubuntu 20.04. 
%\rgmacedo{Do we want to rename the configurations to Deucalion and DGX server, or shall we keep it simple?}

\paragraph{Workloads}
We use three representative workloads from the MLPerf Training Benchmark, with different models and preprocessing pipelines, as described in \cref{subsec:data-prep-pipeline} and Table~\ref{tab:pipeline-steps}.

\paragraph{Data loading frameworks}
The experiments were conducted over three ML data loading frameworks presented in \cref{sec:background-dataloaders}: PyTorch DataLoader~\cite{pytorch_dataloaders_tutorial}, DALI~\cite{dali}, and Pecan~\cite{pecan}.

% ------------

\input{floaters/motivation-persample-preprocessing}

\subsection{Preprocessing Time Variability}
\label{subsec:motivation-preprocessing-variability}

We begin by exploring the variability in preprocessing times across individual data samples with the PyTorch DataLoader.
Figure~\ref{fig:motivation} depicts the preprocessing time of 25 randomly selected samples for two workloads: \emph{Image Segmentation} (Figure~\ref{subfig:variability-image-segmentation}) using the KiTS19 dataset and the 3D-UNet model, and \emph{Object Detection} (Figure~\ref{subfig:variability-object-detection}) using the COCO dataset with the R-CNN model.
While the average preprocessing time is approximately 500ms for image segmentation and 35ms for object detection, individual samples show wide variability, ranging from 10ms to 2.5s in image segmentation and from 10ms to 200ms in object detection, despite undergoing identical transformation pipelines.
Table~\ref{tab:motiv-preproc-times} presents these statistics in more detail for the two workloads across the full datasets. 
The long-tailed latencies, observed in Figure~\ref{fig:motivation} and Table~\ref{tab:motiv-preproc-times} (P90 column), result from a combination of input heterogeneity (\emph{e.g.,} varying resolutions, image sparsity, compression formats), transformation logic, and randomized data augmentations triggered by only a subset of samples.

Despite this variability, existing data loaders treat all samples equally during batch creation. 
For instance, in the Image Segmentation pipeline in Table~\ref{tab:pipeline-steps}, \texttt{RandomCrop} is the first and the slowest transformation ($338ms$ on average). However, because transformations are applied sequentially, \texttt{RandomCrop} will dictate the total time for the entire pipeline, cause head-of-line blocking (\cref{subsec:motivation-hol-blocking}), and exacerbate the variability.

\input{floaters/motivation-preprocessing-table}
\vspace{10pt}
\begin{takeaway} 
    Data loaders should take into account the variability of the preprocessing time across samples for an efficient training pipeline.
    % to organize the training pipeline. 
\end{takeaway}

\subsection{Predicting Sample Processing Time}
\label{subsec:motivation-heuristics}

We now explore the effectiveness of applying heuristics to predict sample processing cost, based on \emph{image size} and \emph{transformation reordering}. Intuitively, image size could be used as a proxy for data preprocessing time. We show that even though in some cases this heuristic is accurate, it does not generalize across workloads. The transformation reordering heuristic is introduced by Pecan~\cite{pecan}.
For these experiments, we extended PyTorch DataLoader with a custom load balancer that implements both heuristics.
Figure~\ref{fig:loadbalancer_heuristic} depicts the CPU and GPU usage over time of each setup.

\input{floaters/motivation-heuristics}
% \begin{figure}[t]
%     \centering
%     \begin{subfigure}[t]{0.48\columnwidth}
%         \centering        
%         \includegraphics[width=\linewidth]{figures/cpu_gpu_objectdetction_imagesizeheurestic (4).pdf}
%         \vspace{-17.5pt}
%         \caption{Image size.}
%         \label{subfig:heuristic-image}
%     \end{subfigure}
%     \hfill
%     \begin{subfigure}[t]{0.48\columnwidth}
%         \centering
%         \includegraphics[width=\linewidth]{figures/cpu_gpu_objectdetction_PECAN (2).pdf}
%         \vspace{-17.5pt}
%         \caption{Transformation reordering~\cite{pecan}.}
%         \label{subfig:heuristic-transformation}
%     \end{subfigure}
%     \vspace{-7.5pt}
%     \caption{CPU and GPU usage of the Object Detection workload when using two heuristics: (a) image size and (b) transformation reordering.}
%     \vspace{-15pt}
%     \label{fig:loadbalancer_heuristic}
% \end{figure}

\paragraph{Image size}
In the image segmentation workload, we observe a strong correlation between input sample size and preprocessing time.
For instance, a 299~MB sample takes 1.4s to preprocess, while a 63~MB sample requires only 0.3s.
This correlation stems from the significant variation in size of 3D medical images, as well as transformations such as \texttt{RandomCrop} and \texttt{RandomBrightness} whose costs scale with the input dimensions.
However, this correlation does not generalize.
In the object detection workload, for instance, a 408~KB image may be preprocessed in just 13ms, while a 220~KB image may take 155ms. 
As shown in Figure~\ref{subfig:heuristic-image}, applying the image size heuristic in this context fails to accurately distinguish between fast and slow samples, leading to increased fluctuations in GPU usage. 

\paragraph{Transformation reordering}
We evaluated a transformation reordering heuristic inspired by Pecan.
Under the object detection workload, the heuristic modifies the position of the \texttt{Resize} transformation based on its effect on sample size: if \texttt{Resize} increases the data size, it is applied at the end in the pipeline; if it reduces the data size, it is placed at the beginning. 
The rationale is to minimize the volume of data processed by subsequent transformations.
However, as shown in Figure~\ref{subfig:heuristic-transformation}, this reordering had limited impact: the GPU utilization is only $\approx$3\% better compared to PyTorch DataLoader (Figure~\ref{fig:cpu-gpu-usage a100}). 
This is because Pecan is designed for disaggregated environments where the primary goal is to reduce training cost across distributed compute nodes. 
The same strategy, however, is not guaranteed to succeed in a single-server, multi-GPU setting, on workloads with high variability in preprocessing times, as the reordering of transformations does not address the batch creation blocking.

\begin{takeaway}
    Even though simple heuristics may work for specific situations, they do not generalize across workloads. More generic mechanisms are needed to distinguish between fast and slow sample processing.
\end{takeaway}
%\vspace{-1.em}

% ------------

\subsection{Head-of-Line Blocking}
\label{subsec:motivation-hol-blocking}

We now analyze head-of-line blocking during data preprocessing. 
We conducted experiments using PyTorch Da\-ta\-Loa\-der, configured with 12 parallel worker threads, while training the image segmentation workload.

Figure~\ref{fig:motiv-gpu-idle} shows that both CPU and GPU resources operate in an inverse relation, where periods of  CPU activity align with GPU idleness. 
Indeed, the GPU frequently remains underutilized, with an average usage of 57.4\%.
This inefficiency stems from the sequential nature of the PyTorch DataLoader preprocessing pipeline, as depicted in Figure~\ref{fig:pytorch-pipeline}. 
Specifically, each worker fetches data samples from persistent storage and applies the corresponding transformations. 
The order in which samples are processed and grouped in batches is predetermined.
However, as observed in \cref{subsec:motivation-preprocessing-variability}, while most samples are processed quickly, a single slow sample can delay the entire batch from being moved to the GPU. 
Because batch construction is synchronous, training cannot proceed until all samples in the batch are processed, causing the GPU to be idle for a significant portion of the training time.

\begin{takeaway}
    Sequential execution of preprocessing and training is inefficient and results in high GPU idleness and, therefore, worse end-to-end training times.
    %due to head-of-line blocking caused by high variability in sample preprocessing times.
\end{takeaway}
% ------------

\subsection{Tuning Prefetching Parameters}
\label{subsec:motivation-prefetching}

\input{floaters/motivation-prefetch-factor}
% \begin{figure}[t]
%     \centering
%     \begin{subfigure}[t]{0.48\columnwidth}
%         \centering
%         \includegraphics[width=\linewidth]{figures/prefetch_factor_pyto.pdf}
%         \vspace{-17.5pt}
%         \caption{Prefetch factor in PyTorch.}
%     \end{subfigure}\hfill
%     \begin{subfigure}[t]{0.48\columnwidth}
%         \centering        
%         \includegraphics[width=\linewidth]{figures/prefetch_queue_depth.pdf}
%         \vspace{-17.5pt}
%         \caption{Prefetch queue depth in DALI.}
%     \end{subfigure}
%     \vspace{-7.5pt}
%     \caption{Impact of prefetch parameter on training time. Increasing the number of batches pre-fetched does not improve the training in both (a) Pytorch and (b) DALI.
%     }
%     \label{fig:pref_factor_impact}
% \end{figure}

To improve data loading performance, data loaders often implement prefetching strategies to load data samples in advance.
In the PyTorch DataLoader, the \texttt{prefetch\_factor} parameter determines how many batches each worker loads in advance from persistent storage.
In DALI, the \texttt{prefetch \_que\-ue\_depth} controls the number of batches buffered between pipeline stages.
To understand the impact of such mechanisms under heavy preprocessing workloads, we varied these parameters across the image segmentation, speech, and object detection workloads.

Figure~\ref{fig:pref_factor_impact} depicts the overall training time under different configurations.
While both mechanisms aim to overlap data loading with training, results show that increasing their values yields limited benefits because neither system reduces the per-sample transformation cost.
In the PyTorch DataLoader, increasing the prefetch factor offers minimal (even almost nil) improvement and may cause out-of-memory (OOM) errors due to excessive buffering. 
In addition, in DALI, where preprocessing is offloaded to the GPU, higher prefetch queue depth increases GPU memory usage and preprocessing load, which can interfere with training computations and ultimately prolong overall training time.

\begin{takeaway} 
    Existing mechanisms, such as prefetching, fail to improve GPU usage and end-to-end training time under sample preprocessing time variability. 
\end{takeaway}

% ------------

\subsection{Moving Preprocessing to GPU}
\label{subsec:motivation-gpu-preprocessing}

To improve the data pipeline throughput, DALI offloads the data preprocessing phase to the GPU. This approach helps maintain high GPU utilization across workloads (as observed in Figure~\ref{fig:cpu-gpu-usage a100}), contrasting with the underutilization often observed in PyTorch DataLoader. However, sharing the GPU between preprocessing and training leads to resource contention, which can interfere with the overall training performance, especially in compute-intensive workloads. While dedicating a separate GPU for preprocessing is a possible workaround, it reduces the number of GPUs available for training, an unfavorable compromise given that training is typically more compute-demanding. 

\begin{takeaway} 
   If possible, it is better to avoid data preprocessing on the GPU, to be able to allocate the more scarce GPU cycles to training.
\end{takeaway}

%----------------------------------------------------
%\input{contribution_rahma}
\section{Design and Implementation}
\label{sec:contrib}

\sys is a general-purpose data loader that proactively prepares training data to enable efficient batch construction, without requiring prior knowledge of the dataset or preprocessing pipeline. 
It is designed for local training scenarios (\emph{i.e.,} single node, multiple GPUs) and is provided as a drop-in replacement of PyTorch DataLoader, following the same API. 
\sys also supports distributed training (\cref{sec:discussion}). 
Next, we present an overview of the \sys architecture (\cref{sec:speedy-summary}), detail the key techniques that underpin its design (\cref{sec:hol-blocking}-\cref{sec:accelerate-training}), and discuss implementation details (\cref{sec:implem}).

\subsection{\sys in a Nutshell}
\label{sec:speedy-summary}

\input{floaters/design-overview}
% \begin{figure}[t]
%     \centering
%     \includegraphics[width=\linewidth]{figures/minatoloader-alternative.pdf}
%     \vspace{-1.5em}
%     \caption{\sys high-level design. It continuously enqueues preprocessed samples into a specified queue based on a load balancer decision. Concurrently, the GPU dequeues preprocessed data for training. We show \sys for one GPU, but it can generalize to multi-GPU settings.}
%     \vspace{-2mm}
%     \label{fig:speedy-asyncdata}
% \end{figure}

Figure~\ref{fig:speedy-asyncdata} depicts the high-level architecture of \textsc{Mi\-na\-to\-Loa\-der}, which is built around three key stages designed to prevent pipeline stalls and maximize GPU utilization: data loading, preprocessing, and batch construction. 
For clarity, we show how \sys operates for one GPU, but it generalizes to multi-GPU settings, as shown in our experiments (\cref{sec:eval}).
First, the data loading stage is identical to PyTorch DataLoader, following the same API.
Second, at the core of the system is a \emph{dynamic, sample-aware load balancer} (\cref{sec:hol-blocking}) designed for mitigating head-of-line blocking caused by slow preprocessing samples (\cref{sec:motiv}) by classifying samples as fast or slow on-the-fly during training. 
To achieve this, it maintains four queues: a \emph{fast queue} for quick preprocessing samples, a \emph{slow queue} for samples exceeding a predefined preprocessing timeout, a \emph{batch queue} that assembles training batches using both fast and slow queues, and a \emph{temp queue} used to temporarily store long samples that need to be preprocessed off the critical path.
\sys maintains one \emph{batch queue} per GPU, and separate \emph{fast}, \emph{slow}, and \emph{temp} queues for each CPU worker.
These queues enable decoupling data preparation from batch construction, allowing \sys to prioritize readily available samples and delay slower ones without stalling the pipeline.
Further, \sys includes a \emph{worker scheduler} (\cref{sec:accelerate-training}) that dynamically tunes the number of CPU preprocessing workers to match the throughput demands of the training workers on the GPUs (omitted in Figure~\ref{fig:speedy-asyncdata} for clarity).

% \blueblock{
% \sys requests samples in the same random order as PyTorch's DataLoader. The difference is that \sys builds batches from samples that complete preprocessing first, without imposing any fixed order. We emphasize that slow samples are not preempted: they continue preprocessing in the background and are added to future batches as soon as they are ready. \sys does not push the preprocessing of slow samples until the last batches. Our strategy leads to randomized batches similar to PyTorch DataLoader, as shown in Section~\ref{sec:eval}.
% }

% \blueblock{
Like PyTorch, \sys requests samples in random order, but it builds batches from whichever samples finish preprocessing first, rather than imposing a strict order.
Additionally, slow samples are never preempted: they continue preprocessing in background and are added to future batches as soon as they are ready.
Importantly, \sys does not defer these samples to the very end, ensuring that its randomized batches remain statistically similar to PyTorch Dataloader, as shown in \cref{subsec:sensitivity-analysis}.
% }

\paragraph{Operation flow}
Figure~\ref{fig:speedy-asyncdata} depicts the end-to-end workflow of \sys.
The process begins with data samples being fetched from persistent storage by preprocessing workers (\ding{202}).
Once loaded, each sample undergoes the model's preprocessing pipeline, where a sequence of transformations is applied in parallel by CPU worker threads (\ding{203}). 
As discussed in \cref{subsec:motivation-preprocessing-variability}, preprocessing times can vary significantly across samples. 
To overcome this, \sys's \emph{load balancer} applies a per-sample timeout to classify samples as fast or slow (\ding{204}). 
At the same time, \sys maintains light-weight profiling to adjust the threshold used to determine whether the samples are fast or slow (\ding{192}, explained in \cref{sec:hol-blocking}).
Samples that complete preprocessing within the timeout period are inserted into the \emph{fast queue}, while those that exceed it are moved to the \emph{temp queue}, where they continue applying the transformations (\ding{205}). 
In the latter, as soon as slow samples finish preprocessing (occurring in parallel with the rest of the pipeline), they are transferred to the \emph{slow queue}, as they are ready to be included in the batch construction (\ding{206}).
After that, a dedicated \emph{batch queue} (one per GPU) constructs training batches by fetching samples from both fast and slow queues (\ding{207}).
Contrary to prior work, this design ensures slow samples do not block batch creation (\emph{i.e.,} head-of-line blocking).
Concurrently, multiple GPU worker threads continuously dequeue ready-to-train batches from the \emph{batch queue} (\ding{207}). 
This process overlaps with data loading and preprocessing to ensure GPUs are constantly fed with data for training, minimizing idle periods. 
Finally, in the background, \sys's \emph{worker scheduler} continuously monitors queue occupancy and CPU utilization, and dynamically adjusts the number of CPU preprocessing workers and GPU workers to sustain high training performance.

%------------------------------------
\subsection{Avoiding Head-of-Line Blocking}
\label{sec:hol-blocking}

\sys integrates the takeaways from \cref{sec:motiv} by designing a load balancer to classify samples. Initially, \sys optimistically assumes that all samples are fast. 
If a sample takes longer than a predefined timeout, it is flagged as slow.
At that point, its preprocessing is paused, and the sample is migrated to the \emph{temp queue}, where its processing continues in the background.

Figure~\ref{fig:speedy-pipeline} illustrates an example of this process, where a slow sample (\emph{e.g.,} \texttt{sample 2}) is migrated during execution to the \emph{slow queue}, and later included in a subsequent training batch.
We show that even though the ordering of the samples in each batch can slightly change when compared to the default PyTorch DataLoader (Figure~\ref{fig:pytorch-pipeline}), the accuracy of the model training is not impacted. 
Intuitively, the training accuracy follows a similar trend when using \sys because, in any case, samples need to be shuffled during training.
This decoupled design enables the GPU to remain consistently busy by eliminating the dependency between the slowest sample and the training pipeline. 

The slow sample timeout is determined through a simple offline workload profiling, used to make an educated guess regarding the cutoff threshold between fast and slow samples. 
However, if the dataset sample used for the offline profiling is not representative of the entire workload, or if the workload drifts with time, \sys is able to automatically adjust the threshold.
Before training begins, \sys performs a warm-up by running the model for a configurable period (\emph{e.g.,} 10 minutes in our configuration). 
During this phase, it collects profiling statistics for each sample, including sample size, preprocessing time per transformation, total preprocessing time, and the number of transformations applied. 
At the end of the warm-up, the load balancer analyzes the recorded data and computes the 75th percentile of the total preprocessing times. 
This percentile is used as the default timeout value $t_{out}$ by the load balancer, moving only the 25\% slowest samples to \emph{temp queue}.
Unlike the median, which would split the dataset evenly, the 75th percentile strikes a better balance across workloads, focusing on true outliers and ensuring that \emph{slow queue} remains smaller compared to \emph{fast queue}.
Although \sys uses the 75th percentile by default,
%work well across workloads, 
the threshold $t_{out}$ is adjustable. 
Indeed, if \sys detects that too many samples are incorrectly classified as slow (\emph{e.g.,} due to a skewed distribution), it can automatically fall back to the 90th percentile. 
Moreover, \sys continues running the profiler in the background, in parallel with the preprocessing, adjusting the threshold as needed. 
This adaptive mechanism allows \sys to dynamically tune its timeout value for robust performance.

\input{floaters/minatoloader-pipeline-diagram}
% \begin{figure}[t]
%     \centering
%     \includegraphics[width=1\columnwidth]{figures/minatoloader_diagram_finale.pdf}
%     \vspace{-15pt}
%     \caption{\sys preprocessing pipeline. Slow sample \#2 does not delay the batch creation and gets migrated to the \emph{temp queue} (red arrow). The fast and slow queues in the batch construction are omitted for brevity.}
%     \label{fig:speedy-pipeline}
% \end{figure}

\begin{algorithm}[t]
    \caption{Load Balancer in \sys}
    \label{alg:hol}
    \small
    
    \KwIn{Transforms $T = [t_0, \dots, t_{n-1}]$, sample $s$, timeout $t_{out}$}
    \SetKwBlock{Queues}{\textbf{Queues}}{end}
    
    \Queues{
        $fast\_queue$ \tcp*[f]{Samples completed within timeout} \\
        $temp\_queue$ \tcp*[f]{Partially processed samples + transformation index} \\
        $slow\_queue$ \tcp*[f]{Completed timed-out samples} \\
        $batch\_queue$ \tcp*[f]{Batches ready for training}
    }
    
    $start \gets$ current time \;
    $partial \gets s$ \tcp*[f]{Initialize sample}
    
    \For{$i = 0$ to $n - 1$}{
        $partial \gets t_i(partial)$ \;
        \If{elapsed time since $start > t_{out}$}{
            $temp\_queue$.put($(partial, i)$ )\tcp*[f]{Timeout}
        }
    }
    
    $fast\_queue$.put($partial$) \tcp*[f]{Finished within timeout}
    \vspace{0.5em}
    
    \SetKwBlock{Background}{\textbf{Background Worker: Resume Timed-out Samples}}{end}
    \Background{
        % \While{not stopped}{
        \While{true}{
            $(partial, i)$ = $temp\_queue$.pop() \;
            \For{$j = i$ to $n - 1$}{
                $partial \gets t_j(partial)$
            }
            $slow\_queue$.put($partial$) \tcp*[f]{Now preprocessed}
        }
    }
    \vspace{0.5em}
    
    \SetKwBlock{BatchBuilder}{\textbf{Batch Construction Thread}}{end}
    \BatchBuilder{
        % \While{not stopped}{
        \While{true}{
            $batch \gets []$ \;
            \While{len($batch$) $<$ batch\_size}{
                \uIf{$fast\_queue$ not empty}{
                    sample $\gets$ fast\_queue.pop() \;
                }
                \ElseIf{$slow\_queue$ not empty}{
                    sample $\gets$ slow\_queue.pop() \;
                }
                \Else{
                    \textbf{sleep(t)};   \tcp*[f]{Not enough samples} \
                    \Continue
                }
                append sample to $batch$ \;
            }
           $batch\_queue$.put($batch$)
        }
    }
    \vspace{0.5em}
    \SetKwBlock{NextCall}{\textbf{\sys \texttt{\_\_next\_\_}() Call}}{end}
    \NextCall{
    %\While{not stopped}{
    \While{true}{
        \If{$batch\_queue$ not empty}{
            $batch \gets$ batch\_queue.pop() \;
            \Return $batch$ \tcp*[f]{Send to GPU}
        }
        \Else{
            \textbf{sleep(t)} ; \tcp*[f]{Batch queue empty} \
        }
    }
}
\end{algorithm}

Algorithm~\ref{alg:hol} illustrates the core logic of \sys's load balancer. 
Given a sample $s$ and a list of transformations $T$, \sys applies the transformations sequentially while monitoring the elapsed time (\texttt{7-10}). 
If all transformations are completed within the timeout budget $t_{out}$, the sample is enqueued into the \texttt{fast\_queue} (\texttt{12}). 
If the timeout $t_{out}$ is exceeded, the system interrupts preprocessing, records the index of the transformation that was in progress, and enqueues it along with its index into the \texttt{temp\_queue} (\texttt{11}). 
Since the last transformation was only partially applied, it must be re-executed to ensure that the sample is in a valid state to carry on the remaining transformations. 
This design avoids restarting the entire data preprocessing pipeline.

Afterwards, a background worker resumes the preprocessing of the slow samples from the recorded index and, once complete, places them into the \texttt{slow\_queue} (\texttt{14-18}).
Meanwhile, a batch construction thread continuously builds batches by pulling from both \texttt{fast\_queue} and \texttt{slow\_queue} (\texttt{22-30}).
Finally, \sys continuously dequeues ready-to-train batches from the \texttt{batch\_queue} (\texttt{32-37}). 
Empirically, we determine 10\,ms as the optimal sleep time (\texttt{28} and \texttt{37}).

\subsection{Keeping the GPUs Busy}
\label{sec:accelerate-training}

In addition to mitigating head-of-line blocking, \textsc{Mi\-na\-to\-Loa\-der} also ensures that the GPUs stay busy by automatically determining the right number of CPU workers needed to keep up with each GPU. Each CPU worker is mapped to one CPU core.
The optimal number of CPU workers can vary across workloads due to differences in data loading and preprocessing complexity. This is because preprocessing time is not uniform; some workloads involve more compute-intensive or variable transformations, which affect how many workers are needed to sustain throughput.

The CPU workers include the number of data loading workers, the slow-task workers, and the batch workers.
The data loading CPU workers are responsible for loading data from disk, preprocessing fast samples, and moving slow samples to the \emph{temp queues}.
Slow-task CPU workers run in the background, without blocking the pipeline.
The batch CPU workers pull samples from the \emph{slow} and \emph{fast queues} to construct batches ahead of time and ensure that the GPU workers always have prebuilt batches to ingest. In addition, \sys incorporates a prefetching mechanism that uses a CUDA stream to transfer the batch $i$ from the \emph{batch queue} to the GPU memory before the GPU has finished executing the batch $i-1$, which helps achieve continuous GPU training.

\sys starts by setting the number of CPU workers to 12 and \texttt{$max\_workers$} to the total number of CPU cores. 
The initialization for the CPU workers can be tuned further depending on system characteristics such as dataset size and data preprocessing complexity.  
While the workload is running, \sys adjusts the number of CPU workers by monitoring average batch queue size and CPU utilization. 
The number of CPU workers is updated using Formula~\ref{eq:dynamic-workers}:
%\vspace{-0.8em}
\begin{equation}
    \texttt{workers} = \min(\texttt{max\_workers}, \max(1,\ \texttt{workers'} + \Delta))
    \label{eq:dynamic-workers}
\end{equation}
% \vspace{-1em}
\begin{equation}
    \Delta = \alpha \cdot \left(1 - \frac{Q_{\text{size}}}{Q_{\text{max}}}\right) + \beta \cdot (C_{\text{usage}} - \theta_c)
    \label{eq:dynamic-workers-adjust}
\end{equation}

\noindent where:
\( Q_{\text{size}} \) is the moving average of the queue size,
\( Q_{\text{max}} \) is the maximum queue capacity,
\( C_{\text{usage}} \) is the average CPU utilization (normalized to [0,1]),
\( \theta_c \) is the CPU threshold (\emph{e.g.}, 0.7),
and \( \alpha \) and \( \beta \) are scaling factors controlling the sensitivity of queue and CPU adjustments, respectively.

Intuitively, Formula~\ref{eq:dynamic-workers} increases the number of CPU workers when the queues are frequently empty \((Q_{\text{size}} \ll Q_{\text{max}})\) and/or CPU utilization is high, indicating a CPU bottleneck. 
Conversely, it reduces the number of workers when the queues are full and CPU usage is low, which indicates over-provisioning. 
Formula~\ref{eq:dynamic-workers-adjust} clips the final value of \(\Delta\) to a small integer range (\emph{e.g.,} \([-2, +2]\)) to avoid instability.

%----------------------------------------------------
\subsection{Implementation} 
\label{sec:implem}

\sys seamlessly integrates with PyTorch DataLoader interface, requiring no annotations or modifications to run, and preserves existing interfaces.
It is implemented using 750 LoCs of Python: 350 LoCs for the data loading implementation and 300 LoCs for the data preprocessing.

\paragraph{Integration with PyTorch}
Our implementation is based on PyTorch 2.7 and uses 
\texttt{torch.multiprocessing.Process} to launch processes, 
and \texttt{torch.multiprocessing.Queue} for the queues. 
Unlike threads, \texttt{torch.multiprocessing} provides isolated memory spaces.
However, PyTorch provides support for sharing CPU tensors via shared memory. 

\texttt{torch.multiprocessing} uses the \texttt{spawn} start method by default (which is safer for CUDA operations) and bypasses the Global Interpreter Lock (GIL). 
This allows full parallelism across cores, especially when executing Python-based data transformations.
Another approach would be an implementation in C or C++ using \texttt{Pybind11} to further accelerate data loading. However, it requires taking the GIL to access PyTorch preprocessing libraries, effectively introducing a synchronization bottleneck~\cite{GIL_problems}. A final approach would be to reimplement all preprocessing operations in C/C++ and bypass PyTorch altogether; however, while likely improving performance, we argue that this approach is not practical for end users who typically call PyTorch libraries when defining their data preprocessing pipelines.  

The \texttt{torch.multiprocessing.Queue} provides a process-safe, first-in-first-out (FIFO) communication channel between multiple producers and consumers. 
When multiple producers concurrently invoke \texttt{Queue.put()}, the operations are synchronized via internal locks, ensuring atomic insertion while preserving ordering. 
Similarly, multiple consumers calling \texttt{Queue.get()} block until items are available, with the operating system's scheduler determining which consumer retrieves the next item. 

%----------------------------------------------------
%\input{evaluations}
\section{Evaluation}
\label{sec:eval}

Our evaluation answers the following questions:

\begin{itemize}[leftmargin=*]
    \item \textbf{Q1:} Can \sys improve training time and GPU utilization across different workloads (\cref{sec:eval-training-throughput}--\cref{sec:eval-gpu-usage})?
    
    \item \textbf{Q2:} How does \sys scale under varying hardware configurations (\cref{sec:eval-varying-gpu})?
    
    \item \textbf{Q3:} How does \sys behave under memory-cons\-trai\-ned environments (\cref{sec:eval-mem-constraint})?
    
    \item \textbf{Q4:} Does \sys impact model accuracy (\cref{subsec:sensitivity-analysis})?
    
    \item \textbf{Q5:} How does \sys behave under varying proportions of slow samples (\cref{subsec:sensitivity-analysis})?
\end{itemize}

%----------------------------------
\subsection{Experimental Environment}
\label{sec:eval-setup}

\paragraph{Hardware and OS configurations}
We used the same hardware and OS configurations described in \cref{sec:motiv-testbed}.

\paragraph{Data loaders}
Experiments were executed using the PyTorch ML framework, configured under four distinct data loaders (\cref{sec:background-dataloaders}) -- PyTorch DataLoader~\cite{pytorch_dataloaders_tutorial}, DALI~\cite{dali}, Pecan~\cite{pecan}, and \sys.
While Pecan was originally developed for TensorFlow, we reimplemented its AutoOrder policy (\cref{sec:background-dataloaders}) in PyTorch to ensure a consistent and fair comparison across all systems.
As this paper targets single-node multi-GPU scenarios, we have not used Pecan's AutoPlacement policy.

\paragraph{Workloads and data loaders tuning}
Experiments were conducted over four workloads from the MLPerf Training Benchmark~\cite{mlperf-training-results}, including image segmentation (3d-UNet), object detection (R-CNN), and speech recognition (RNN-T) (\cref{subsec:data-prep-pipeline}).
The original pipeline of each workload is described in Table~\ref{tab:pipeline-steps}.
Table~\ref{tab:model-config-train-time} depicts the training configurations used for each model.
We did our best to tune each system according to the specific characteristics of each workload's pipeline, as described below (unless stated otherwise).

\input{floaters/eval-model-configs-table}

\input{floaters/eval-throughput.tex}
% \begin{figure*}[t]
%     \centering
%      \begin{subfigure}[t]{0.24\textwidth}
%         \includegraphics[width=\linewidth, height=.65\textwidth]{figures/imseg_A100.pdf}
%         \caption{Img. Seg. (3D-UNet)}
%         \label{subfig:a100-throughput-image}
%     \end{subfigure}
%      \begin{subfigure}[t]{0.24\textwidth}
%         \includegraphics[width=\linewidth, height=.65\textwidth]{figures/throughput_object_detection_A100_1.pdf}
%         \caption{Obj. Det. (R-CNN)}
%         \label{subfig:a100-throughput-object}
%     \end{subfigure}
%      \begin{subfigure}[t]{0.24\textwidth}
%         \includegraphics[width=\linewidth, height=.65\textwidth]{figures/throughput_speech_3s_A100.pdf}
%         \caption{Speech-3s (RNN-T)}
%         \label{subfig:a100-throughput-speech-3}
%     \end{subfigure}
%      \begin{subfigure}[t]{0.24\textwidth}
%         \includegraphics[width=\linewidth, height=.65\textwidth]{figures/throughput_speech_10s_A100.pdf}
%         \caption{Speech-10s (RNN-T)}
%         \label{subfig:a100-throughput-speech-10}
%     \end{subfigure}
   
%    % \vspace{-.7em}
%     \caption{Throughput (MB/s) of PyTorch DataLoader, Pecan, DALI, and \sys on four ML workloads using A100 GPUs (\emph{Config. A}). 
%     Solid vertical lines represent the end of the training for each data loader.}
%     \label{fig:throughput-a100}
% \end{figure*}

\emphparagraph{PyTorch DataLoader}
We set the number of worker threads to 12 and used a \texttt{pre\-fet\-ch\_factor} of 2, as increasing beyond these values had little effect on training time (\cref{subsec:motivation-prefetching}).

\emphparagraph{DALI}
We enabled DALI's \texttt{exec\_pipelined} and \texttt{exec\_async} flags, which overlap computation stages through buffering and enable asynchronous execution between the DALI backend and the Python front end, respectively.
Similarly to PyTorch DataLoader, we used the default \texttt{pre\-fet\-ch\_que\-ue\_dep\-th} of $2$.
The number of worker threads was set to the total number of CPU cores available on the machine. 
Furthermore, to account for GPU-accelerated preprocessing and ensure a fair comparison, we measured the average execution time of speech workload transformations (\emph{e.g.,} spectrogram, normalization) in both PyTorch DataLoader and DALI under identical conditions. 
We found that DALI was consistently ~10$\times$ faster. 
Based on this, we scaled down the \emph{LightStep} and \emph{HeavyStep} transformations (Table~\ref{tab:pipeline-steps}) by a factor of 10 for DALI: \emph{LightStep} was set to 0.05s, and \emph{HeavyStep} to 0.3s (Speech-3s) and 1s (Speech-10s). 

\emphparagraph{Pecan}
In the speech recognition workload, Pecan's AutoOrder policy moves the \texttt{Pad} transformation to the end of the pipeline, as it is an inflationary transformation.
In the object detection workload, \texttt{Resize} is either inflationary or deflationary depending on the input size; thus, Pecan moves it to the end of the pipeline if it inflates the sample, or leaves it in the original position otherwise. 
For the image segmentation workload, the AutoOrder algorithm is not applied because the transformations are already optimally ordered.

\emphparagraph{\sys} 
The initial worker configuration is of 1 GPU worker per GPU and 12 CPU workers per GPU worker. 
We use a prefetch factor of 2, and set the sample timeout threshold to the 75th percentile of all samples' preprocessing time. 
All maximum queue sizes are set to 100. 

\paragraph{Collected metrics}
In all experiments, we measured the overall training time (in seconds) and model throughput (in MB/s), GPU and CPU utilization, memory usage, and disk read bandwidth.
We represent the model throughput as the cumulative size in MB of data samples trained per second.
We used the \texttt{nvidia-smi} tool from the NVIDIA Management Library~\cite{nvml} to monitor GPU utilization, and \texttt{dstat}~\cite{dstat} to collect CPU, memory, and disk usage.

%----------------------------------------------
\subsection{End-to-end Training Performance}
\label{sec:eval-training-throughput}

We start by evaluating the impact of \sys on end-to-end training performance.
Figures~\ref{fig:throughput-a100} and \ref{fig:training-time} depict the throughput and training time of all systems.
Due to space constraints, we omit the throughput plots for the \emph{Config.B} testbed, composed of V100 GPUs. 
For the image segmentation workload, since the transformations are already ordered, Pecan's results are identical to those of PyTorch DataLoader and therefore omitted.

Results show that \sys consistently achieves the highest throughput across all workloads.
In the image segmentation workload (Figure~\ref{subfig:a100-throughput-image}), \sys improves throughput by 2.5$\times$ over PyTorch DataLoader and $1.3\times$ over DALI.
For object detection (Figure~\ref{subfig:a100-throughput-object}), \sys increases the throughput of PyTorch DataLoader and Pecan up to 2$\times$ and 1.6$\times$ of DALI.
In the speech recognition workloads (Figures~\ref{subfig:a100-throughput-speech-3} and \ref{subfig:a100-throughput-speech-10}), which comprise the most time-consuming transformations, \sys outperforms PyTorch DataLoader and Pecan by 3.5$\times$ to 5.5$\times$, and DALI by 2$\times$ on average.
These performance gains stem from its ability to eliminate head-of-line blocking in the pipeline by prioritizing fast samples and minimizing GPU idleness.

\input{floaters/eval-resource-usage}

Moreover, as depicted in Figure~\ref{fig:training-time}, \sys achieves the shortest training times across all workloads.
Over the \emph{Config.A} testbed (4$\times$A100 GPUs), \sys improves training time up to 7.5$\times$, 4.9$\times$, and 3$\times$, when compared to PyTorch DataLoader, Pecan, and DALI, respectively.
Over \emph{Config.B} testbed (8$\times$V100 GPUs), using an older GPU architecture, \sys still achieves substantial gains, improving training time up to 4.9$\times$ compared to PyTorch DataLoader and Pecan, and 2.6$\times$ over DALI.

%---------------------------------

\subsection{Compute Resources Usage}
\label{sec:eval-gpu-usage}

%\input{floaters/eval-resource-usage}
% \begin{figure}[t]
%     \centering
%     %\vspace{-.5em}
%     \includegraphics[width=\linewidth]{figures/cpu_gpu_usage_allworkloads-cropped-4.pdf}
%     \caption{CPU and GPU usage for all systems across all workloads using 4$\times$A100 GPUs.}
%     \label{fig:cpu-gpu-usage a100}
%     \vspace{-3mm}
% \end{figure}

\input{floaters/eval-training-time-gpus}

We now compare the GPU and CPU utilization of \sys against the baselines. 
Due to space constraints, we report results for \emph{Config.A} (A100 GPUs) in Figure~\ref{fig:cpu-gpu-usage a100}; results for \emph{Config.B} follow similar trends.
Further, we omit Pecan from this analysis as its utilization closely mirrors that of PyTorch DataLoader, thus drawing similar conclusions.

Among the baselines, DALI achieves the highest GPU utilization by performing both preprocessing and training on the GPU, effectively maximizing usage. 
Similarly, \sys achieves a near-100\% GPU usage, averaging $90.45\%$ across all workloads.
% This is a $97\%$ improvement compared to PyTorch DataLoader.
However, unlike DALI, \sys achieves this without offloading preprocessing to the GPU, whose usage exclusively reflects training activity.
This highlights the effectiveness of \sys's load balancer and worker scheduler components in mitigating head-of-line blocking and minimizing GPU idleness. 

Moreover, we observe drops in DALI's GPU usage during the first epoch of the image segmentation workload, also manifested in its throughput (Figure~\ref{subfig:a100-throughput-image}). 
These drops stem from I/O contention on the shared filesystem in \emph{Config. A} server.
Specifically, as DALI aggressively loads and decodes data directly into GPU memory, it can temporarily saturate I/O bandwidth during the initial batch loading phase, creating read contention and pipeline stalls. 
Once data is cached, GPU usage stabilizes.
As for PyTorch DataLoader, it experiences poor GPU utilization (averaging 46.4\%) due to its synchronous behavior, causing head-of-line blocking. Finally, as expected, \sys exhibits slightly higher CPU utilization than PyTorch DataLoader (up to 20\% utilization), due to its load balancer and adaptive worker scheduler. 

%--------------------------------------

\subsection{Scalability}
\label{sec:eval-varying-gpu}

We now evaluate \sys's scalability by varying the number of GPUs from 1$\times$ to 4$\times$A100 (\emph{Config. A}, Figures~\ref{fig:training-time}a--\ref{fig:training-time}d), and from 2$\times$ to 8$\times$V100 (\emph{Config. B}, Figures~\ref{fig:training-time}e--\ref{fig:training-time}h).

As expected, training time decreases with more GPUs in all data loaders. 
However, \sys consistently outperforms all systems across all workloads and testbeds.
Moreover, some baselines fail to scale efficiently.
For instance, on 1$\times$A100, DALI trains Speech-3s 25.6\% faster than Pecan, but Pecan catches up on 3$\times$A100. 
In contrast, \sys not only maintains its improvement as GPU count increases, but even achieves comparable or better training performance with 1$\times$A100 than the baselines configured with 4$\times$A100, effectively addressing the dominant bottleneck.
For example, in the image segmentation workload, \sys on a single A100 reduces training time by up to 60\% compared to PyTorch DataLoader and DALI using all GPUs. 

%--------------------------------
\subsection{Performance Under Memory Constraints}
\label{sec:eval-mem-constraint}

\input{floaters/eval-memory-constraints}

% \begin{figure}[t]
%     \centering
%     \includegraphics[width=.9\columnwidth]{figures/mem constraint-cropped-5.pdf}
%     \vspace{-.6em}
%     \caption{CPU and GPU usage (left) and disk read (right) of the image segmentation workload with PyTorch DataLoader ($\approx$650s), DALI ($\approx$500s), and \sys ($\approx$330s) when training a 230GB dataset under an 80GB memory constraint.}
%     \label{fig:mem-constraint}
% \end{figure}

We now evaluate the effectiveness of \sys's design on memory-constrained environments.
We generated a 230GB dataset from KiTS19 by replicating it. 
We trained the image segmentation workload (3D-UNet) for 10 epochs, while limiting \emph{Config.B}'s memory to 80GB using Linux \texttt{cgro\-ups} (\emph{i.e.,} approximately one-third of the dataset size), forcing all data loaders to access persistent storage.
We measured the GPU and CPU usage and disk read bandwidth. 
Figure~\ref{fig:mem-constraint} presents the results.
Again, Pecan is omitted from these results because it behaves similarly to PyTorch. 

The PyTorch DataLoader (top row) experiences frequent memory pressure, resulting in continuous but volatile disk reads, reduced GPU usage (averaging around 57\%), and prolonged training time ($\approx$650 seconds). 
DALI (middle row) improves GPU usage (81.2\% on average) but exhibits multiple drops below 50\%, leading to a training time of around 500 seconds. 
To avoid substantial GPU usage drops, DALI performs data loading from persistent storage on the CPU.
On the other hand, \sys (bottom row) maintains consistently high GPU utilization ($82.1\%$ on average) and completes training in 330 seconds. 
Disk I/O activity remains stable and high (maximizing the NVMe bandwidth) with \sys, only showing periodic drops at the end of every epoch due to model validation. 
This demonstrates that \sys can handle memory-constrained scenarios thanks to its design that leverages multiple queues to pipeline data preprocessing and training.

% -------------------------------------

%-----------------------------------

\subsection{Sensitivity Analysis}
\label{subsec:sensitivity-analysis}

We now evaluate distinct features of \sys.

\input{floaters/eval-accuracy-batch-composition.tex}

\paragraph{Accuracy Preserving with \sys}
To demonstrate that \sys does not affect the model performance, we conducted experiments measuring the accuracy of the image segmentation (3D-UNet) and object detection (Mask R-CNN) workloads. 
Due to space limitations and because speech recognition was primarily used as a microbenchmark, we omit the training results.
Achieving good accuracy generally requires training for a large number of iterations.
For example, training the Mask R-CNN model requires between 90,000 and 180,000 iterations to reach convergence~\cite{massa2018mrcnn}, which is $\approx$14 days of training on our testbed.
Since the goal of \sys is not to improve the accuracy, it suffices to show here that it preserves the same model accuracy trend as other data loaders, while accelerating training.
We trained the Mask R-CNN model for $45,000$ iterations using PyTorch DataLoader and \sys. 
With \sys, the model reached the same 6\% accuracy but 60\% faster, completing in 5 hours and 12 minutes, compared to 13 hours and 55 minutes with PyTorch DataLoader. 
The results, shown in Figure~\ref{subfig:accuracy}, confirm that \sys maintains a similar accuracy to PyTorch DataLoader.

For image segmentation, we trained the model until accuracy stabilized, which required approximately 500 epochs. 
This took 8 hours and 2 minutes with PyTorch DataLoader and only 3 hours and 52 minutes with \sys. 
As illustrated in Figure~\ref{subfig:accuracy}, \sys preserves the full accuracy trend and reaches the same final accuracy of 58\%.

\paragraph{Batch composition analysis}
To complement the accuracy results, we conducted a batch composition study to validate that \sys's strategy does not introduce bias.
Specifically, we evaluated (i) the distribution of batches by the number of slow samples they contain (Figure~\ref{subfig:batchcomp}), \emph{i.e.}, the proportion of batches with a given number of samples, and (ii) the proportion of slow samples over training iterations (Figure~\ref{subfig:slow-samples}). 
Experiments were conducted on the object detection (R-CNN) and image segmentation (3D-UNet) workloads with a batch size of 4.

Results show that \sys constructs mixed batches comparable to PyTorch DataLoader and does not introduce systematic bias.
Figure~\ref{subfig:batchcomp} shows a similar distribution of batches across systems for both workloads, indicating that \sys preserves the natural ratio of slow samples.
Figure~\ref{subfig:slow-samples} further shows that slow samples are incorporated into batches as soon as they are ready, rather than being deferred or preempted. 
For example, PyTorch DataLoader yields an average proportion of slow samples of 0.15 and 0.23  under the object detection and image segmentation workloads, respectively, while \sys presents 0.17 and 0.24. 
% \rgmacedo{@rahma: please fill X and Y}
These results confirm that \sys preserves fairness in sample usage and maintains a batch distribution closely aligned with PyTorch DataLoader.
% Supporting the accuracy results (\ref{subfig:accuracy}), we conducted a batch composition study for two workloads, object detection and image segmentation. Specifically, we evaluated (i) the fraction of slow samples over training iterations (Figure~\ref{subfig:slow-samples}), and (ii) the distribution of batches by the number of slow samples they contain (Figure~\ref{subfig:batchcomp})—that is, the proportion of batches with a given number of slow samples. These experiments were run with a batch size of 4.

% The results, shown in Figure~\ref{fig:sensitivity-analysis}, confirm that \sys constructs mixed batches comparable to PyTorch and does not introduce any strong bias in how batches are formed. In particular, the batch composition plots~\ref{subfig:batchcomp} for both workloads show similar distributions across systems, indicating that our scheduling mechanism preserves the ratio of slow samples. Moreover, the timeline plots~\ref{subfig:slow-samples} demonstrate that slow samples are incorporated as soon as they are ready, rather than being deferred or preempted. This behavior ensures that \sys maintains fairness in sample usage while closely matching PyTorch. For object detection, the average fraction of slow samples, as shown in Figure~\ref{subfig:slow-samples} (left), is 0.15 for PyTorch and 0.17 for \sys, which further confirms that our approach does not change the distribution of batches.

\paragraph{Cluster of slow samples}
We now evaluate \sys under varying proportions of slow and fast samples.
To this end, we modified the Speech-3s workload so that the \emph{HeavyStep} transformation is applied to a configurable proportion of the dataset, rather than every five samples.
Experiments were conducted with increasing proportion of slow samples from 0\% up to 100\%.
Figure~\ref{fig:cluster-slow-samples} depicts the training time of PyTorch DataLoader, DALI, Pecan, and \sys.
As expected, in the edge cases (\emph{i.e.,} 0\% and 100\%), \sys performs similarly to PyTorch DataLoader and Pecan, since all samples have uniform preprocessing cost. 
The benefits of \sys appear in the intermediate range (25\%--75\%), whose strategy leverages the variability across data samples, outperforming existing solutions up to $2.4\times$.

% \rgmacedo{@rahma: please fill X.}

% To evaluate the impact of slow samples on training performance, we design a microbenchmark that builds on the speech recognition workload described in Section~\ref{subsec:data-prep-pipeline}, using the Speech-3s configuration in Table~\ref{tab:pipeline-steps}. Unlike Speech-3s, which applies the \emph{HeavyStep} transformation deterministically every 5 samples, our microbenchmark applies \emph{HeavyStep} to a configurable fraction of the dataset. By varying this fraction (0\%, 25\%, 50\%, 75\%, and 100\%), we introduce different levels of slow samples and measure the corresponding training times across systems. Figure~\ref{fig:cluster-slow-samples} presents the results: as expected, \sys performs comparably to PyTorch and Pecan in the edge cases (0\% and 100\%), but achieves lower training time at intermediate fractions (25\%, 50\%, and 75\%), demonstrating its robustness under different preprocessing conditions.

\input{floaters/eval-slow-samples.tex}

% \begin{figure}[t]  
%     \includegraphics[width=1\columnwidth]{figures/training_time_slowsamples.pdf}
%   \caption{Training time across different percentages of slow samples. Each bar shows the end-to-end training time for PyTorch, Pecan, DALI, and \sys.}

%     \label{fig:cluster-slow-samples}
% \end{figure}

\section{Discussion}
\label{sec:discussion}

% Our work on optimizing the training pipeline to reduce the cost of ML training opens up several future directions.
% \rgmacedo{I would remove this sentence, since we are not necessarily discussing future work.}

\paragraph{Order-sensitive scenarios} 
Some training scenarios require strict sample ordering. 
In multimodal models (\emph{e.g.,} image-text or audio-text), it is essential to maintain alignment between paired modalities. 
In the speech recognition workload (\cref{subsec:data-prep-pipeline}), \sys orders samples based on audio preprocessing time but always processes the audio–text pair together, preventing mismatches. 
This approach generalizes to other multimodal settings. 
Similarly, in curriculum learning~\cite{soviany2022curriculum}, where the global sample order is semantically important (\emph{e.g.,} easier examples preceding harder ones), \sys can be configured to disable reordering. 
In this mode, it behaves like PyTorch DataLoader, preserving the required sample order. 
While this disables \sys’s reordering advantage, it guarantees correctness in scenarios that depend on strict order.

\paragraph{Distributed training}
While our evaluation focused on a single server, \sys's design generalizes for distributed training with multiple nodes and GPUs. 
%can be deployed for distributed training with multiple nodes with multiple GPUs. 
In such scenarios, each server has a PyTorch instance accessing data from local or shared storage, with \sys integrated into each instance. 
Under data and model parallelism, \sys retains its preprocessing and batch construction benefits, and performs similarly to PyTorch's DataLoader during the data loading phase, as shown in Figure~\ref{fig:speedy-asyncdata}.

\paragraph{Suitable workloads} 
\sys provides the largest benefits for workloads where preprocessing is heavy, such as image, speech, or video training pipelines. 
In contrast, text-based NLP workloads, including large language models, are typically not input-bound, as their preprocessing is lightweight (\emph{e.g.,} tokenization) or performed offline. 
In these cases, \sys offers limited additional benefit and behaves similarly to the default PyTorch DataLoader. 
A key future challenge for efficient ML preprocessing lies in multimodal workloads~\cite{liang2024survey}, where different data modalities impose distinct preprocessing requirements, making this stage critical for training performance. % preprocessing remains critical for training performance, as data from different domains introduce distinct requirements.

% We expect \sys to provide benefits in these multimodal scenarios.

%----------------------------------------------------
%\input{related_work}
\section{Related Work}
\label{sec:rw}
 
Numerous systems aim to improve ML training efficiency by optimizing various stages of the data pipeline. Some focus on the input pipeline~\cite{pecan, tfdata, um2023fastflow, kuchnik2019progressive, zhao2022understanding, recd}, while others cache preprocessed data~\cite{graur2022cachew, mohan2020analyzing, lee2021refurbish}. Some offload preprocessing to GPUs or FPGAs~\cite{dali, kim2023fusionflow, park2020trainbox}, and others use storage-centric solutions to accelerate data preprocessing~\cite{lee2024presto, zhao2023tectonic}. Finally, there is work on tools to characterize the performance of data preprocessing for ML~\cite{bachkaniwala2024lotus, balmau2022characterizing}, as well as big data systems techniques that address data preprocessing~\cite{spark2019spark}.

\paragraph{Input pipeline optimizations} 
FastFlow~\cite{um2023fastflow} and tf.data \cite{tfdata} alleviate input data stalls by offloading preprocessing tasks to remote CPU workers in disaggregated environments. While these systems aim to improve training throughput, they treat the preprocessing step as a black box and do not analyze its internal overhead, unlike \sys. Pecan~\cite{pecan} is the most closely related to our work. Pecan introduces a transformation reordering policy and a policy to distribute preprocessing to worker threads in disaggregated scenarios. In contrast, \sys operates on a single machine with multiple GPUs and dynamically adapts to per-sample preprocessing variability.

\paragraph{Caching optimizations} 
Cachew~\cite{graur2022cachew} and CoorDL~\cite{mohan2020analyzing} focus on caching preprocessed data during training to reduce input pipeline stalls. 
CoorDL provides an analysis of the preprocessing pipeline and characterizes sources of data stalls. 
Unlike \sys, neither system adapts dynamically to per-sample preprocessing time variability during training.

\paragraph{Using accelerators for preprocessing} 
Meta's Presto~\cite{lee2024presto} proposes in-storage preprocessing for recommender systems. Other approaches, including DALI~\cite{dali}, FusionFlow~\cite{kim2023fusionflow}, and TrainBox~\cite{park2020trainbox}, offload preprocessing to specialized hardware such as GPUs or FPGAs. While this can accelerate transformations, it poses challenges when handling user-defined functions, which are often difficult to port to these environments. In contrast, \sys performs preprocessing on the CPU and generalizes to a broad range of tasks.

\paragraph{Tools for bottleneck diagnostics in preprocessing}  
Lotus~\cite{bachkaniwala2024lotus} and the MLPerf Storage tool~\cite{balmau2022characterizing} are dedicated profilers for identifying bottlenecks in data preprocessing pipelines and are orthogonal to this work.

\paragraph{Big data systems} 
\sys targets the preprocessing computation and batch construction phases, whereas techniques applied by traditional big data systems (\emph{e.g.,} data skew/straggler, dynamic repartitioning, workload estimation) target how data is organized, being complementary. 
Speculative execution is not applicable in ML preprocessing. 
Adaptive scheduling is the closest technique to \sys. 
However, while ML preprocessing has simpler dependencies between samples and workers, it has stricter latency requirements than adaptive scheduling. % is not applicable out-of-the-box, as the latency requirements are much more stringent for ML. 
Prior work~\cite{tfdata} also notes this difference. 
For example, Spark Streaming~\cite{spark_streaming_guide} recommends batch granularities of at least 50ms, while ML training requires step times below 1ms. 
Further, in production, Spark is used earlier in the pipeline for ETL tasks~\cite{zhao2022understanding}, with outputs written to storage and later consumed by PyTorch. 
This separation of concerns reflects current design choices: Spark addresses offline preprocessing, while PyTorch performs online transformations.

%----------------------------------------------------
%\input{conclusion}
\section{Conclusion}
\label{sec:conclusion}

We presented \sys, a general-purpose data loader that accelerates machine learning training by addressing head-of-line blocking in the data preprocessing pipeline. Unlike existing solutions, \sys dynamically adapts to per-sample processing variability. Through a sample-aware load balancer and adaptive worker scheduling, \sys sustains high throughput without requiring any prior knowledge of the dataset or transformations. Our evaluation across diverse workloads and hardware platforms demonstrates that \sys consistently improves training time, achieves up to 7.5$\times$ speedup over PyTorch DataLoader and Pecan, all while preserving model accuracy.

% \blueblock{
\section*{Acknowledgments}

We thank our shepherd, Muhammad Ali Gulzar, and the anonymous reviewers for their valuable feedback. 
This research was supported by the Natural Sciences and Engineering Research Council of Canada (NSERC) CREATE 584767-2024; a gift from MLCommons; by computational resources from FCT IP at Deucalion supercomputer, jointly funded by EuroHPC JU and Portugal; and by the European Regional Development Fund (ERDF) through the Innovation and Digital Transition Programme (COMPETE 2030) under Portugal 2030, within the scope of the project CDMS, reference 17409 (COMPETE2030-FEDER-01193000).
% \balance 

% \newpage

%-------------------------------------------------------------------------------
\bibliographystyle{ACM-Reference-Format}
\bibliography{main}
%-------------------------------------------------------------------------------

\input{appendix}
\end{document}

%% file: configs/commands.tex
\usepackage{thmtools}
\usepackage[framemethod=TikZ]{mdframed}
\mdfsetup{skipabove=1em,skipbelow=0em, innertopmargin=5pt, innerbottommargin=6pt}
\theoremstyle{definition}

\definecolor{DimGray}{rgb}{0.41, 0.41, 0.41}
\colorlet{thmgreencolor}{DimGray!70!black}

\makeatletter

\declaretheoremstyle[
  headfont=\bfseries, bodyfont=\normalfont\itshape,
  mdframed={
    innertopmargin=2pt,
    innerleftmargin=8pt,
    innerrightmargin=8pt,
    innerbottommargin=5pt,
    linewidth=2pt,
    rightline=false, topline=false, bottomline=false,
    linecolor=black, backgroundcolor=DimGray!5,
  }
  ]{thmgraybox}
  
\makeatother

\declaretheorem[style=thmgraybox, numbered=yes, name=Takeaway]{takeaway}

\AtBeginEnvironment{takeaway}{
  \captionsetup{labelfont={color=thmgraycolor}}
}

\definecolor{Gray}{gray}{0.9}
\newcolumntype{?}{!{\vrule width 1.5pt}}

\newcommand{\emphparagraph}[1]{\vspace{.2\baselineskip}\noindent\emph{#1.}}

%% file: floaters/motivation-pytorch-dataloader-pipeline.tex
\begin{figure}[t]
    \centering
    % Subfigure 1: PyTorch DataLoader pipeline inefficiency
    \begin{subfigure}[t]{\columnwidth}
        \centering
        \includegraphics[width=1\linewidth]{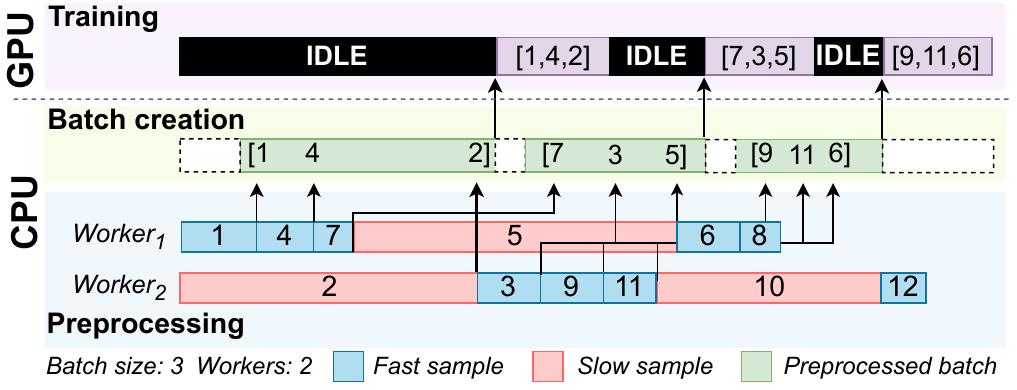}
        %\caption{Pipeline inefficiency of preprocessing and training with PyTorch DataLoader. The diagram shows how a slow sample delays batch construction.}
        \vspace{-15pt}
        \caption{Data preprocessing and training pipeline of PyTorch DataLoader.}
        \label{fig:pytorch-pipeline}
    \end{subfigure}
    
    % Subfigure 2: CPU–GPU usage over time
    \begin{subfigure}[t]{\columnwidth}
        \vspace{5pt}
        \centering
        \includegraphics[width=1\linewidth]{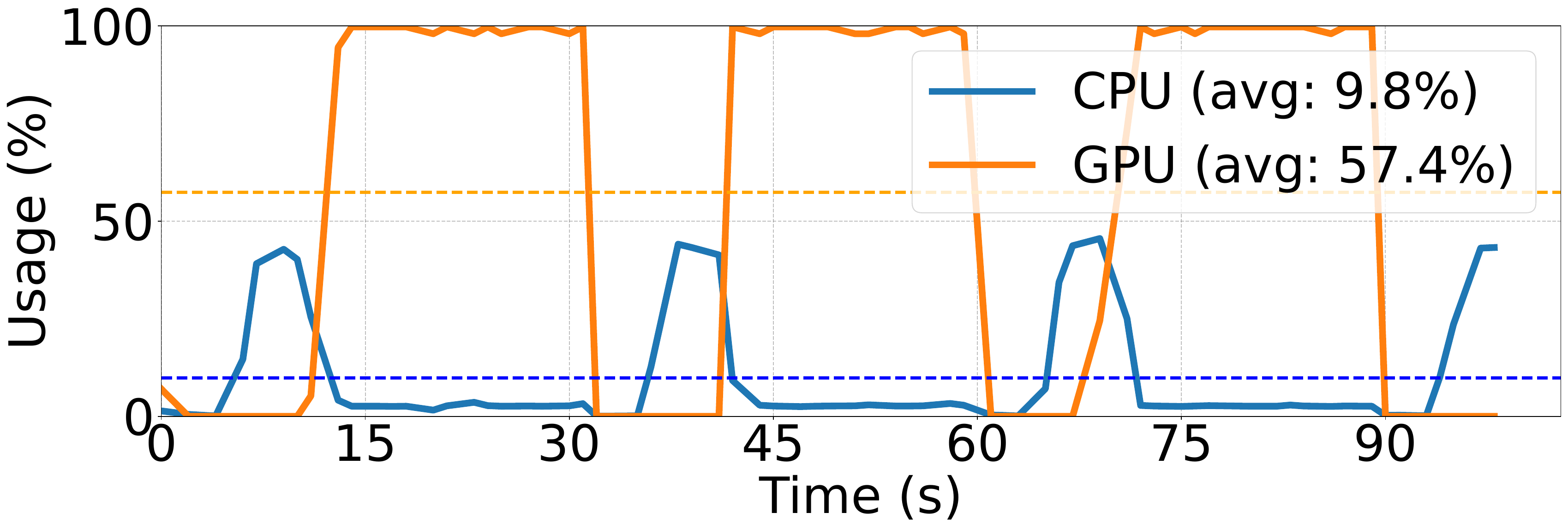}
        % \caption{CPU and GPU usage over time during 3D‑UNet training on 4$\times$ A100 GPUs with PyTorch DataLoader.}
        \vspace{-15pt}
        \caption{CPU and GPU usage of PyTorch DataLoader during 3D‑UNet training.}
        \label{fig:motiv-gpu-idle}
    \end{subfigure}
    \vspace{-.7em}
    \caption{Inefficient PyTorch DataLoader pipeline. Slow data samples delay the batch construction process, resulting in GPU under-utilization and poor training performance. }
    
    % \oana{inconsistency: change pre-processing to preprocessing in Fig a, in the legend and in the blue box}}
    \label{fig:combined-pytorch-analysis}
    \vspace{-10pt}
\end{figure}

%% file: floaters/preprocessing-pipelines-table.tex
\begin{table*}[t]
    \caption{Preprocessing pipelines for object detection, image segmentation, and speech recognition (\emph{Speech-$X$}) workloads. 
    The most time-consuming steps are highlighted in \textbf{bold}. 
    \emph{LightStep} simulates lightweight preprocessing (\emph{e.g.,} volume normalization, frame splicing).
    \emph{HeavyStep} simulates compute-intensive steps that may include complex filtering or augmentation (\emph{e.g.,}long-context time-stretching, multi-pass spectrogram enhancement) and is performed every 5 samples.
    }
    \label{tab:pipeline-steps}
    \vspace{-1em}
    \small
    \renewcommand{\arraystretch}{1.0}  % Reduce row height
    \begin{center}
    \begin{tabular}{ll}  % Slightly narrower left column
        \toprule
        \textbf{Workload} & \textbf{Preprocessing Pipeline} \\
        \midrule
        Object Detection & \textbf{Resize} $\rightarrow$ RandomHorizontalFlip $\rightarrow$ \textbf{ToTensor} $\rightarrow$ Normalize \\
        Image Segmentation & \textbf{RandomCrop} $\rightarrow$ RandomFlip $\rightarrow$ RandomBrightness $\rightarrow$ GaussianNoise $\rightarrow$ Cast \\
        Speech-3s & Pad $\rightarrow$ SpecAugment $\rightarrow$ FilterBank $\rightarrow$ FrameSplicing $\rightarrow$ PermuteAudio  $\rightarrow$ LightStep (0.5s)  $\rightarrow$ \textbf{HeavyStep (3s)}\\
        Speech-10s & Pad $\rightarrow$ SpecAugment $\rightarrow$ FilterBank $\rightarrow$ FrameSplicing $\rightarrow$ PermuteAudio $\rightarrow$ LightStep (0.5s) $\rightarrow$ \textbf{HeavyStep (10s)} \\
        \bottomrule
    \end{tabular}
    \end{center}
    %\vspace*{-5pt}
\end{table*}

%% file: floaters/motivation-persample-preprocessing.tex
\begin{figure}[t]
    \centering
    \begin{subfigure}[t]{0.48\columnwidth}
        \centering
        \includegraphics[width=\linewidth]{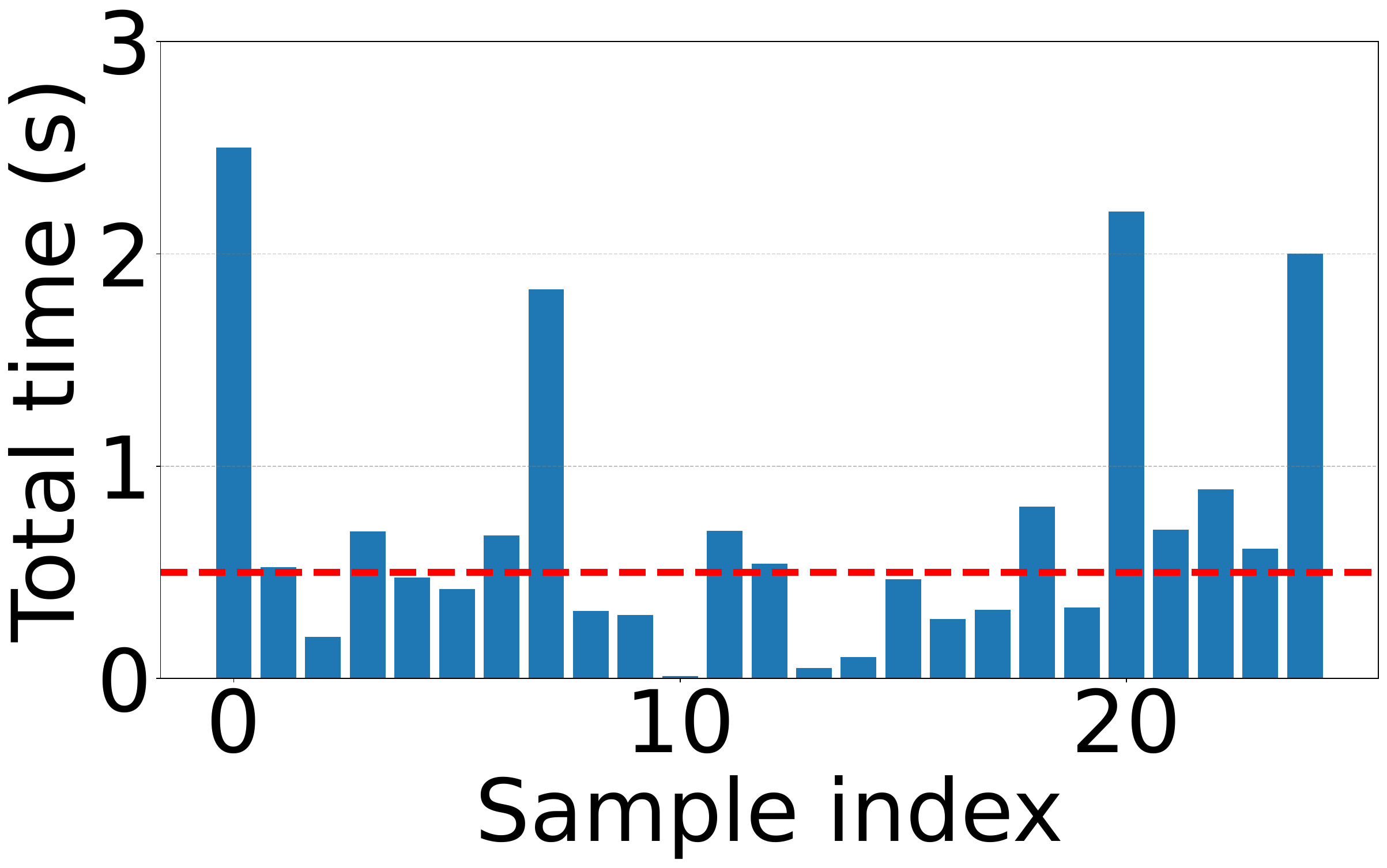}
        \vspace{-15pt}
        \caption{Image segmentation (3D-UNet).}
        \label{subfig:variability-image-segmentation}
    \end{subfigure}\hfill
    \begin{subfigure}[t]{0.48\columnwidth}
        \centering        
        \includegraphics[width=\linewidth]{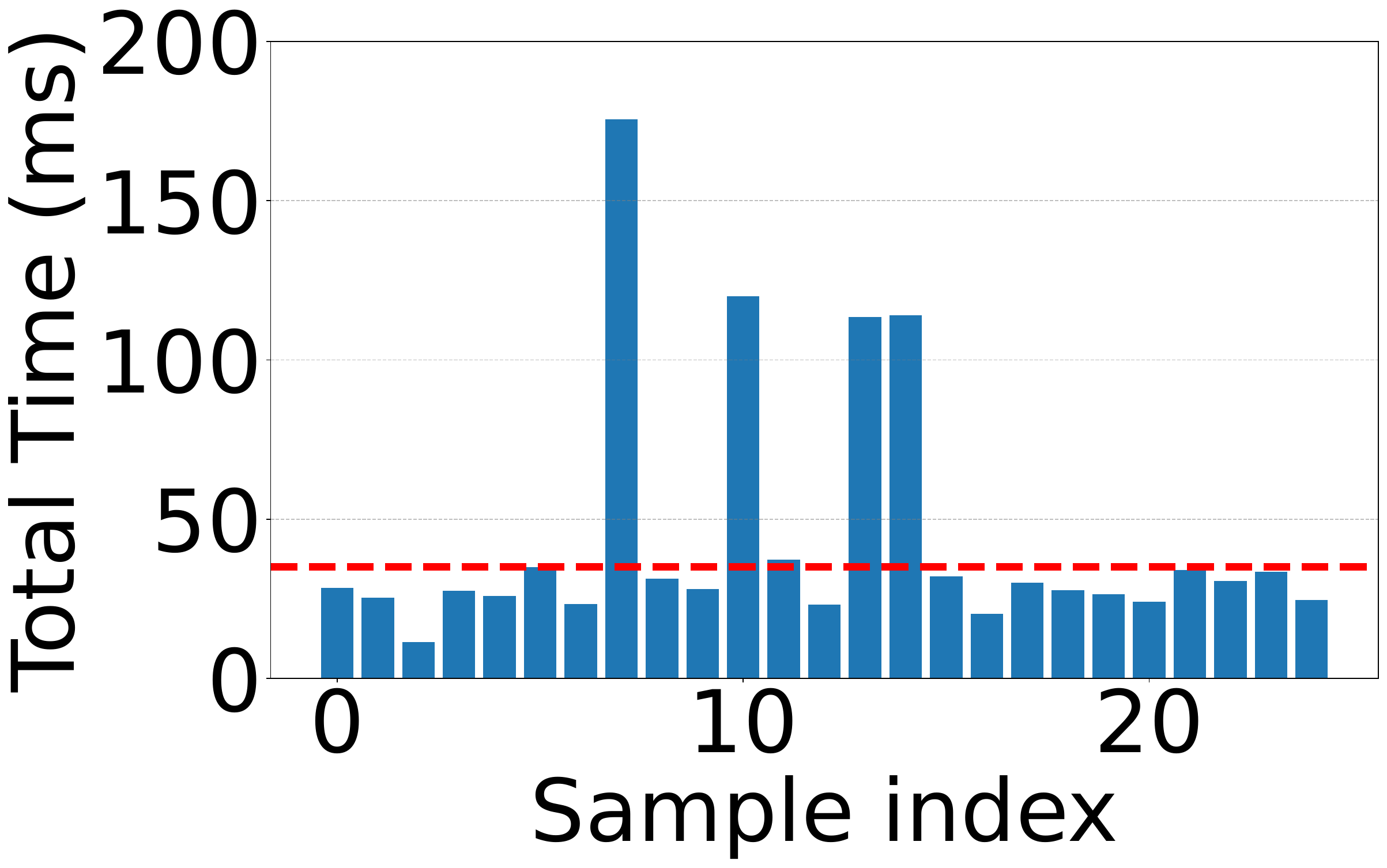}
        \vspace{-15pt}
        \caption{Object detection (R-CNN).}
        \label{subfig:variability-object-detection}
    \end{subfigure}
    \vspace{-5pt}
    \caption{%The causes and effects of Head‑of‑line blocking. 
   Variability in per-sample preprocessing time for image segmentation and object detection workloads. 
   The red dashed lines depict the average preprocessing time across all samples -- 0.5s in (a) and 35ms in (b).
    }
    \label{fig:motivation}
    \vspace{-10pt}
\end{figure}

%% file: floaters/motivation-preprocessing-table.tex
\begin{table}[t]
    \centering
    \caption{Preprocessing time (in \emph{ms}) for each workload.}
    \label{tab:motiv-preproc-times}
    \vspace{-0.75em}
    {\small
    \renewcommand{\arraystretch}{0.95}
    \begin{tabular}{lccccc}
        \toprule
        \textbf{Workload} & \textbf{Avg} & \textbf{Med.} & \textbf{P75} & \textbf{P90} & \textbf{Min–Max–Std} \\
        \midrule
        Obj. Det.   & 31   & 28  & 30   & 35    & 11–176–19 \\
        Img. Seg.   & 500  & 470 & 630  & 750   & 10–2230–197 \\
        Speech-3s   & 998  & 508 & 509  & 3008  & 502–3017–992 \\
        Speech-10s  & 2351 & 508 & 509  & 10008 & 502–10014–3757 \\
        \bottomrule
    \end{tabular}
    }
    \vspace*{-5pt}
\end{table}

%% file: floaters/motivation-heuristics.tex
\begin{figure}[t]
    \centering
    \begin{subfigure}[t]{0.48\columnwidth}
        \centering        
        \includegraphics[width=\linewidth]{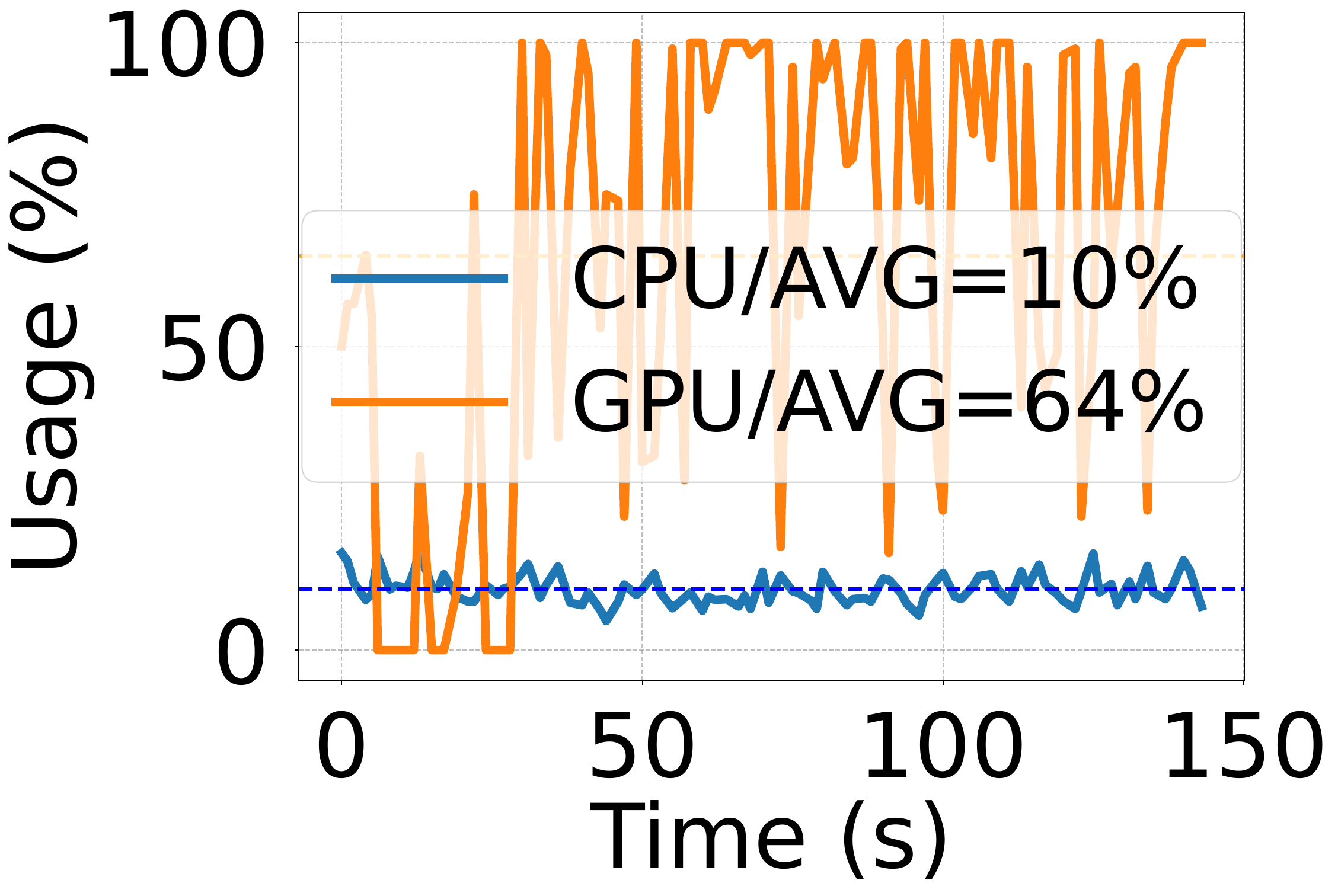}
        \vspace{-17.5pt}
        \caption{Image size.}
        \label{subfig:heuristic-image}
    \end{subfigure}
    \hfill
    \begin{subfigure}[t]{0.48\columnwidth}
        \centering
        \includegraphics[width=\linewidth]{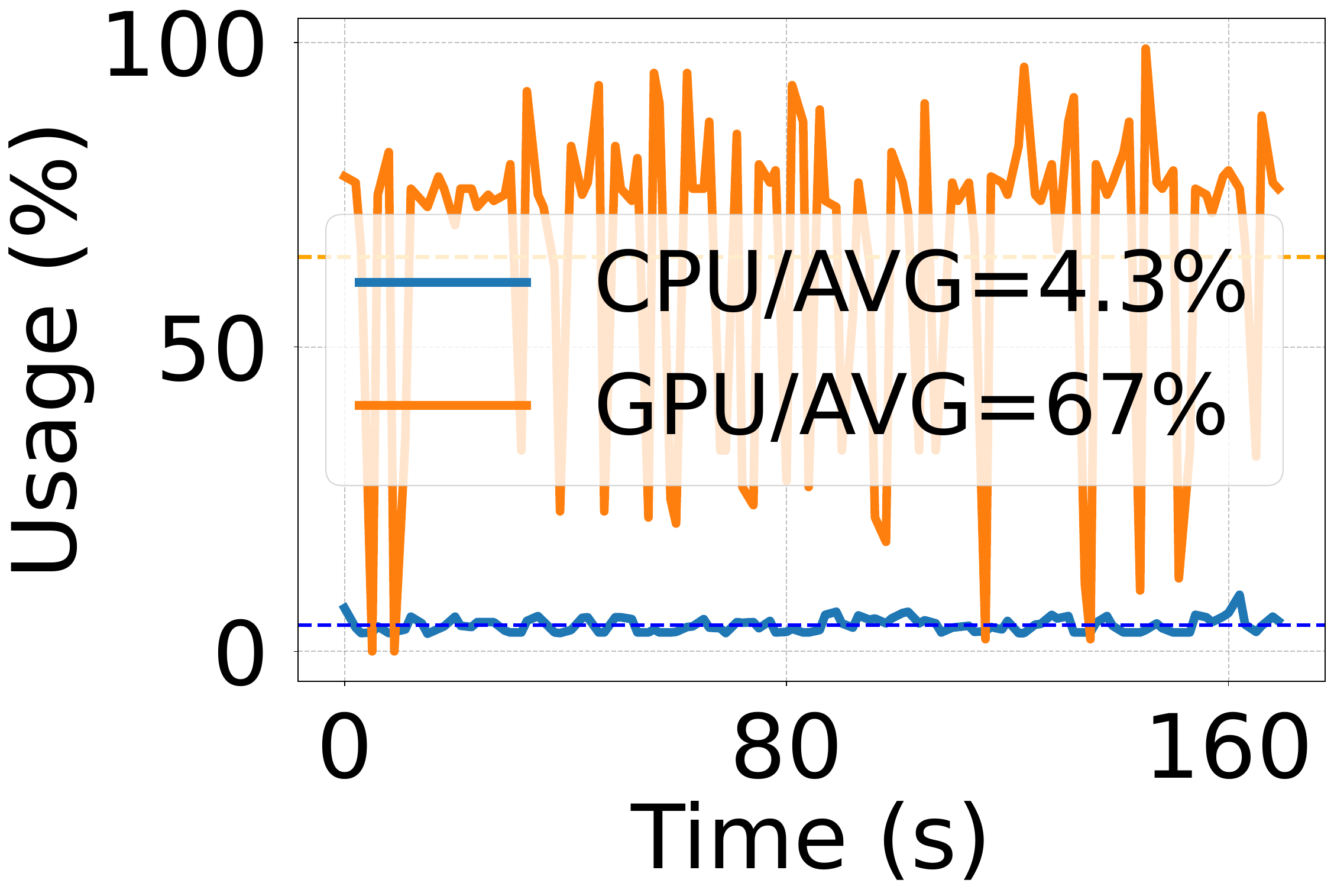}
        \vspace{-17.5pt}
        \caption{Transformation reordering~\cite{pecan}.}
        \label{subfig:heuristic-transformation}
    \end{subfigure}
    \vspace{-7.5pt}
    \caption{CPU and GPU usage of the Object Detection workload when using two heuristics: (a) image size and (b) transformation reordering.}
    \vspace{-10pt}
    \label{fig:loadbalancer_heuristic}
\end{figure}

%% file: floaters/motivation-prefetch-factor.tex
\begin{figure}[t]
    \centering
    \begin{subfigure}[t]{0.48\columnwidth}
        \centering
        \includegraphics[width=\linewidth]{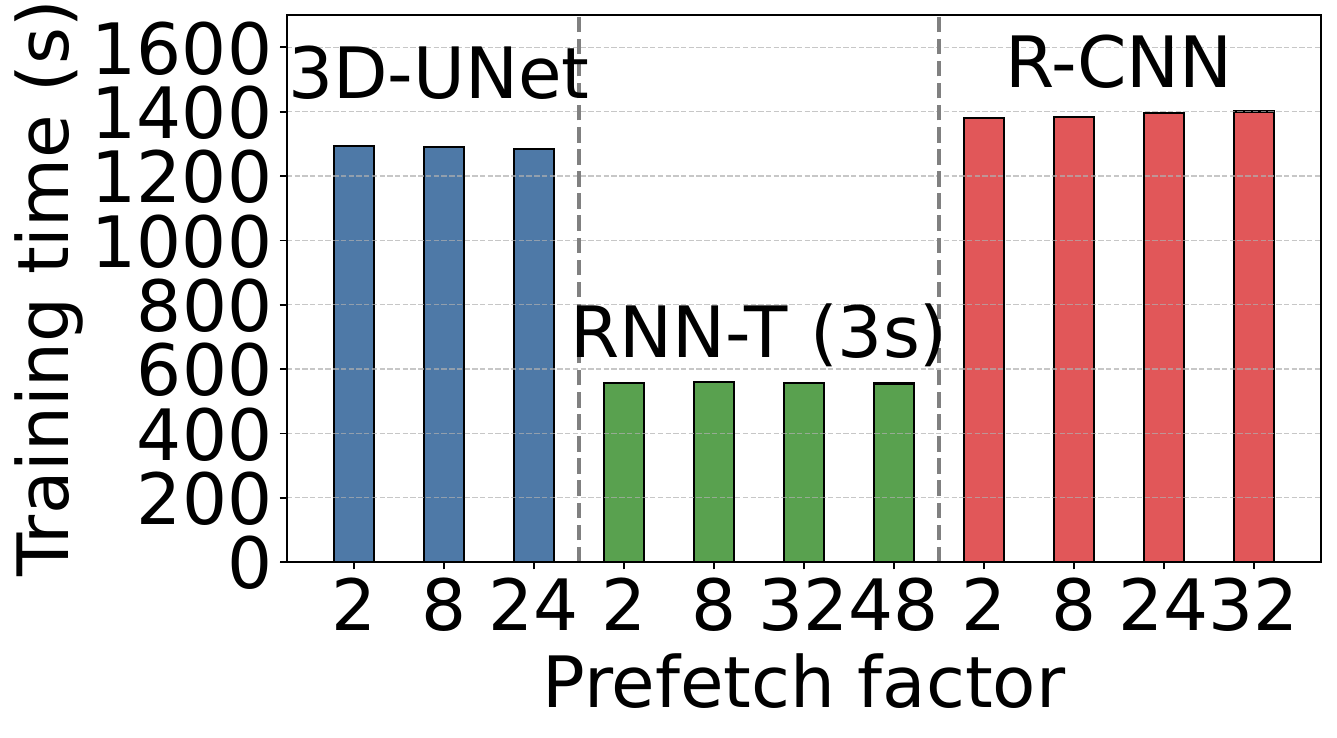}
        \vspace{-17.5pt}
        \caption{Prefetch factor in PyTorch.}
    \end{subfigure}\hfill
    \begin{subfigure}[t]{0.48\columnwidth}
        \centering        
        \includegraphics[width=\linewidth]{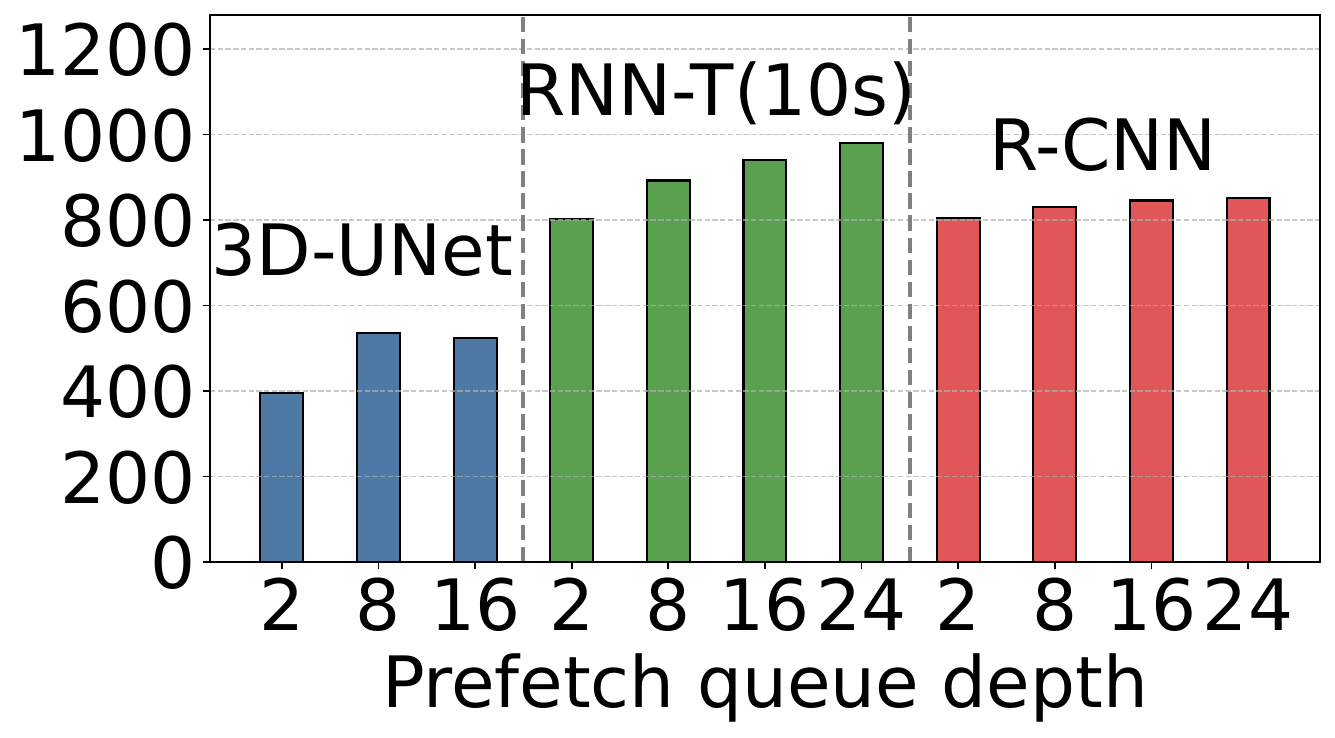}
        \vspace{-17.5pt}
        \caption{Prefetch queue depth in DALI.}
    \end{subfigure}
    \vspace{-7.5pt}
    \caption{Impact of prefetch parameter on training time. Increasing the number of batches pre-fetched does not improve the training in both (a) Pytorch and (b) DALI. %\rgmacedo{Is there a reason behind having different prefetch factors across the workloads? Also, why are we using different speech workloads between PyTorch and DALI?} \rahma{I mentioned in the text that it is because it runs OOM till a the last one, as for having different speech, I thought showing the two is not a problem and it shows that we are not only focusing on three workloads not all four. }
    }
    \label{fig:pref_factor_impact}
    \vspace*{-10pt}
\end{figure}

%% file: floaters/design-overview.tex
\begin{figure}[t]
    \centering
    %\vspace{-.5em}
    % \includegraphics[width=\linewidth]{figures/design_async_new_withcolors-cropped.pdf}
    % \includegraphics[width=\linewidth]{figs/speedy-loader-overview.pdf}
    \includegraphics[width=\linewidth]{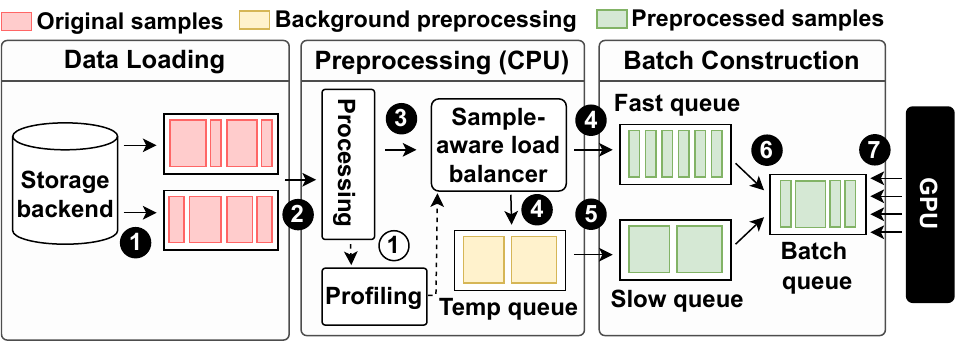}
    \vspace{-1.5em}
    \caption{\sys high-level design. It continuously enqueues preprocessed samples into a specified queue based on a load balancer decision. Concurrently, the GPU dequeues preprocessed data for training. We show \sys for one GPU, but it generalizes to multi-GPU settings.}
    \vspace{-15pt}
    \label{fig:speedy-asyncdata}
\end{figure}

%% file: floaters/minatoloader-pipeline-diagram.tex
\begin{figure}[t]
    \centering
    \includegraphics[width=1\columnwidth]{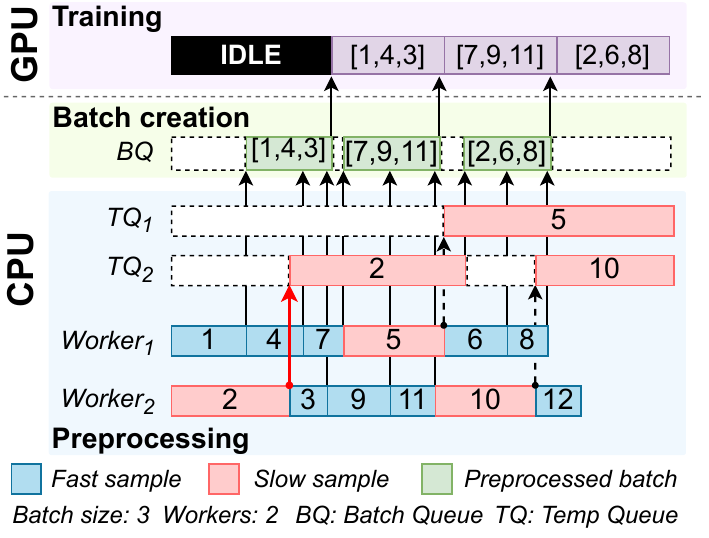}
    \vspace{-15pt}
    \caption{\sys preprocessing pipeline. Slow sample \#2 does not delay the batch creation and gets migrated to the \emph{temp queue} (red arrow). The fast and slow queues in the batch construction are omitted for brevity.}
    \label{fig:speedy-pipeline}
    \vspace*{-10pt}
\end{figure}

%% file: floaters/eval-model-configs-table.tex
\begin{table}[t]
    \centering
    % \caption{Workload configurations used to evaluate the training time of \sys compared to the baselines.}
    \caption{Training configurations used for each workload.}
    \label{tab:model-config-train-time}
    \vspace{-0.75em}
    {\small
    \renewcommand{\arraystretch}{0.95}
    \begin{tabular}{lccccc}
        \toprule
        \textbf{Workload} & \textbf{\#Epochs} & \textbf{\#Iterations} & \textbf{Batch Size} \\
        \midrule
        Image segmentation  & 50    & -     & 3  \\
        Object detection    & -     & 1000  & 48 \\
        Speech recognition  & -     & 1000  & 24 \\
        \bottomrule
    \end{tabular}
    }
    %\vspace*{-10pt}
\end{table}

%% file: floaters/eval-throughput.tex
\begin{figure*}[t]
    \centering
     \begin{subfigure}[t]{0.24\textwidth}
        \includegraphics[width=\linewidth, height=.65\textwidth]{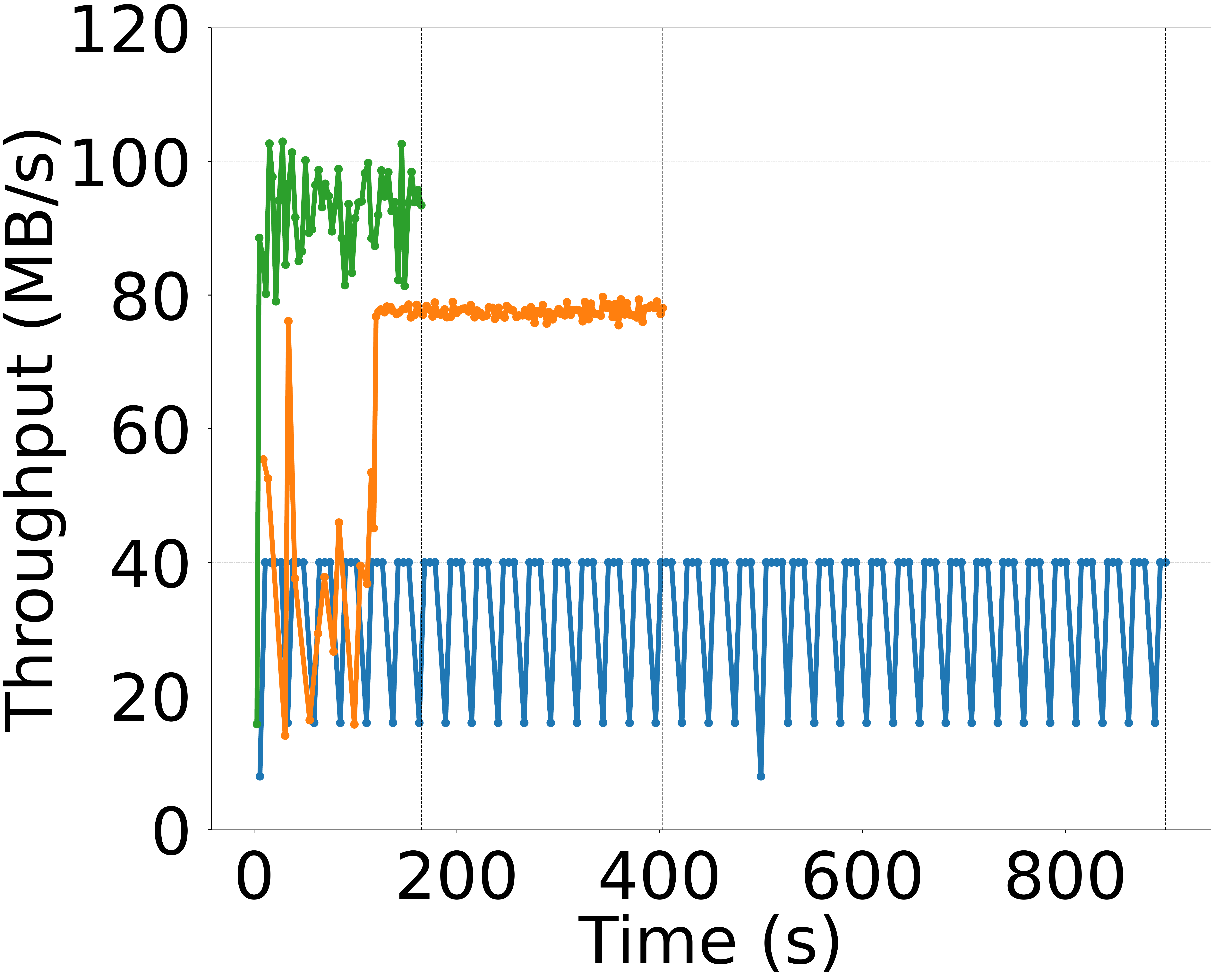}
        \caption{Img. Seg. (3D-UNet)}
        \label{subfig:a100-throughput-image}
    \end{subfigure}
     \begin{subfigure}[t]{0.24\textwidth}
        \includegraphics[width=\linewidth, height=.65\textwidth]{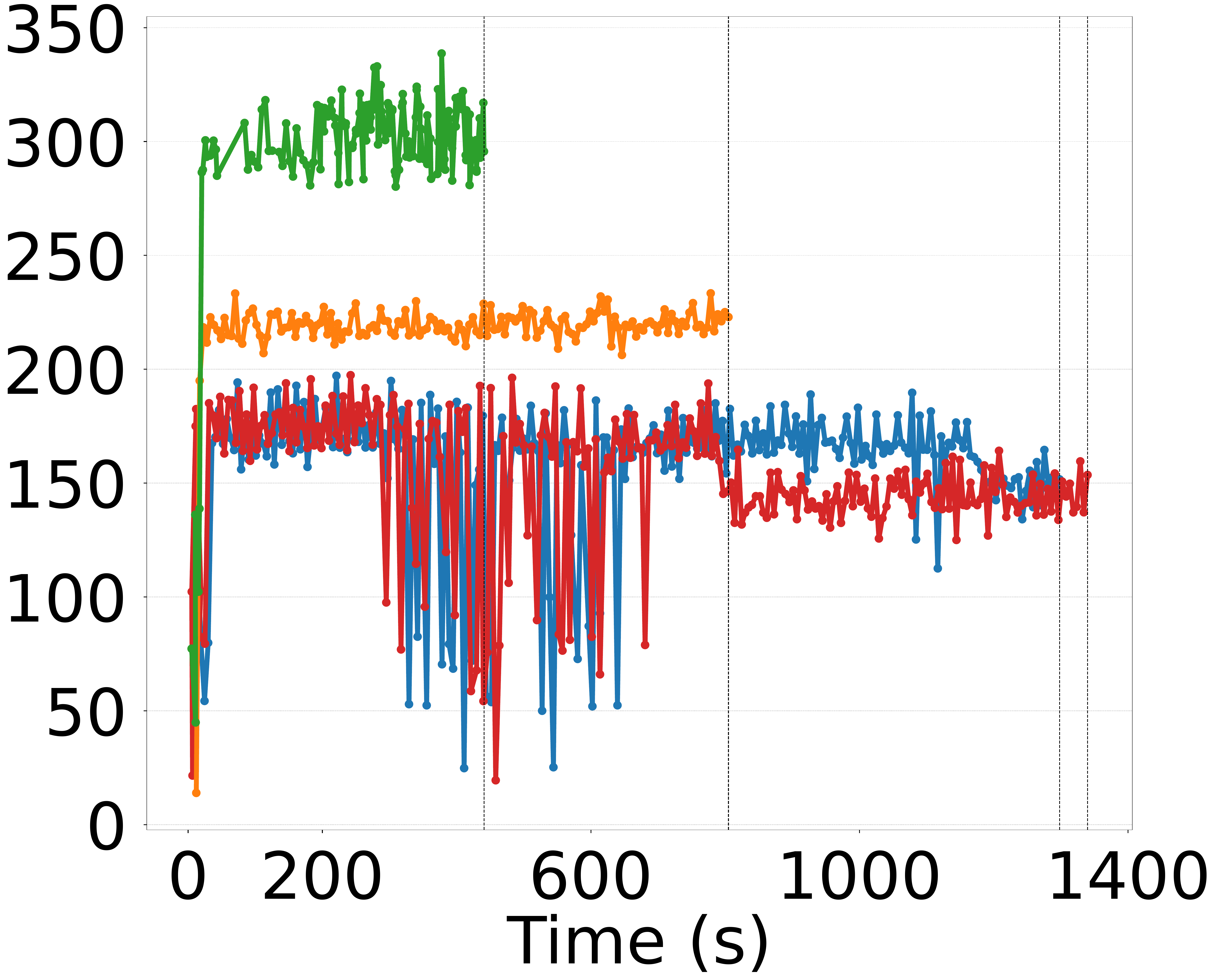}
        \caption{Obj. Det. (R-CNN)}
        \label{subfig:a100-throughput-object}
    \end{subfigure}
     \begin{subfigure}[t]{0.24\textwidth}
        \includegraphics[width=\linewidth, height=.65\textwidth]{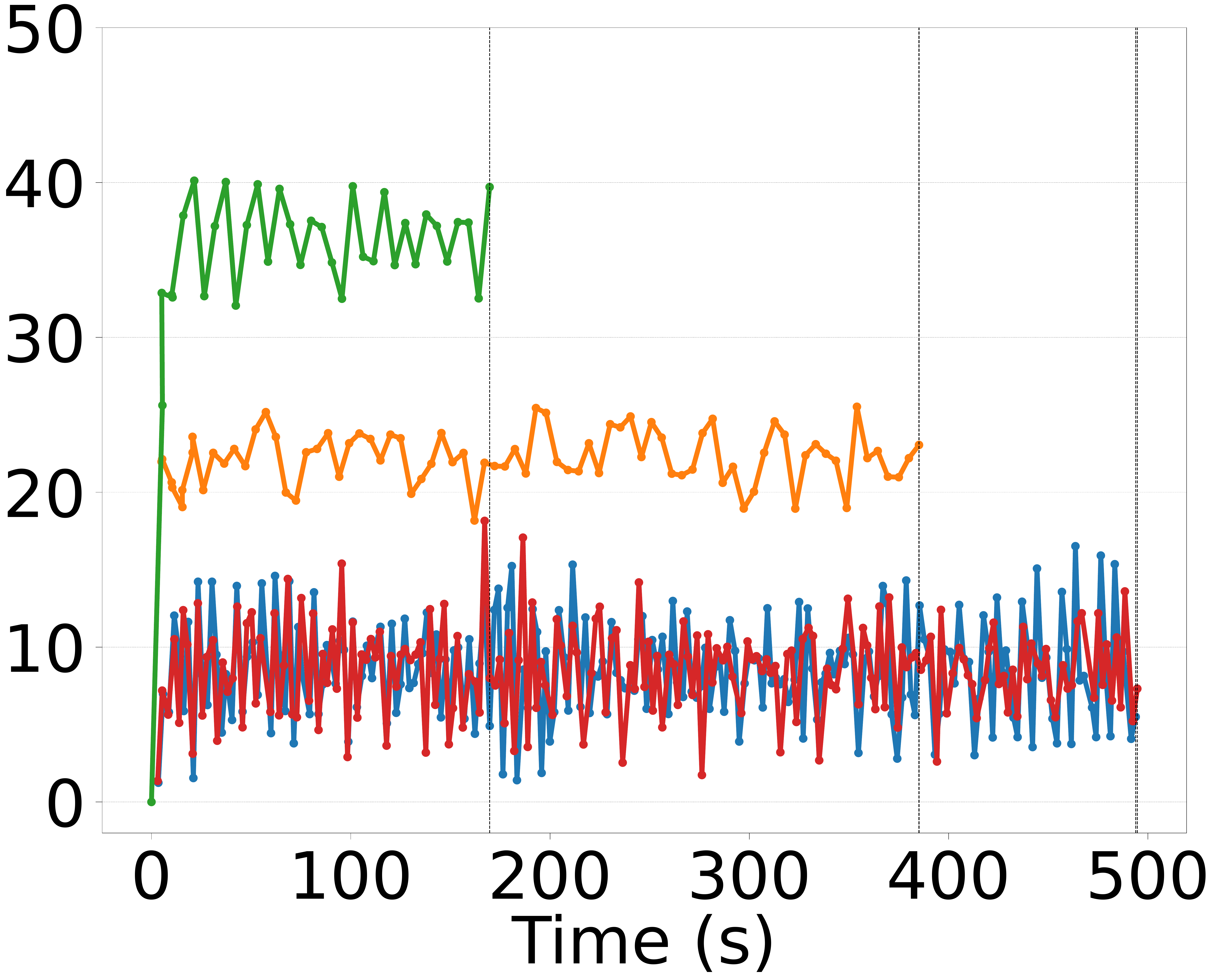}
        \caption{Speech-3s (RNN-T)}
        \label{subfig:a100-throughput-speech-3}
    \end{subfigure}
     \begin{subfigure}[t]{0.24\textwidth}
        \includegraphics[width=\linewidth, height=.65\textwidth]{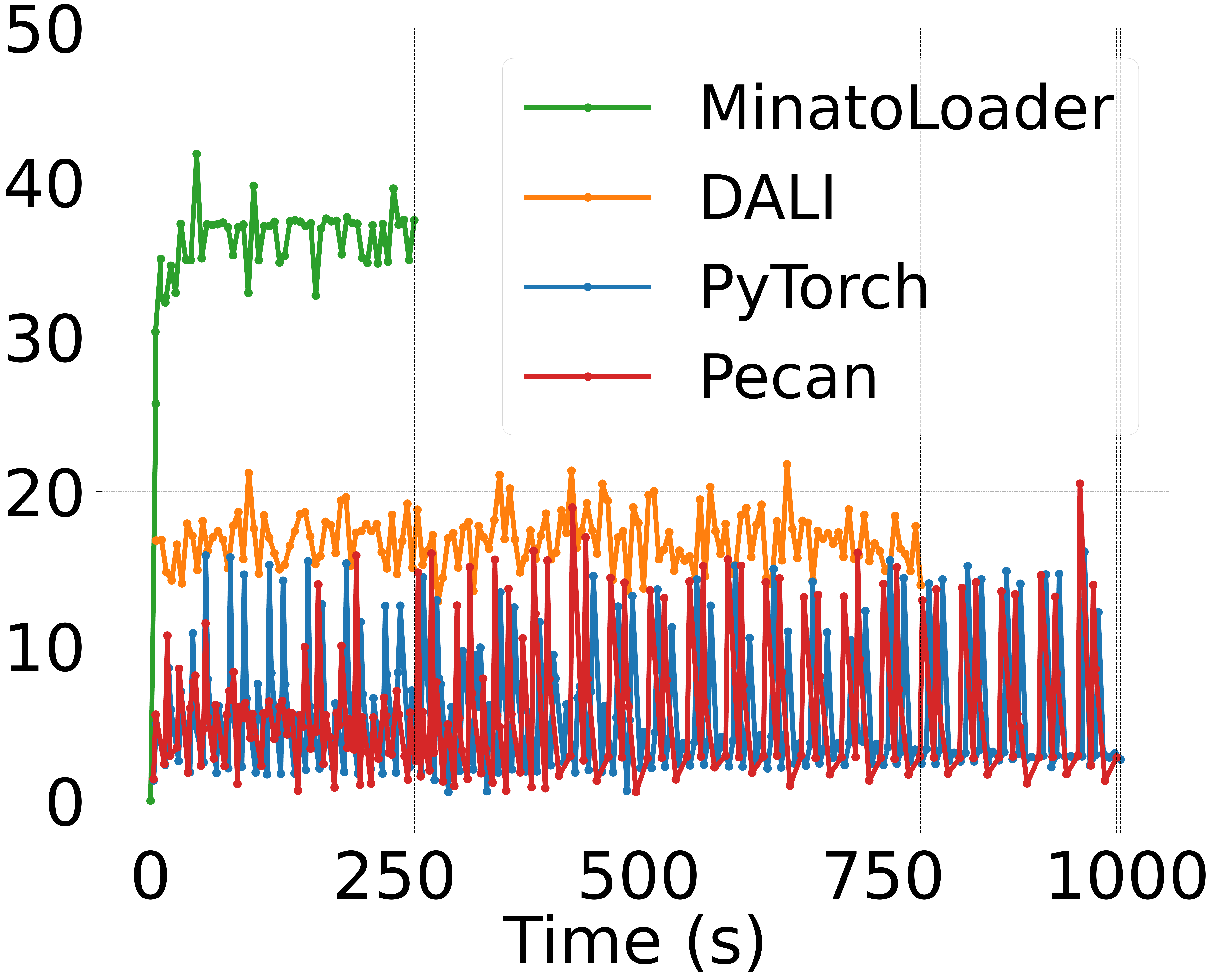}
        \caption{Speech-10s (RNN-T)}
        \label{subfig:a100-throughput-speech-10}
    \end{subfigure}
   
    \vspace{-5pt}
    
    \caption{Throughput (MB/s) of PyTorch DataLoader, Pecan, DALI, and \sys on four ML workloads using A100 GPUs (\emph{Config. A}). 
    Solid vertical lines represent the end of the training for each data loader.}
    \label{fig:throughput-a100}
    %\vspace{-7.5pt}
\end{figure*}

%% file: floaters/eval-resource-usage.tex
\begin{figure}[t]
    \centering
    %\vspace{-.5em}
    \includegraphics[width=\linewidth]{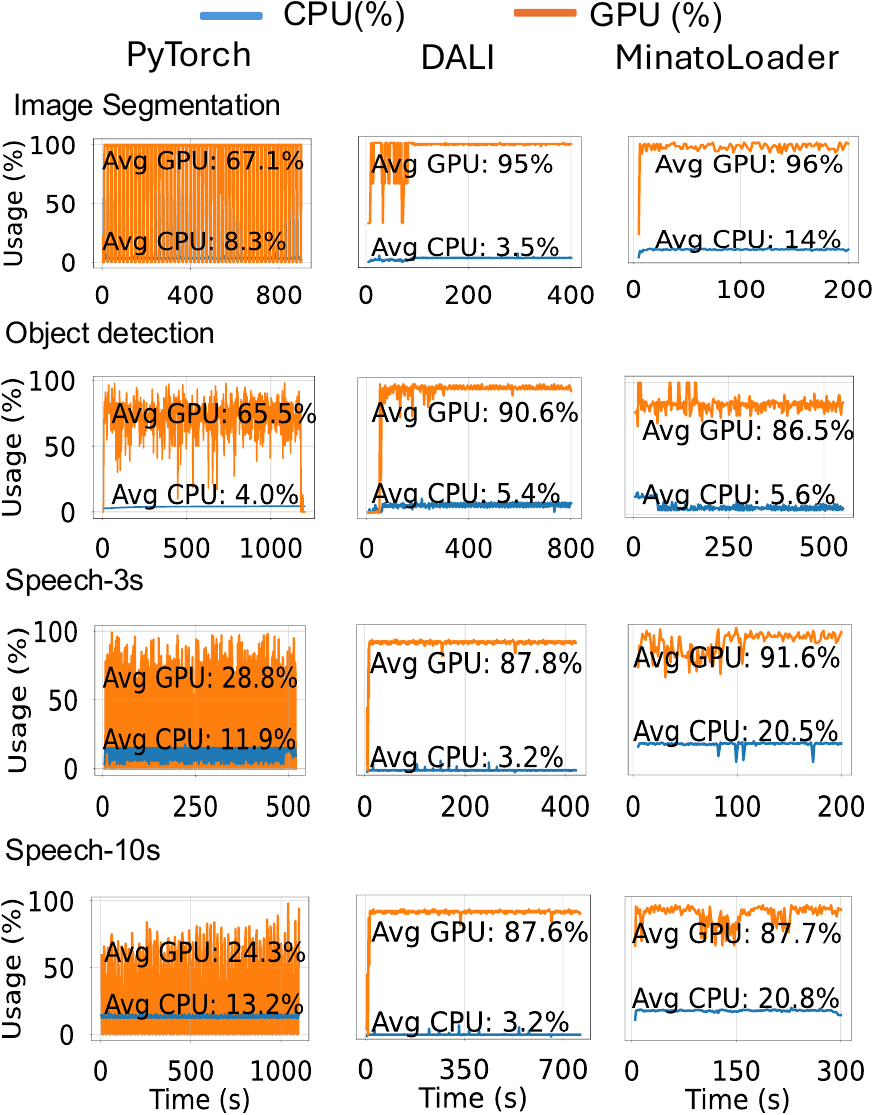}
    \caption{CPU and GPU usage for all systems across all workloads using 4$\times$A100 GPUs.}
    \label{fig:cpu-gpu-usage a100}
    \vspace{-10pt}
\end{figure}

%% file: floaters/eval-training-time-gpus.tex
%%%%%%%%%%%%%%%%%%%%%%%%%%%%%%%%%%%%%%%%%%%%%%%%%

\begin{figure*}[t]
    \centering
    % Top row: A100 GPUs
    \begin{subfigure}[t]{0.23\textwidth}
        \centering
        \includegraphics[width=\linewidth, height=.6\textwidth]{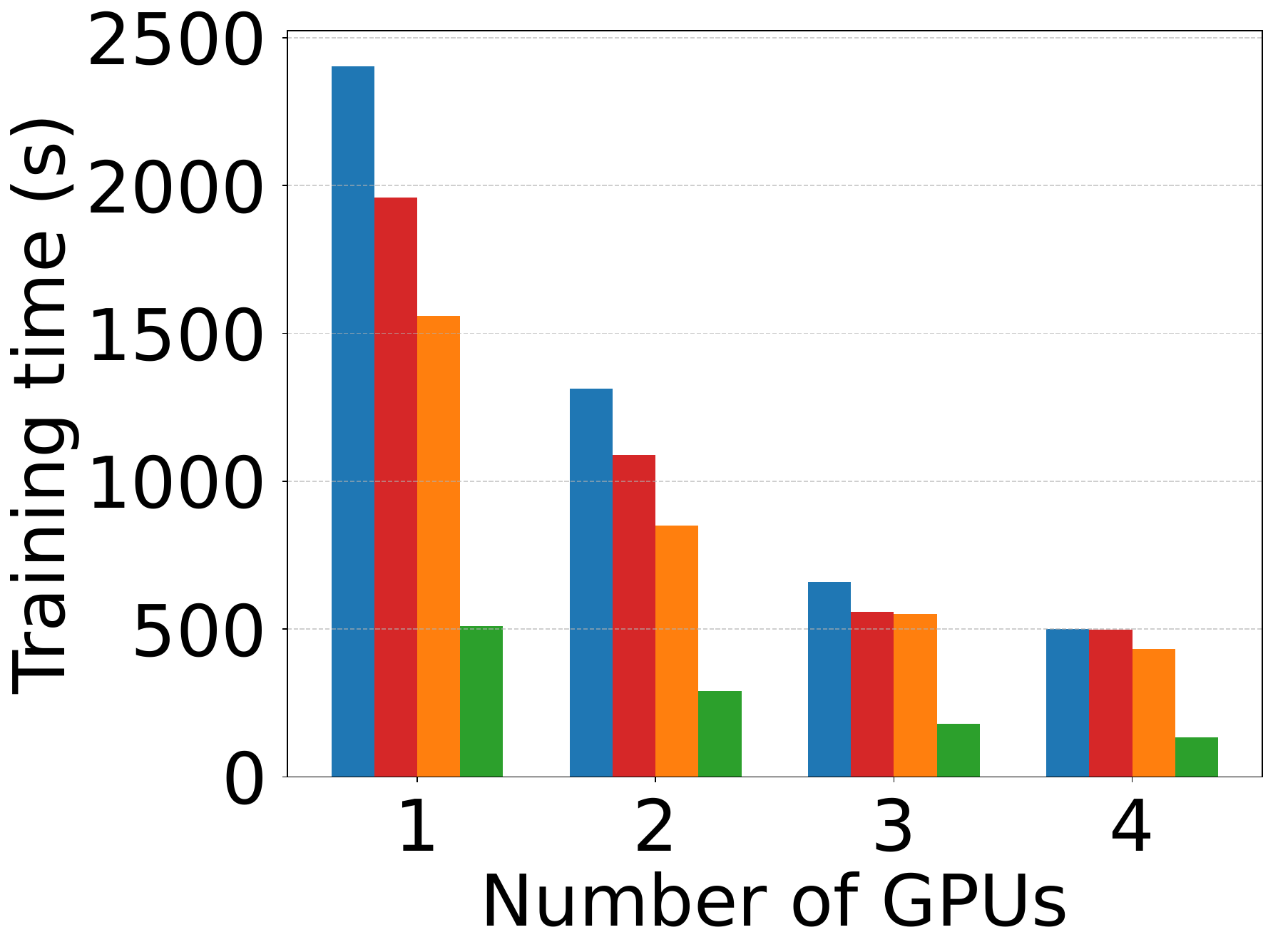}
        \caption{Speech-3s (RNN-T) - A100}
    \end{subfigure}
    \hfill
    \begin{subfigure}[t]{0.23\textwidth}
        \centering
        \includegraphics[width=\linewidth, height=.6\textwidth]{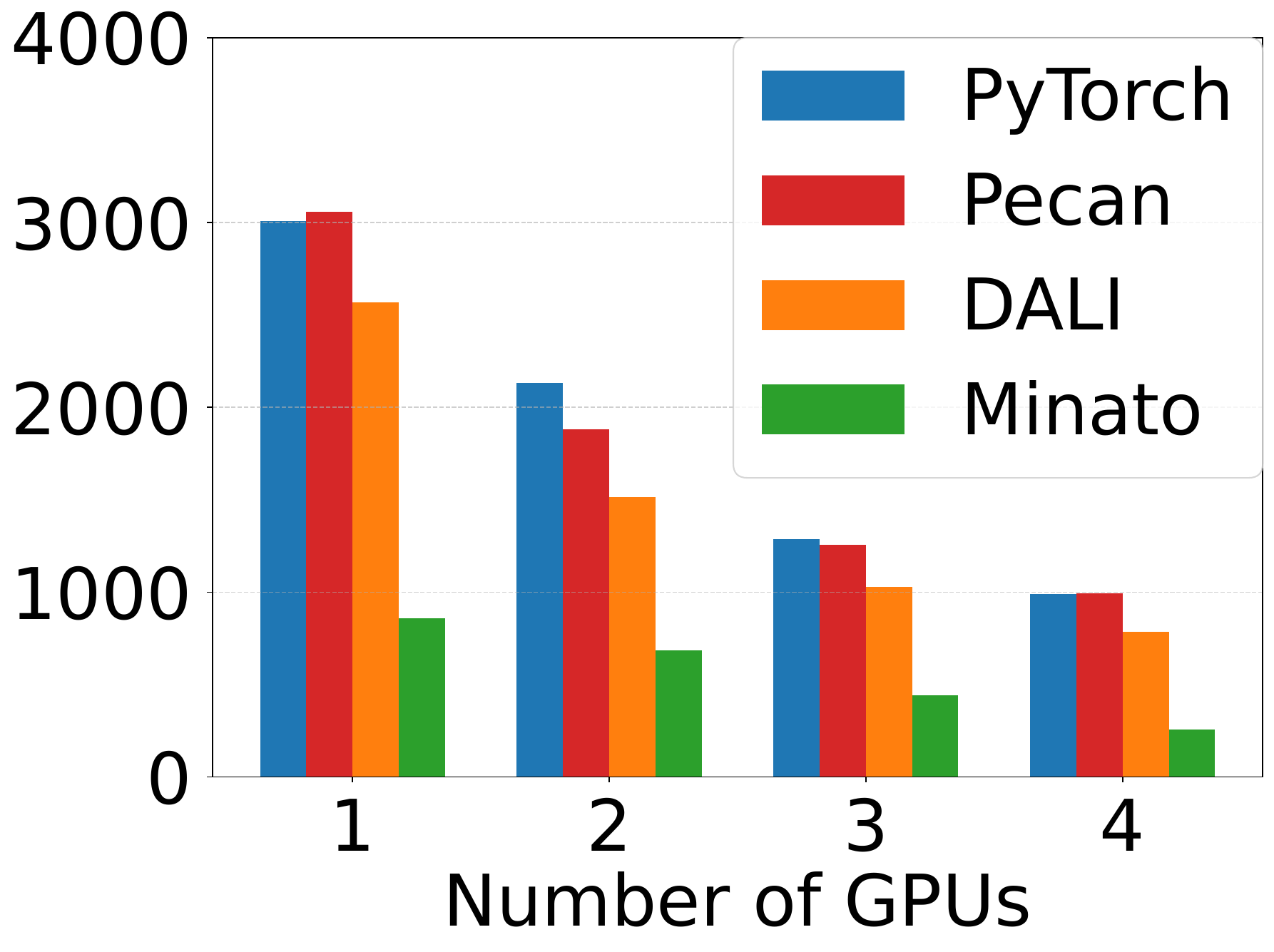}
        \caption{Speech-10s (RNN-T) - A100}
    \end{subfigure}
    \hfill
    \begin{subfigure}[t]{0.23\textwidth}
        \centering
        \includegraphics[width=\linewidth, height=.6\textwidth]{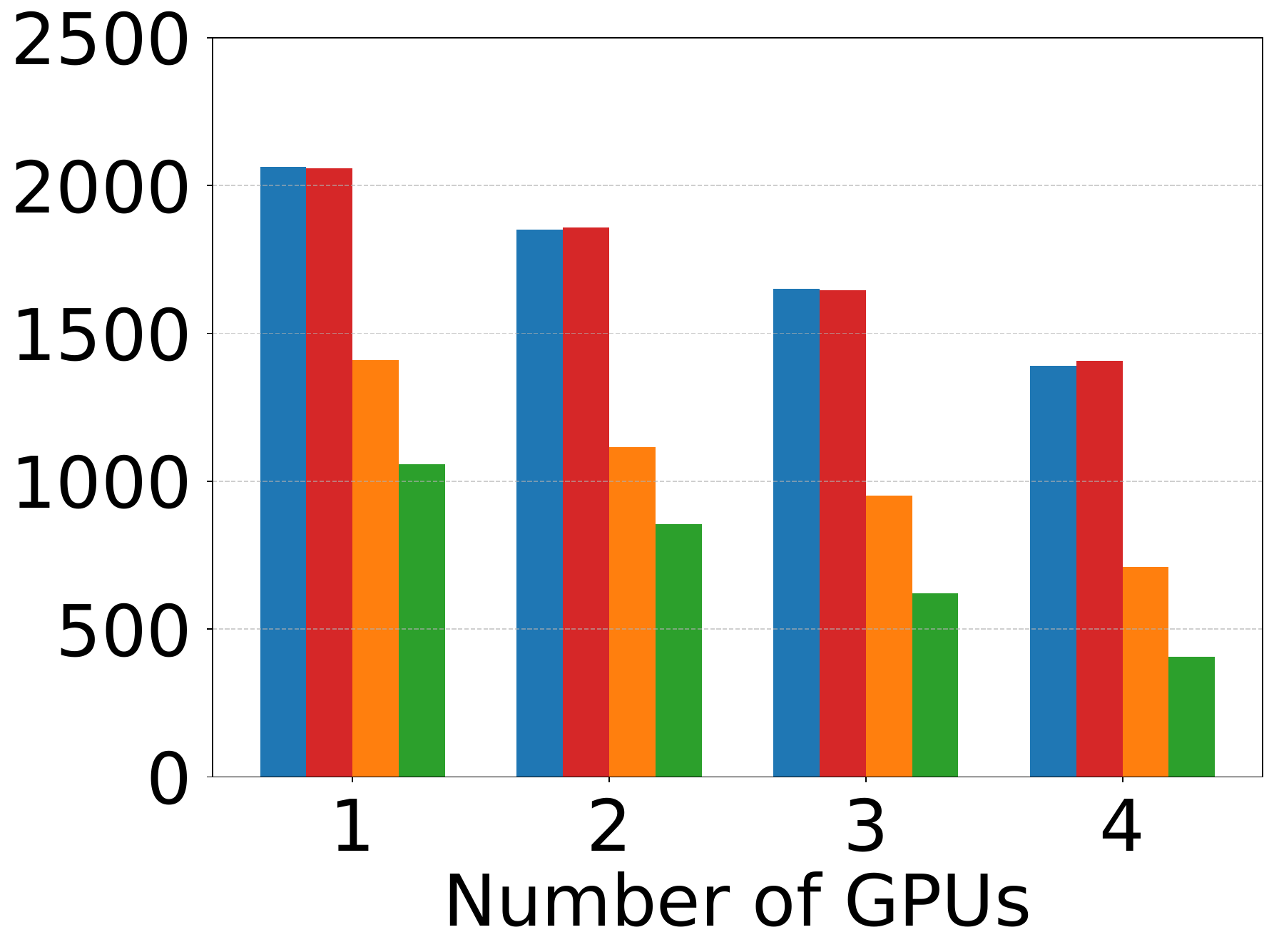}
        \caption{Obj. Det. (R-CNN) - A100}
    \end{subfigure}
    \hfill
    \begin{subfigure}[t]{0.23\textwidth}
        \centering
    \includegraphics[width=\linewidth, height=.6\textwidth]{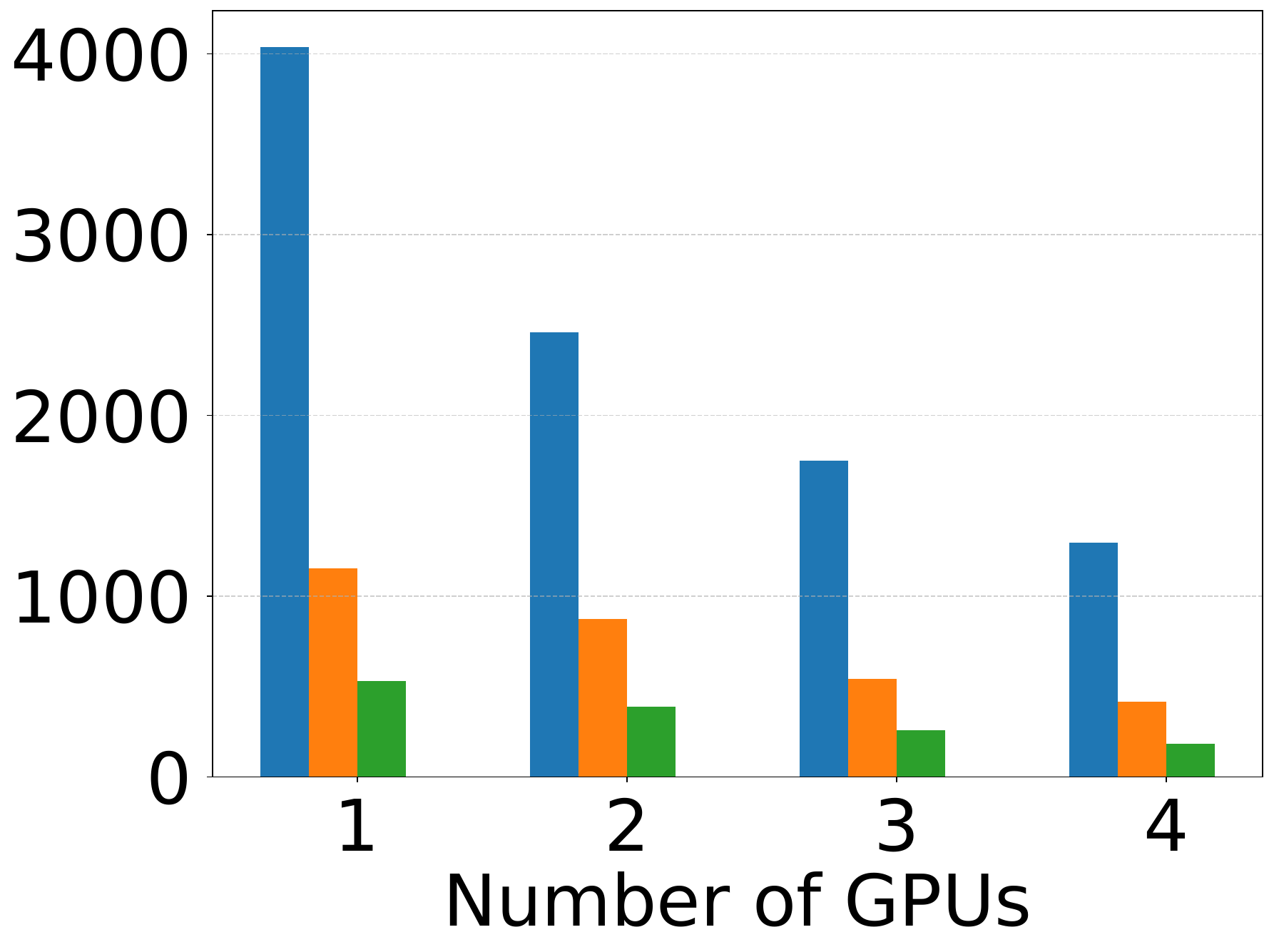}
        \caption{Img. Seg. (3D-UNet) - A100}
    \end{subfigure}

    %\vspace{-0.5em}
    %\caption*{A100 GPUs}
    \vspace{0.5em}

    % Bottom row: V100 GPUs
    \begin{subfigure}[t]{0.23\textwidth}
        \centering
        \includegraphics[width=\linewidth, height=.6\textwidth]{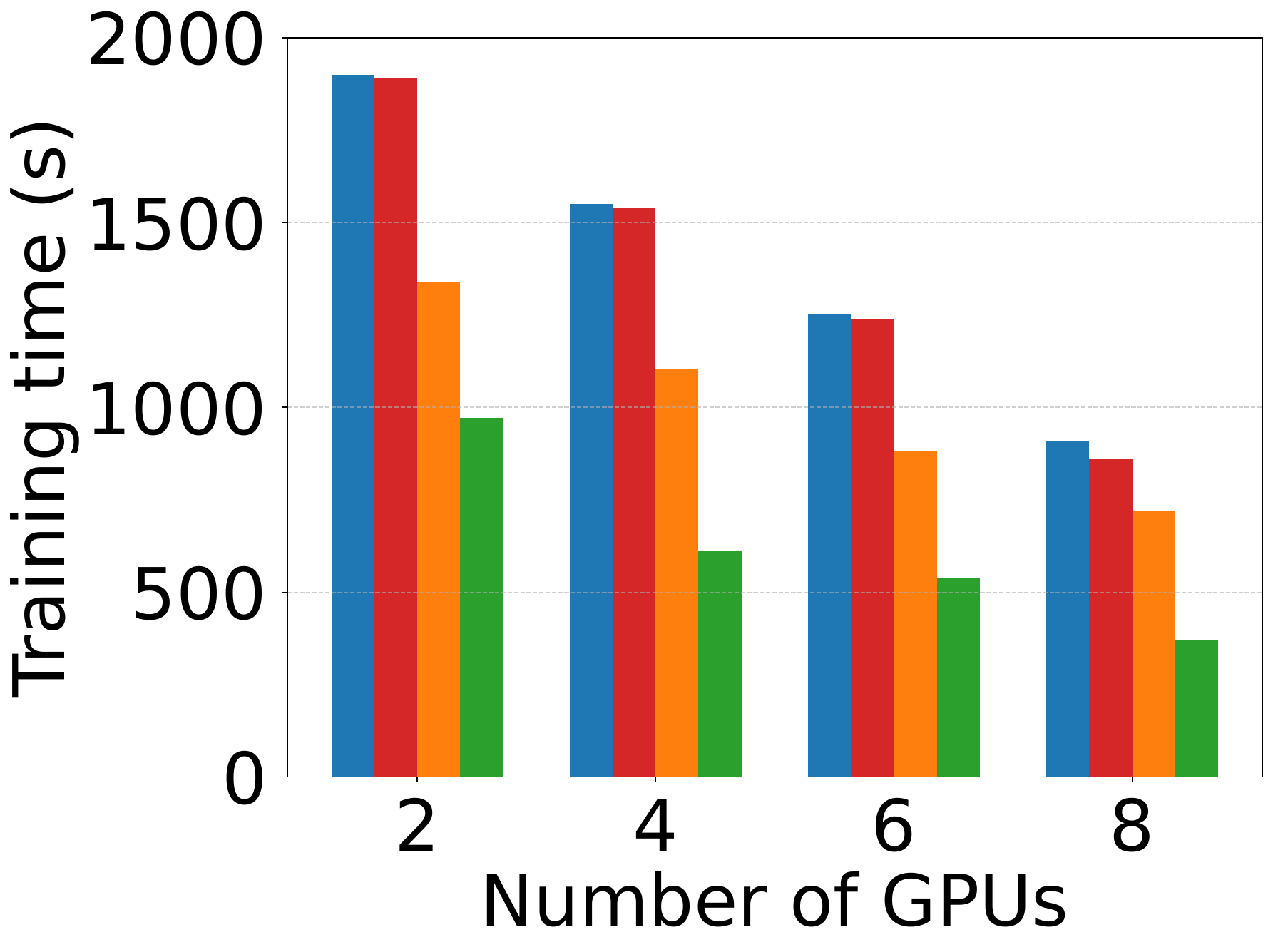}
        \caption{Speech-3s (RNN-T) - V100}
    \end{subfigure}
    \hfill
    \begin{subfigure}[t]{0.23\textwidth}
        \centering
        \includegraphics[width=\linewidth, height=.6\textwidth]{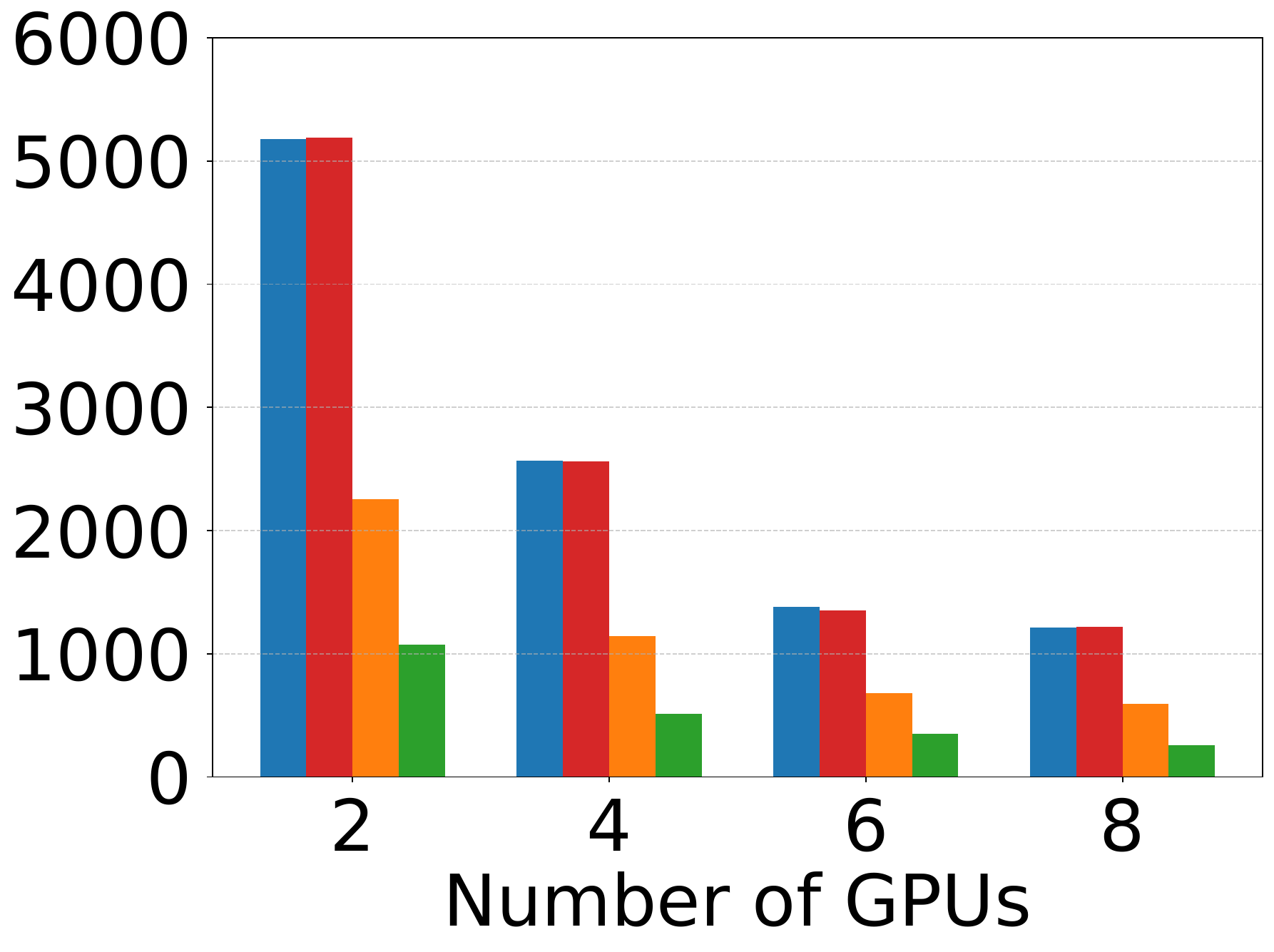}
        \caption{Speech-10s (RNN-T) - V100}
    \end{subfigure}
    \hfill
    \begin{subfigure}[t]{0.23\textwidth}
        \centering
        \includegraphics[width=\linewidth, height=.6\textwidth]{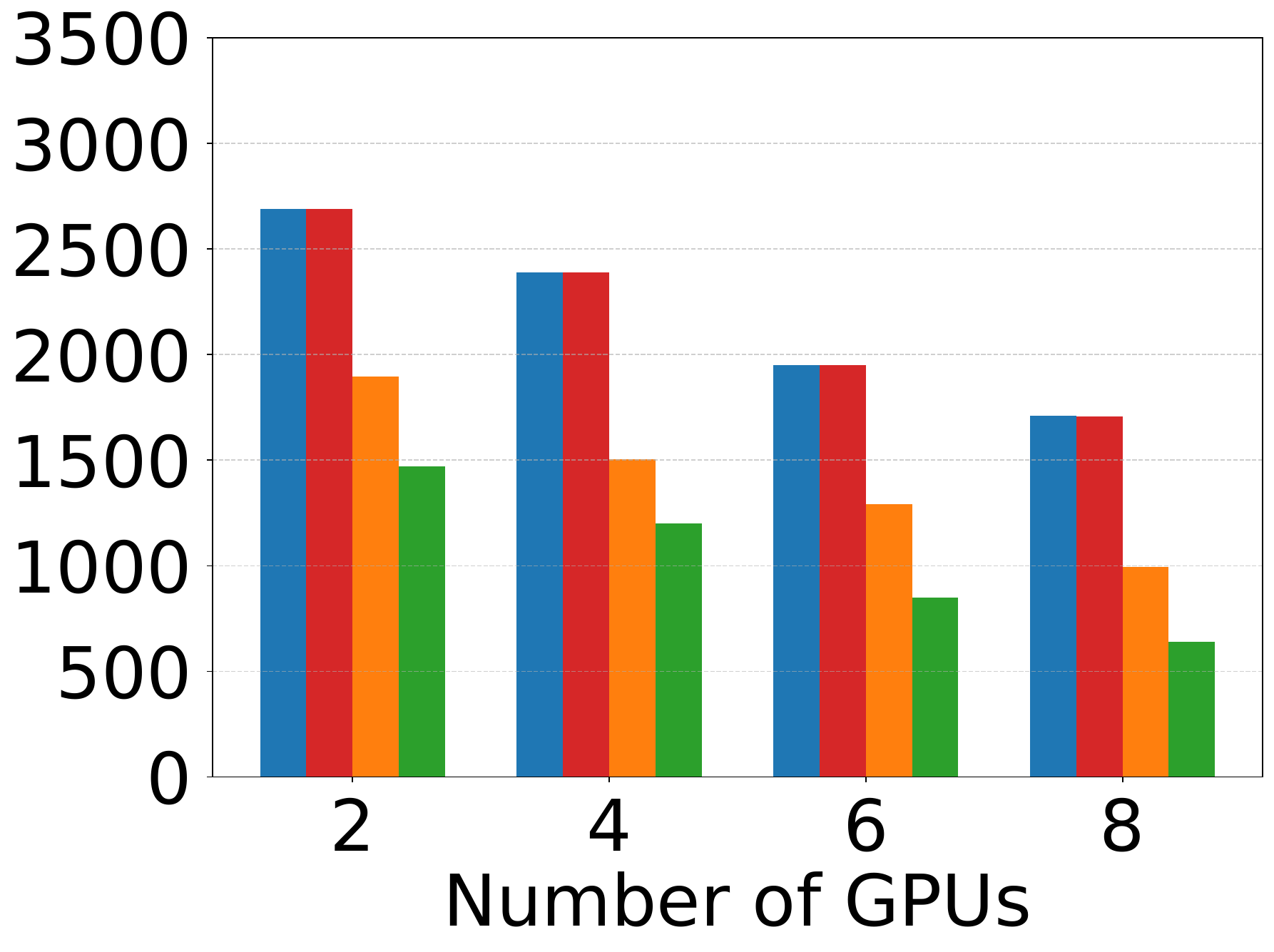}
        \caption{Obj. Det. (R-CNN) - V100}
    \end{subfigure}
    \hfill
    \begin{subfigure}[t]{0.23\textwidth}
        \centering
        \includegraphics[width=\linewidth, height=.6\textwidth]{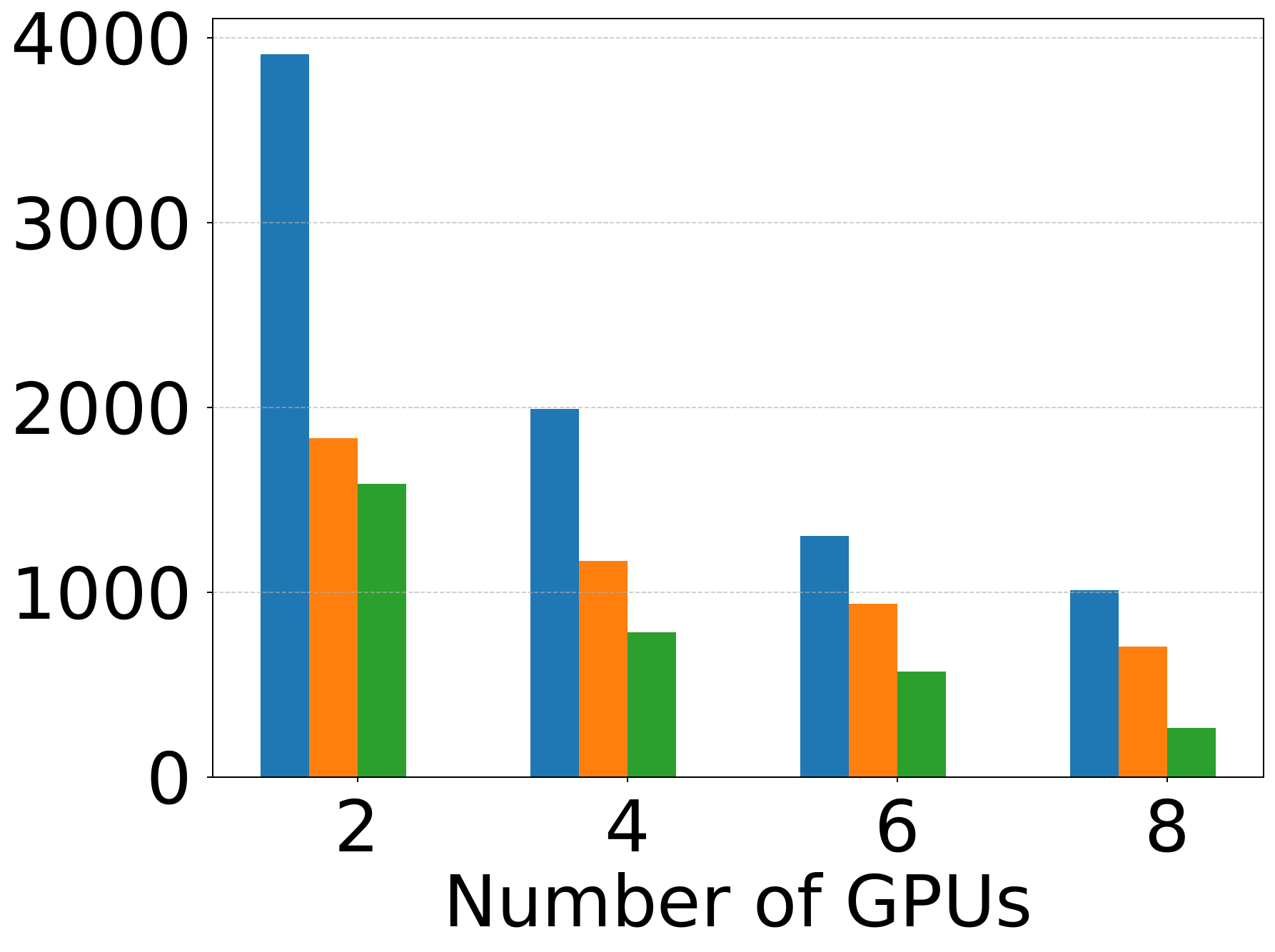}
        \caption{Img. Seg. (3D-UNet) - V100}
    \end{subfigure}

    \vspace{-0.5em}
    %\caption*{V100 GPUs}

    \caption{Training time (in seconds) of PyTorch DataLoader, Pecan, DALI, and \sys on four ML workloads executed over a varying number of A100 (top) and V100 GPUs (bottom).}
    \label{fig:training-time}
    %\vspace{-10pt}
\end{figure*}

%% file: floaters/eval-memory-constraints.tex
\begin{figure}[t]
    \centering
    \includegraphics[width=.9\columnwidth]{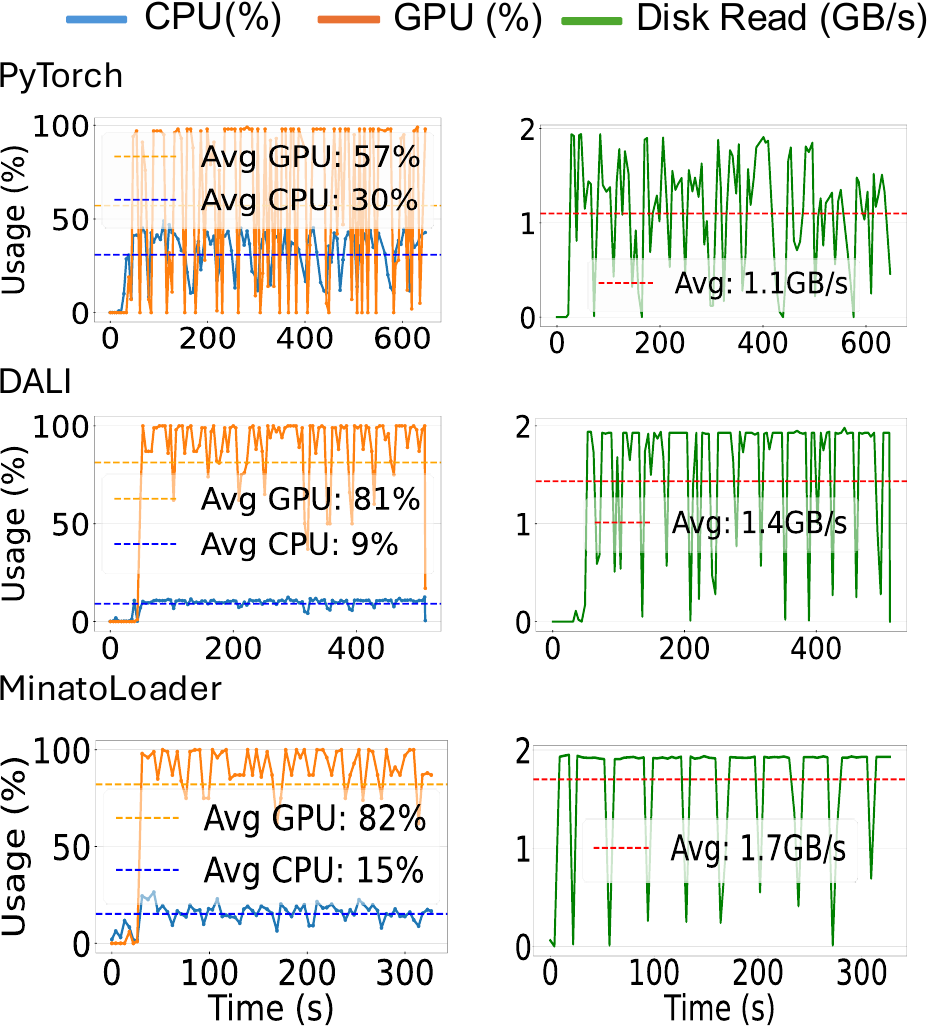}
    \vspace{-.6em}
    \caption{CPU and GPU usage (left) and disk read (right) of the image segmentation workload with PyTorch DataLoader ($\approx$650s), DALI ($\approx$500s), and \sys ($\approx$330s) when training a 230GB dataset under an 80GB memory limit.}
    \label{fig:mem-constraint}
\end{figure}

%% file: floaters/eval-accuracy-batch-composition.tex
\begin{figure}[t]
    \centering
    % --- Row 1: Accuracy ---
    \begin{subfigure}[t]{1\columnwidth}
        \begin{subfigure}[t]{0.48\columnwidth}
            \centering
            \includegraphics[width=\linewidth]{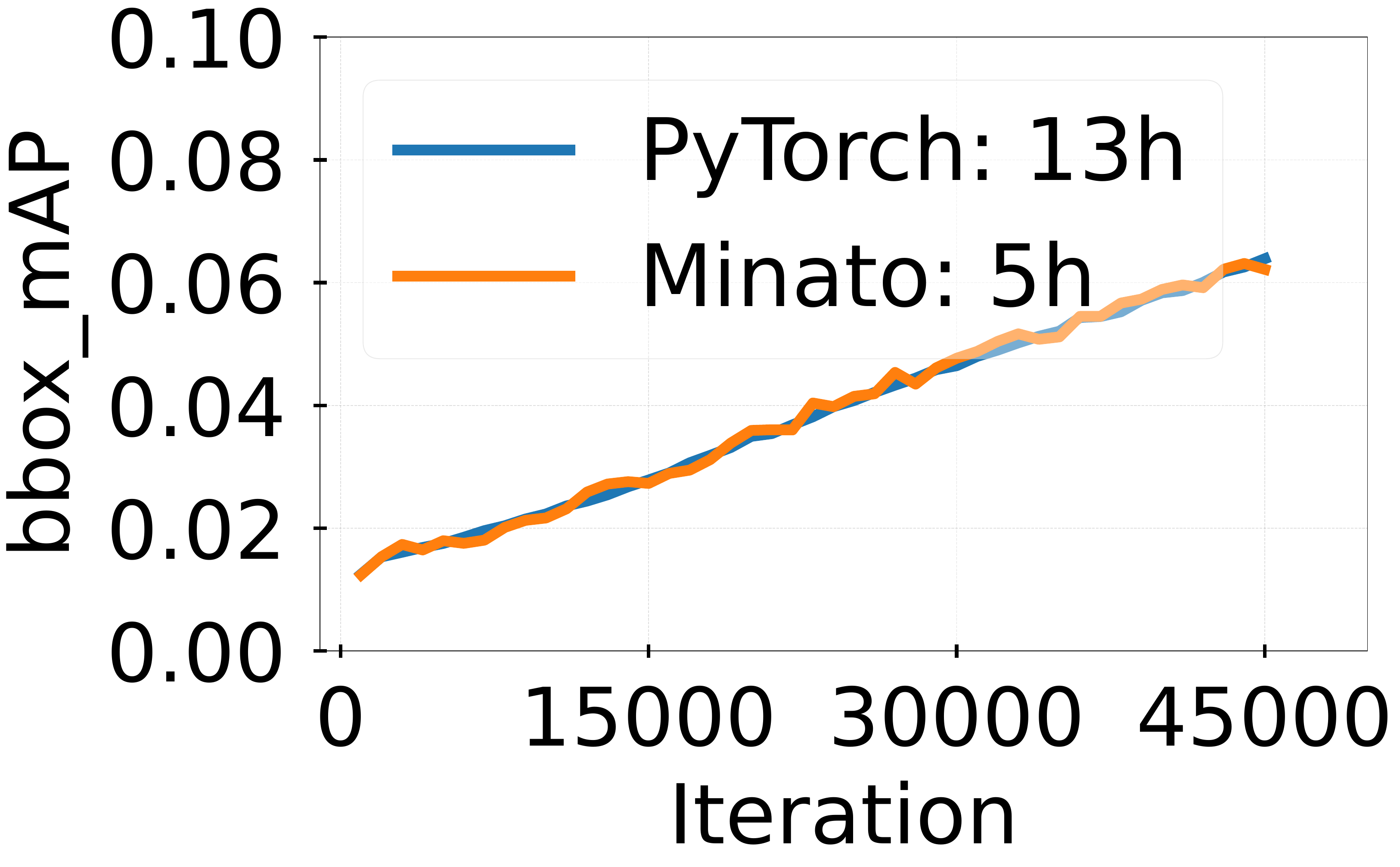}
            %\caption{Accuracy (Obj.Det)}
            %\label{subfig:accuracy-object}
        \end{subfigure}
        \hfill
        \begin{subfigure}[t]{0.48\columnwidth}
            \centering
            \includegraphics[width=\linewidth]{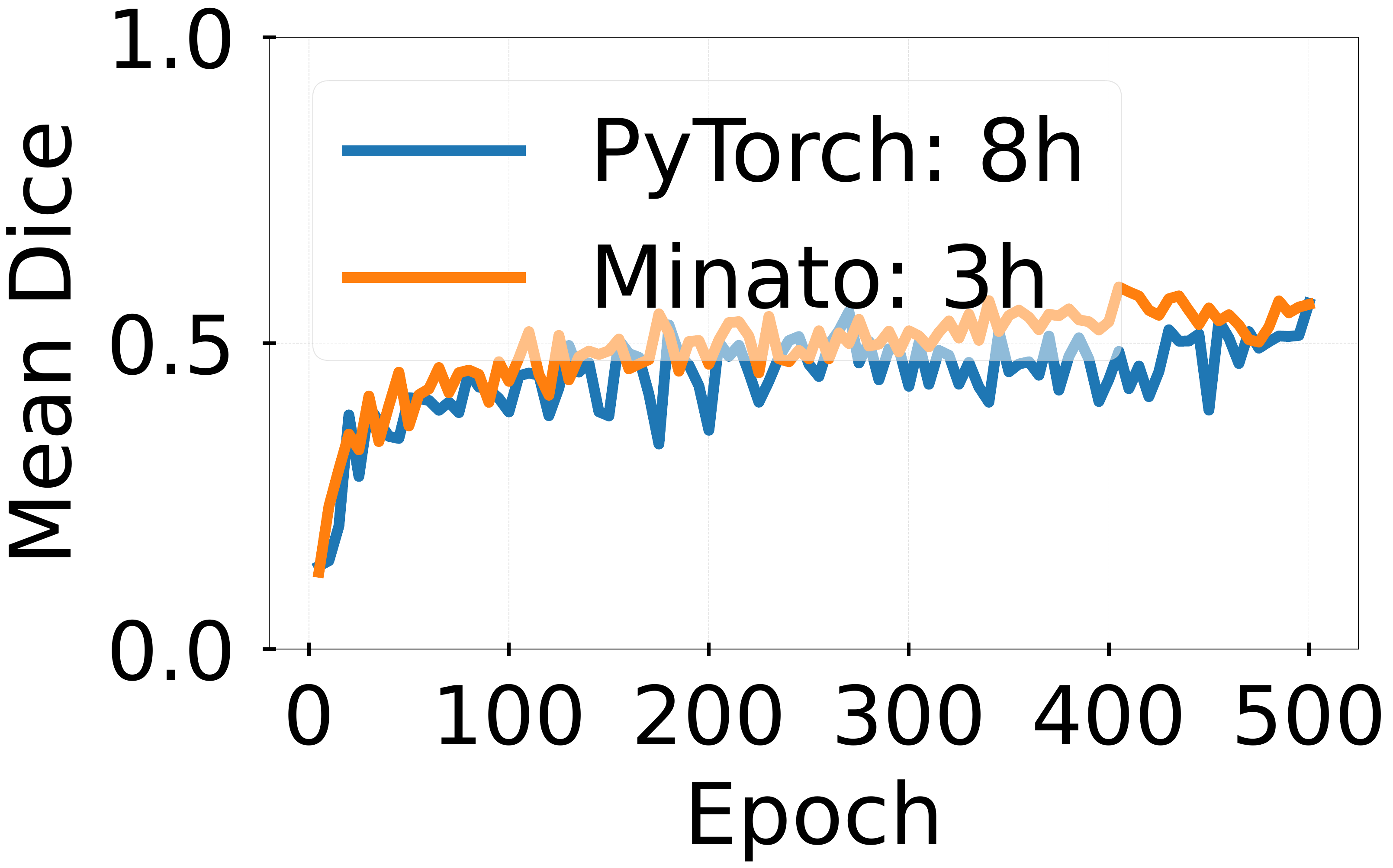}
            %\caption{Accuracy (Im.Seg)}
            %\label{subfig:accuracy-3dunet}
        \end{subfigure}
        \caption{Accuracy of PyTorch DataLoader and \sys.} % in the object detection (left) and image segmentation (right) workloads.}
        \label{subfig:accuracy}
    \end{subfigure}

    % --- Row 2: Batch composition ---
    \begin{subfigure}[t]{1\columnwidth}
        \begin{subfigure}[t]{0.48\columnwidth}
            \centering
            \includegraphics[width=\linewidth]{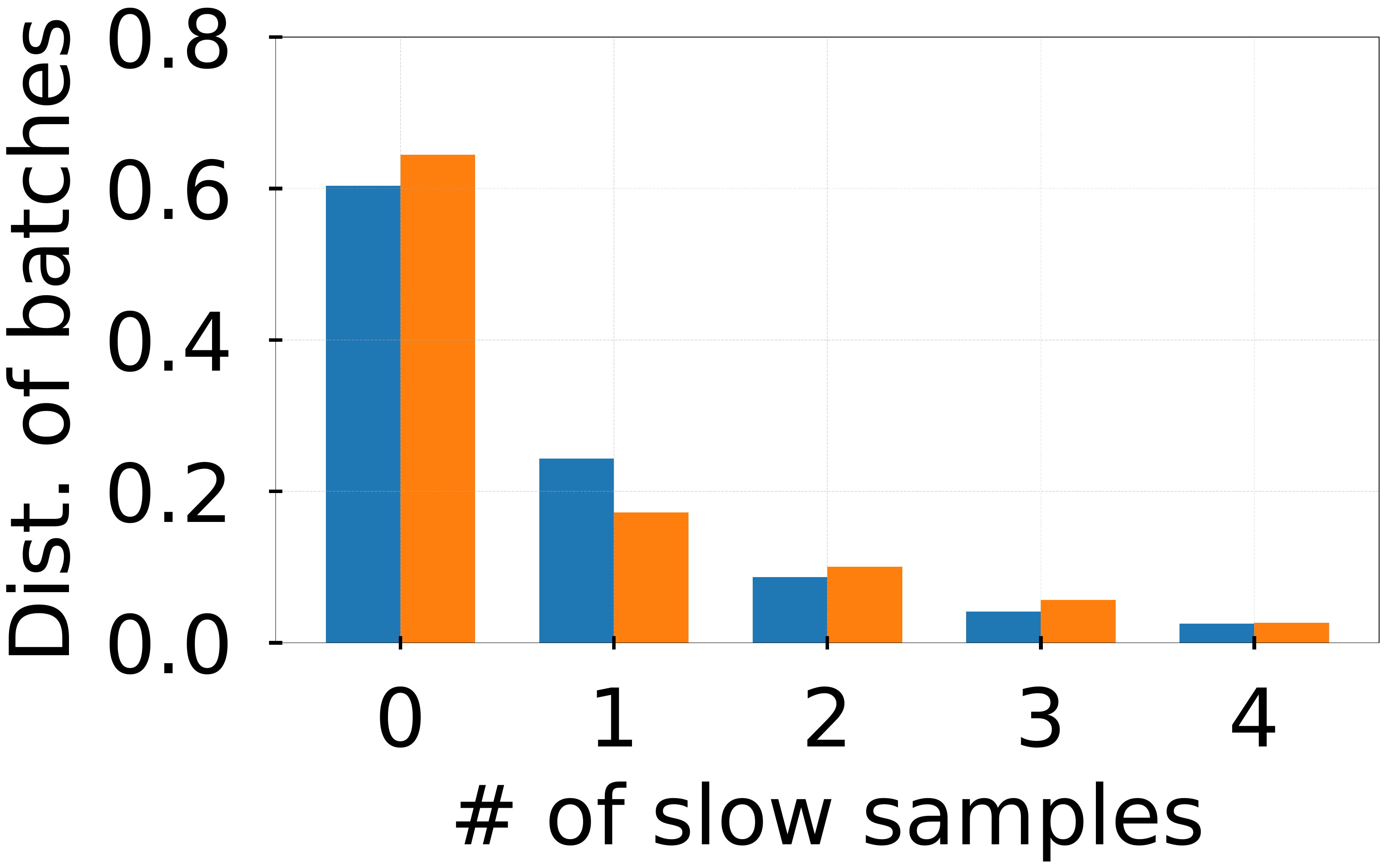}
            %\caption{Fraction of batches (Obj.Det)}
            %\label{subfig:batchcomp-object}
        \end{subfigure}
        \hfill
        \begin{subfigure}[t]{0.48\columnwidth}
            \centering
            \includegraphics[width=\linewidth]{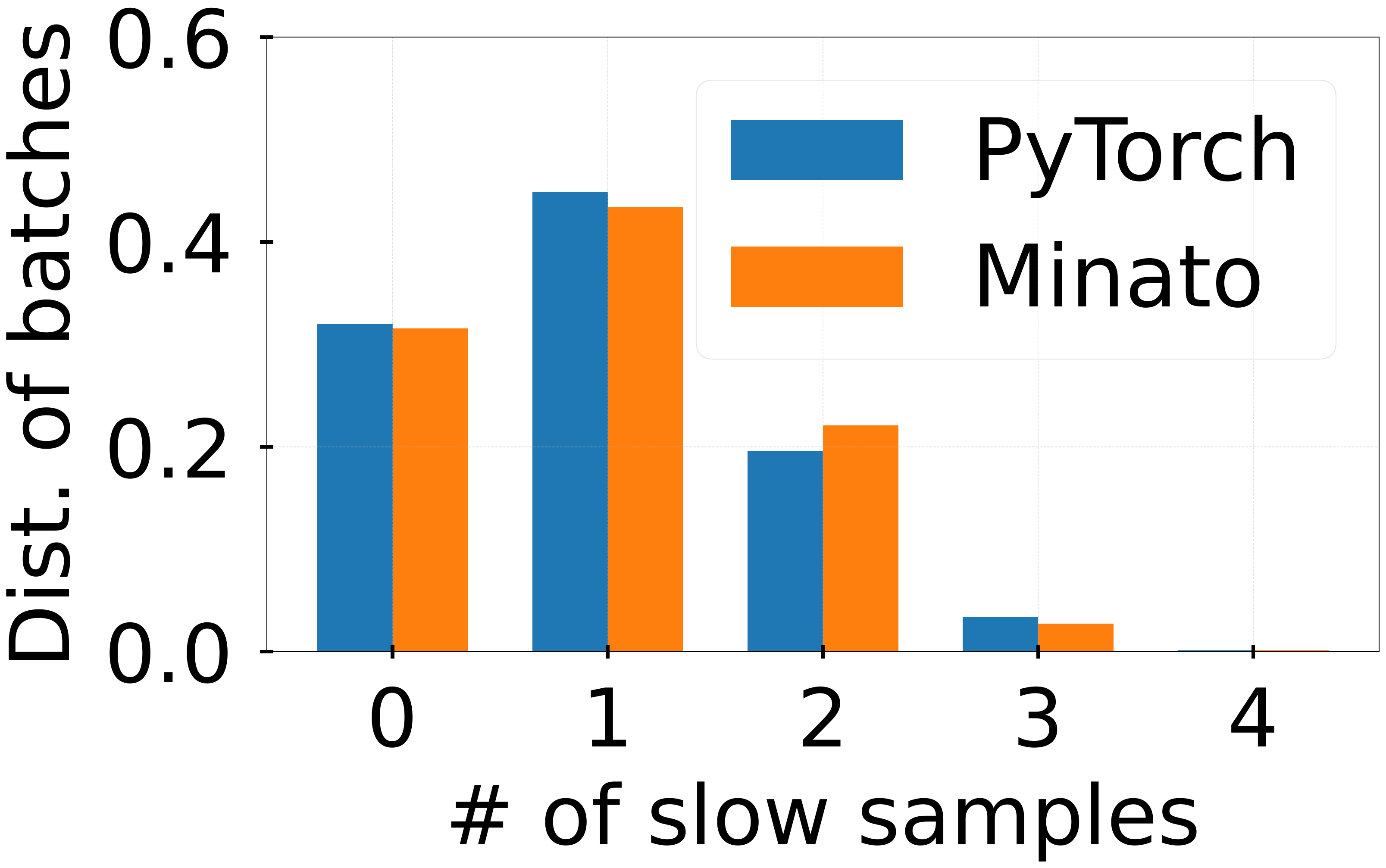}
            %\caption{Fraction of batches (Im.Seg)}
            %\label{subfig:batchcomp-3dunet}
        \end{subfigure}
        \caption{Distribution of batches by the number of slow samples.}% for PyTorch DataLoader and \sys in the object detection (left) and image segmentation (right) workloads.}
        \label{subfig:batchcomp}
    \end{subfigure}
    
    % --- Row 3: Fraction of slow samples ---
    \begin{subfigure}[t]{1\columnwidth}
        \begin{subfigure}[t]{0.48\columnwidth}
            \centering
            \includegraphics[width=\linewidth]{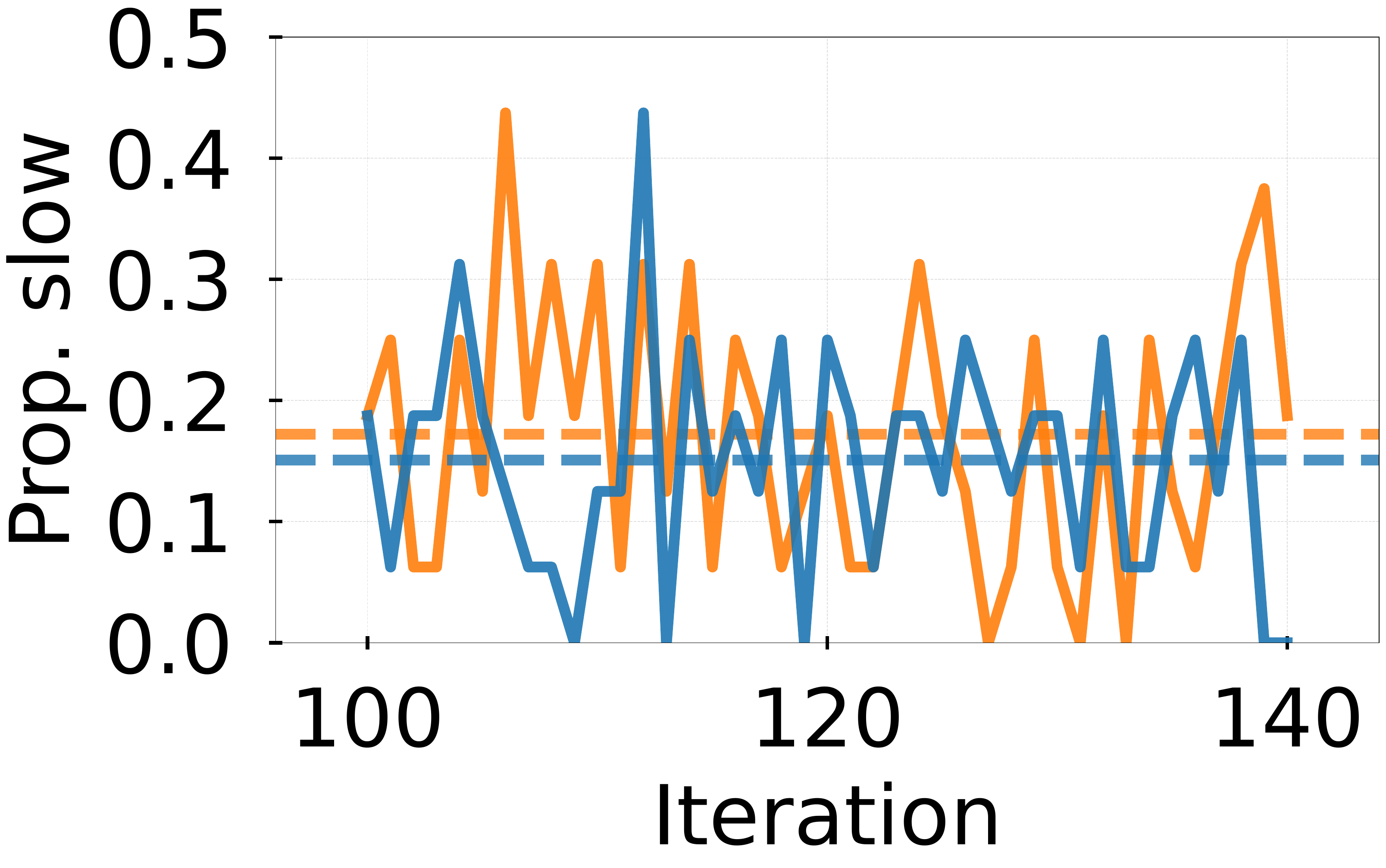}
            % \caption{Fraction of slow samples. (Obj.Det)}
            % \label{subfig:slow-object}
        \end{subfigure}
        \hfill
        \begin{subfigure}[t]{0.48\columnwidth}
            \centering
            \includegraphics[width=\linewidth]{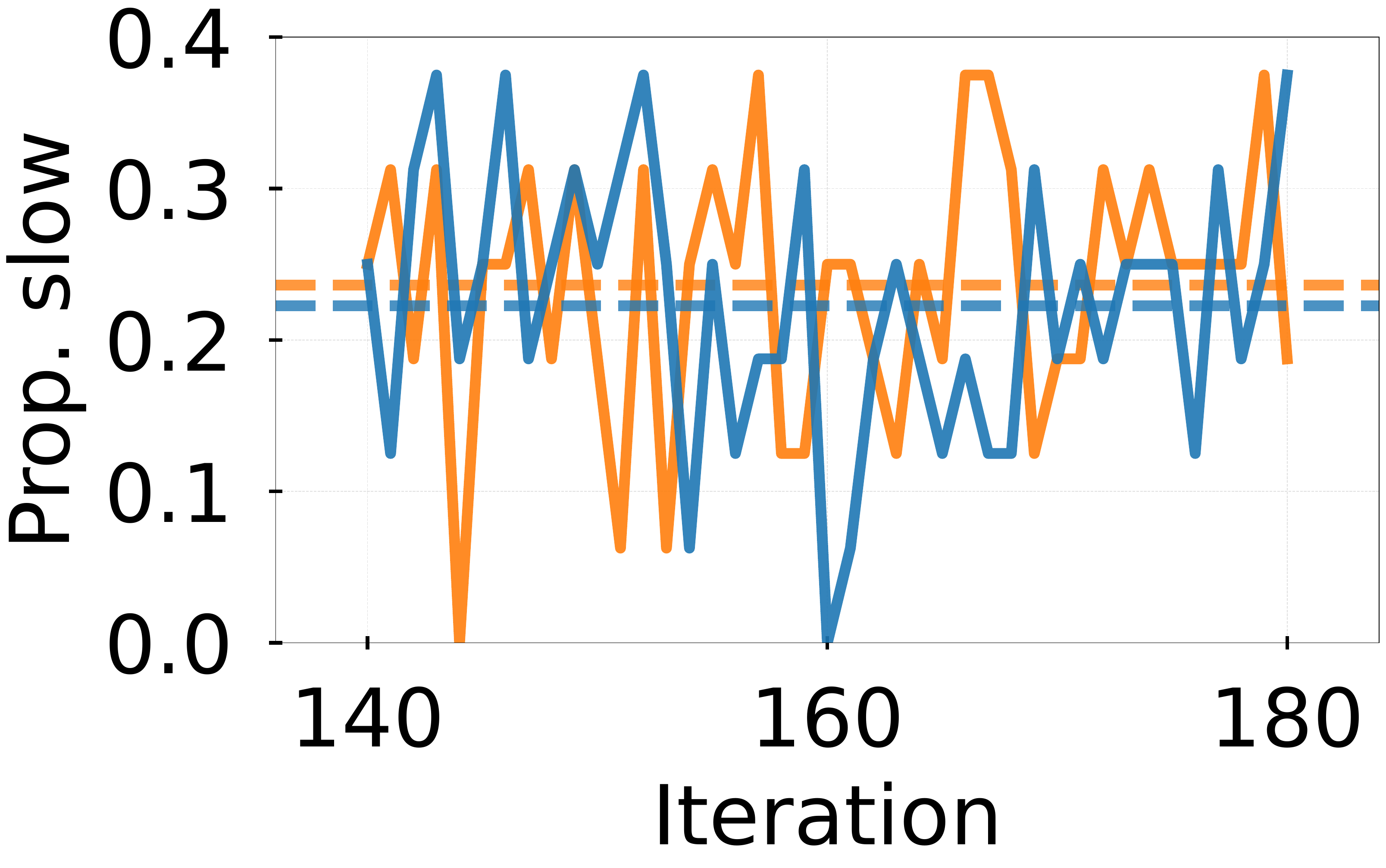}
            % \caption{Fraction of slow samples. (Im.Seg)}
            % \label{subfig:slow-3dunet}
        \end{subfigure}
        \caption{Proportion of slow samples over training iterations. % for PyTorch DataLoader and \sys in the object detection (left) and image segmentation (right) workloads. 
        Dashed lines show the average fraction of slow samples. 
        For readability, only a subset of the execution is shown.}
        \label{subfig:slow-samples}
    \end{subfigure}

    \vspace{-.8em}
%     \caption{Sensitivity analysis results for two workloads namely Object detection (Obj.Det) and Image segmentation (Img.Seg). 
%     Row 1: Accuracy; Row 2: Fraction of batches per number of samples; 
% Row 3: Timeline of the slow-sample fraction over iterations (dashed line shows the 
% average fraction of slow samples: orange for \sys, blue for PyTorch).
% .}
    \caption{Sensitivity analysis of PyTorch DataLoader (in blue) and \sys (in orange) with object detection (left) and image segmentation (right) workloads.}
    \vspace{-4.5mm}
    \label{fig:sensitivity-analysis}
\end{figure}

%% file: floaters/eval-slow-samples.tex
\begin{figure}[t]  
    \includegraphics[width=0.9\columnwidth]{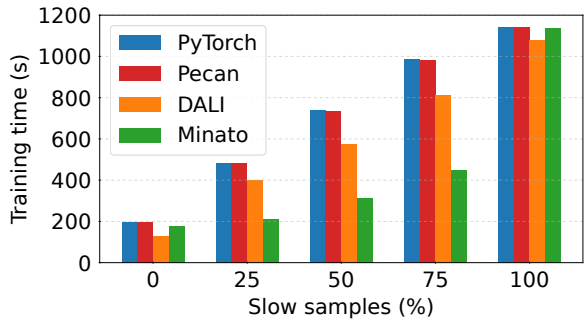}
    \caption{Training time of PyTorch DataLoader, Pecan, DALI, and \sys across different proportions of slow samples.} %Each bar shows the end-to-end training time for PyTorch, Pecan, DALI, and \sys.}
    \vspace{-5mm}
    \label{fig:cluster-slow-samples}
\end{figure}

%% file: appendix.tex
\appendix

\section{Artifact Appendix} 

\subsection{Abstract}
{\em This artifact provides a comparative evaluation of three systems - \sys, PyTorch DataLoader, and NVIDIA DALI - on the 3D-UNet workload.}

%%%%%%%%%%%%%%%%%%%%%%%%%%%%%%%%%%%%%%%%%%%%%%%%%%%%%%%%%%%%%%%%%%%%%
\subsection{Description \& Requirements}

\subsubsection{How to access}

\textit{The artifact is publicly available at:  
\url{https://github.com/Rahm-no/MinatoLoader}
and \url{https://doi.org/10.5281/zenodo.17048007}.
The repository contains all the necessary code, Dockerfile, scripts, and instructions to reproduce a minimal example from the paper.}
% Note: This evaluation do not mandate the use of specific public repositories, so institutional repositories, or open commercial repositories are acceptable. In any case, repositories used to archive the artifact should have a declared plan to enable permanent accessibility.

\subsubsection{Hardware dependencies}

\textit{The experiments in this artifact were run using a single node machine with 8 × NVIDIA Tesla V100-SXM2-32GB, 503 GB of RAM, and 446 GB SSD. 
If the reviewers want access to our setup, we will grant them.}

\subsubsection{Software dependencies}

\textit{All software dependencies (CUDA 12.6, cuDNN 8.9.7, PyTorch 2.4.1, DALI, Python 3.8, system libraries) are packaged inside the Docker image. The Docker version used is 28.1.1.}

\subsubsection{Benchmarks} 
\textit{The artifact experiments use the 2019 Kidney Tumor Segmentation Challenge (KiTS19) dataset~\cite{kits19}. The workload is a 3D-UNet model variant derived from MLCommons Training Image Segmentation~\cite{mlcommons-github}. Additional evaluations in the paper used COCO~\cite{cocodataset} (object detection) and LibriSpeech~\cite{librispeech}(speech recognition).}

%%%%%%%%%%%%%%%%%%%%%%%%%%%%%%%%%%%%%%%%%%%%%%%%%%%%%%%%%%%%%%%%%%%%%
\subsection{Set-up}

\begin{enumerate}[leftmargin=*]
  \item \textbf{Clone the repository:}
  \begin{lstlisting}[language=bash]
  git clone https://github.com/Rahm-no/MinatoLoader.git
  cd MinatoLoader
  \end{lstlisting}

  \item \textbf{Build the Docker image ($\approx$ 5 minutes):}
  \begin{lstlisting}[language=bash]
  docker build -t minato:latest .
  \end{lstlisting}

  \item \textbf{Download the KiTS19 dataset ($\approx$ 48 minutes, 27 GB):}
  \begin{lstlisting}[language=bash]
  cd raw-data-dir/kits19
  pip3 install -r requirements.txt
  python3 -m starter_code.get_imaging
  \end{lstlisting}

  \item \textbf{Start the container:}
  \begin{lstlisting}[language=bash]
  ./start_container.sh
  \end{lstlisting}

  \item \textbf{Preprocess the dataset ($\approx$ 13 minutes, 29 GB):}
  \begin{lstlisting}[language=bash]
  python3 preprocess_dataset.py \
      --data_dir /raw_data \
      --results_dir /data
  \end{lstlisting}
\end{enumerate}

%%%%%%%%%%%%%%%%%%%%%%%%%%%%%%%%%%%%%%%%%%%%%%%%%%%%%%%%%%%%%%%%%%%%%
\subsection{Evaluation workflow}
% {\em [Mandatory]} \textit{This section should include all the operational steps and experiments which must be performed to evaluate your artifact is functional and to validate your paper's key results and claims. For that purpose, we ask you to use the two following subsections and cross-reference the items therein as explained next.}
\subsubsection{Major Claims}

\begin{itemize}[leftmargin=*]
 \item (C1): \sys reduces end-to-end training time compared to PyTorch DataLoader and NVIDIA DALI. This is validated by runtime measurements across the three systems (\textbf{\S Experiment (E1))}.
    \item (C2): \sys improves GPU utilization compared to PyTorch DataLoader. This is demonstrated by utilization traces showing that PyTorch suffers from significant under-utilization, while MinatoLoader sustains consistently higher GPU usage (\textbf{\S Experiment(E2))}.
\end{itemize}

%     \item \textit{(C1): \sys achieves the same accuracy of the state-of-the-art systems for a task X while saving 2x storage resources. This is proven by the experiment (E1) described in [refer to your paper's sections] whose results are illustrated/reported in [refer to your paper's plots, tables, sections or the sort].}
%     \item \textit{(C2): System\_name has been used to uncover new bugs in the Y software. This is proven by the experiments (E2) and (E3) in [ibid].}
% \end{itemize}

\subsubsection{Experiments}
Both experiments are executed automatically when running the \texttt{run\_all.sh} script, which sequentially launches each system and collects training throughput and training time, and GPU/CPU utilization statistics.
% \textit{Link explicitly the description of your experiments to the items you have provided in the previous subsection about Major Claims. We also highly encourage you to provide your estimates of human- and compute-time for each of the listed experiments. Follows an example:}
% ~\\\\ 

\noindent
\textit{\textbf{\S Experiment (E1)}: [Training Time] [\textasciitilde 10 minutes ]: Measures and compares the time-to-train for PyTorch, NVIDIA DALI, and MinatoLoader on the 3D-UNet workload.}

\noindent
\textit{\textbf{[Preparation]}}
Build the Docker image using the provided \texttt{Dockerfile} and start the container with \begin{lstlisting}[language=bash]
./start_container.sh
\end{lstlisting}

\noindent
\textit{\textbf{[Execution]}} Run all systems sequentially (PyTorch, DALI, MinatoLoader) on 8 GPUs for 10 epochs using:  
\begin{lstlisting}[language=bash]
./scripts/run_all.sh NUM_GPUs
\end{lstlisting}
% Alternatively, run individual systems with \texttt{./run\_SYSTEM.sh NUM\_GPUs}, 
% where \texttt{SYSTEM} is \texttt{pytorch}, \texttt{dali}, or \texttt{minato} 
% under each system directory.

\noindent\textit{\textbf{[Results]}} 
After execution, training times are automatically appended 
to \texttt{results/results\_allsystems.csv}. 

\noindent
Expected runtimes on 8$\times$ V100 GPUs (10 epochs):  
PyTorch: $\approx$210 sec, DALI: $\approx$151 sec, and \sys: $\approx$81 sec.  

These results are consistent with the paper claim (C1) that \sys provides a speedup of $2.6\times$ compared to PyTorch and $1.9\times$ compared to DALI.  
They are lower than the numbers reported in Section~\ref{sec:eval} because we used a minimal setup here (\emph{i.e.,} 10 epochs instead of 50).

The CSV file allows direct comparison across systems. The expected outcome is that MinatoLoader achieves the lowest training time due to reduced GPU stalls, whereas PyTorch shows the worst performance due to poor utilization. You can choose to plot training time results using
\begin{lstlisting}[language=bash]
python3 scripts/plot_figure.py
\end{lstlisting}

\balance
\noindent
\textit{\textbf{\S Experiment (E2):} [Resource Utilization Analysis] [\textasciitilde 10 minutes]: 
Evaluates the average CPU and GPU utilization of PyTorch, NVIDIA DALI, and MinatoLoader 
during training of the 3D-UNet workload.}

\noindent
\textit{\textbf{[Preparation \& Execution]}} Same as \textbf{E1}.
\noindent
\textit{\textbf{[Results]}} 
Once execution completes successfully, three CSV files (one per system) are generated in the results folder.  
Each file contains the average resource utilization over time for the corresponding system.  
You can plot the results to visualize differences in utilization between the three data loaders by running:  

\begin{lstlisting}[language=bash]
python3 scripts/plot_usage.py
\end{lstlisting}

The plots show that NVIDIA DALI consistently achieves high GPU utilization because preprocessing is also performed on the GPU.  
By contrast, PyTorch exhibits frequent GPU idle periods accompanied by CPU peaks, indicating that while the CPU is busy with preprocessing, the GPU is left waiting.  
For \sys, the GPU remains consistently and highly utilized throughout training, demonstrating efficient overlap of preprocessing and computation.

\balance

%% file: revision.bbl
%%% -*-BibTeX-*-
%%% Do NOT edit. File created by BibTeX with style
%%% ACM-Reference-Format-Journals [18-Jan-2012].

\begin{thebibliography}{48}

%%% ====================================================================
%%% NOTE TO THE USER: you can override these defaults by providing
%%% customized versions of any of these macros before the \bibliography
%%% command.  Each of them MUST provide its own final punctuation,
%%% except for \shownote{}, \showDOI{}, and \showURL{}.  The latter two
%%% do not use final punctuation, in order to avoid confusing it with
%%% the Web address.
%%%
%%% To suppress output of a particular field, define its macro to expand
%%% to an empty string, or better, \unskip, like this:
%%%
%%% \newcommand{\showDOI}[1]{\unskip}   % LaTeX syntax
%%%
%%% \def \showDOI #1{\unskip}           % plain TeX syntax
%%%
%%% ====================================================================

\ifx \showCODEN    \undefined \def \showCODEN     #1{\unskip}     \fi
\ifx \showDOI      \undefined \def \showDOI       #1{#1}\fi
\ifx \showISBNx    \undefined \def \showISBNx     #1{\unskip}     \fi
\ifx \showISBNxiii \undefined \def \showISBNxiii  #1{\unskip}     \fi
\ifx \showISSN     \undefined \def \showISSN      #1{\unskip}     \fi
\ifx \showLCCN     \undefined \def \showLCCN      #1{\unskip}     \fi
\ifx \shownote     \undefined \def \shownote      #1{#1}          \fi
\ifx \showarticletitle \undefined \def \showarticletitle #1{#1}   \fi
\ifx \showURL      \undefined \def \showURL       {\relax}        \fi
% The following commands are used for tagged output and should be
% invisible to TeX
\providecommand\bibfield[2]{#2}
\providecommand\bibinfo[2]{#2}
\providecommand\natexlab[1]{#1}
\providecommand\showeprint[2][]{arXiv:#2}

\bibitem[kit({[n.\,d.]})]%
        {kits19}
 \bibinfo{year}{[n.\,d.]}\natexlab{}.
\newblock \bibinfo{title}{KiTS19 Challenge Dataset}.
\newblock \bibinfo{howpublished}{\url{https://kits19.grand-challenge.org/data/}}.
\newblock
\newblock
\shownote{Accessed: [Jan 12, 2025]}.


\bibitem[mlp({[n.\,d.]})]%
        {mlperf-training-results}
 \bibinfo{year}{[n.\,d.]}\natexlab{}.
\newblock \bibinfo{title}{MLPerf Training Benchmark Suite V3.1 Results}.
\newblock \bibinfo{howpublished}{\url{https://mlcommons.org/benchmarks/training/}}.
\newblock
\newblock
\shownote{Accessed: [May 5, 2025]}.


\bibitem[num({[n.\,d.]})]%
        {numpy}
 \bibinfo{year}{[n.\,d.]}\natexlab{}.
\newblock \bibinfo{title}{NumPy - The fundamental Package {for scientific computing with Python}}.
\newblock \bibinfo{howpublished}{\url{https://numpy.org/}}.
\newblock
\newblock
\shownote{Accessed: [May 05, 2025]}.


\bibitem[dal({[n.\,d.]})]%
        {dali}
 \bibinfo{year}{[n.\,d.]}\natexlab{}.
\newblock \bibinfo{title}{NVIDIA Data Loading Library (DALI)}.
\newblock \bibinfo{howpublished}{\url{https://developer.nvidia.com/dali}}.
\newblock
\newblock
\shownote{Accessed: [May 5, 2025]}.


\bibitem[pan({[n.\,d.]})]%
        {pandas}
 \bibinfo{year}{[n.\,d.]}\natexlab{}.
\newblock \bibinfo{title}{Pandas: Powerful Python Data Analysis Toolkit}.
\newblock \bibinfo{howpublished}{\url{https://pypi.org/project/pandas/}}.
\newblock
\newblock
\shownote{Accessed: [May 05, 2025]}.


\bibitem[sci({[n.\,d.]})]%
        {scikit}
 \bibinfo{year}{[n.\,d.]}\natexlab{}.
\newblock \bibinfo{title}{Scikit-learn - Machine Learning in Python}.
\newblock \bibinfo{howpublished}{\url{https://scikit-learn.org/stable/}}.
\newblock
\newblock
\shownote{Accessed: [May 05, 2025]}.


\bibitem[nvi(2020)]%
        {nvidia_a100}
 \bibinfo{year}{2020}\natexlab{}.
\newblock \bibinfo{booktitle}{\emph{{NVIDIA A100 Tensor Core GPU}}}.
\newblock
\urldef\tempurl%
\url{https://www.nvidia.com/en-us/data-center/a100/}
\showURL{%
\tempurl}
\newblock
\shownote{Accessed: May 11, 2025}.


\bibitem[nvm(2020)]%
        {nvml}
 \bibinfo{year}{2020}\natexlab{}.
\newblock \bibinfo{booktitle}{\emph{{NVIDIA Managemnet Library (NVML)}}}.
\newblock
\urldef\tempurl%
\url{https://developer.nvidia.com/management-library-nvml}
\showURL{%
\tempurl}
\newblock
\shownote{Accessed: May 11, 2025}.


\bibitem[pyt(2025)]%
        {pytorch_dataloaders_tutorial}
 \bibinfo{year}{2025}\natexlab{}.
\newblock \bibinfo{title}{{Datasets \& DataLoaders}}.
\newblock \bibinfo{howpublished}{\url{https://pytorch.org/tutorials/beginner/basics/data_tutorial.html}}.
\newblock
\newblock
\shownote{Accessed: May 14, 2025}.


\bibitem[dst(2025)]%
        {dstat}
 \bibinfo{year}{2025}\natexlab{}.
\newblock \bibinfo{booktitle}{\emph{{dstat - Versatile Tool for Generating System Resource Metrics}}}.
\newblock
\urldef\tempurl%
\url{https://linux.die.net/man/1/dstat}
\showURL{%
\tempurl}
\newblock
\shownote{Accessed: May 14, 2025}.


\bibitem[GIL(2025)]%
        {GIL_problems}
 \bibinfo{year}{2025}\natexlab{}.
\newblock \bibinfo{title}{{Global Interpreter Lock}}.
\newblock \bibinfo{howpublished}{\url{https://pybind11.readthedocs.io/en/stable/advanced/misc.html}}.
\newblock
\newblock
\shownote{Accessed: May 14, 2025}.


\bibitem[spa(2025)]%
        {spark_streaming_guide}
 \bibinfo{year}{2025}\natexlab{}.
\newblock \bibinfo{title}{Spark Streaming Programming Guide}.
\newblock \bibinfo{howpublished}{\url{https://spark.apache.org/docs/latest/streaming-programming-guide.html}}.
\newblock


\bibitem[Bachkaniwala et~al\mbox{.}(2024)]%
        {bachkaniwala2024lotus}
\bibfield{author}{\bibinfo{person}{Rajveer Bachkaniwala}, \bibinfo{person}{Harshith Lanka}, \bibinfo{person}{Kexin Rong}, {and} \bibinfo{person}{Ada Gavrilovska}.} \bibinfo{year}{2024}\natexlab{}.
\newblock \showarticletitle{Lotus: Characterization of Machine Learning Preprocessing Pipelines via Framework and Hardware Profiling}. In \bibinfo{booktitle}{\emph{IEEE International Symposium on Workload Characterization}}. \bibinfo{publisher}{{IEEE}}, \bibinfo{pages}{30--43}.
\newblock
\urldef\tempurl%
\url{https://doi.org/10.1109/IISWC63097.2024.00013}
\showDOI{\tempurl}


\bibitem[Balmau(2022)]%
        {balmau2022characterizing}
\bibfield{author}{\bibinfo{person}{Oana Balmau}.} \bibinfo{year}{2022}\natexlab{}.
\newblock \showarticletitle{Characterizing {I/O} in Machine Learning with MLPerf Storage}.
\newblock \bibinfo{journal}{\emph{{SIGMOD} Rec.}} \bibinfo{volume}{51}, \bibinfo{number}{3} (\bibinfo{year}{2022}), \bibinfo{pages}{47--48}.
\newblock
\urldef\tempurl%
\url{https://doi.org/10.1145/3572751.3572765}
\showDOI{\tempurl}


\bibitem[{\c{C}}i{\c{c}}ek et~al\mbox{.}(2016)]%
        {10.1007/978-3-319-46723-8_49}
\bibfield{author}{\bibinfo{person}{{\"{O}}zg{\"{u}}n {\c{C}}i{\c{c}}ek}, \bibinfo{person}{Ahmed Abdulkadir}, \bibinfo{person}{Soeren~S. Lienkamp}, \bibinfo{person}{Thomas Brox}, {and} \bibinfo{person}{Olaf Ronneberger}.} \bibinfo{year}{2016}\natexlab{}.
\newblock \showarticletitle{3D U-Net: Learning Dense Volumetric Segmentation from Sparse Annotation}. In \bibinfo{booktitle}{\emph{19th International Conference on Medical Image Computing and Computer-Assisted Intervention}}, Vol.~\bibinfo{volume}{9901}. \bibinfo{pages}{424--432}.
\newblock
\urldef\tempurl%
\url{https://doi.org/10.1007/978-3-319-46723-8\_49}
\showDOI{\tempurl}


\bibitem[Gao et~al\mbox{.}(2024)]%
        {gao2024empirical}
\bibfield{author}{\bibinfo{person}{Yanjie Gao}, \bibinfo{person}{Yichen He}, \bibinfo{person}{Xinze Li}, \bibinfo{person}{Bo Zhao}, \bibinfo{person}{Haoxiang Lin}, \bibinfo{person}{Yoyo Liang}, \bibinfo{person}{Jing Zhong}, \bibinfo{person}{Hongyu Zhang}, \bibinfo{person}{Jingzhou Wang}, \bibinfo{person}{Yonghua Zeng}, \bibinfo{person}{Keli Gui}, \bibinfo{person}{Jie Tong}, {and} \bibinfo{person}{Mao Yang}.} \bibinfo{year}{2024}\natexlab{}.
\newblock \showarticletitle{An Empirical Study on Low {GPU} Utilization of Deep Learning Jobs}. In \bibinfo{booktitle}{\emph{46th IEEE/ACM International Conference on Software Engineering}}. \bibinfo{publisher}{{ACM}}, \bibinfo{pages}{96:1--96:13}.
\newblock
\urldef\tempurl%
\url{https://doi.org/10.1145/3597503.3639232}
\showDOI{\tempurl}


\bibitem[Graur et~al\mbox{.}(2022)]%
        {graur2022cachew}
\bibfield{author}{\bibinfo{person}{Dan Graur}, \bibinfo{person}{Damien Aymon}, \bibinfo{person}{Dan Kluser}, \bibinfo{person}{Tanguy Albrici}, \bibinfo{person}{Chandramohan~A. Thekkath}, {and} \bibinfo{person}{Ana Klimovic}.} \bibinfo{year}{2022}\natexlab{}.
\newblock \showarticletitle{Cachew: Machine Learning Input Data Processing as a Service}. In \bibinfo{booktitle}{\emph{2022 USENIX Annual Technical Conference}}. \bibinfo{publisher}{{USENIX} Association}, \bibinfo{pages}{689--706}.
\newblock
\urldef\tempurl%
\url{https://www.usenix.org/conference/atc22/presentation/graur}
\showURL{%
\tempurl}


\bibitem[Graur et~al\mbox{.}(2024)]%
        {pecan}
\bibfield{author}{\bibinfo{person}{Dan Graur}, \bibinfo{person}{Oto Mraz}, \bibinfo{person}{Muyu Li}, \bibinfo{person}{Mohammad~Sepehr Pourghannad}, \bibinfo{person}{Chandramohan~A. Thekkath}, {and} \bibinfo{person}{Ana Klimovic}.} \bibinfo{year}{2024}\natexlab{}.
\newblock \showarticletitle{Pecan: Cost-Efficient {ML} Data Preprocessing with Automatic Transformation Ordering and Hybrid Placement}. In \bibinfo{booktitle}{\emph{2024 USENIX Annual Technical Conference}}. \bibinfo{publisher}{{USENIX} Association}, \bibinfo{pages}{649--665}.
\newblock
\urldef\tempurl%
\url{https://www.usenix.org/conference/atc24/presentation/graur}
\showURL{%
\tempurl}


\bibitem[Gupta et~al\mbox{.}(2022)]%
        {9744492}
\bibfield{author}{\bibinfo{person}{Udit Gupta}, \bibinfo{person}{Young~Geun Kim}, \bibinfo{person}{Sylvia Lee}, \bibinfo{person}{Jordan Tse}, \bibinfo{person}{Hsien{-}Hsin~S. Lee}, \bibinfo{person}{Gu{-}Yeon Wei}, \bibinfo{person}{David Brooks}, {and} \bibinfo{person}{Carole{-}Jean Wu}.} \bibinfo{year}{2022}\natexlab{}.
\newblock \showarticletitle{Chasing Carbon: The Elusive Environmental Footprint of Computing}.
\newblock \bibinfo{journal}{\emph{{IEEE} Micro}} \bibinfo{volume}{42}, \bibinfo{number}{4} (\bibinfo{year}{2022}), \bibinfo{pages}{37--47}.
\newblock
\urldef\tempurl%
\url{https://doi.org/10.1109/MM.2022.3163226}
\showDOI{\tempurl}


\bibitem[Heller et~al\mbox{.}(2020)]%
        {heller2019kits19}
\bibfield{author}{\bibinfo{person}{Nicholas Heller}, \bibinfo{person}{Niranjan Sathianathen}, \bibinfo{person}{Arveen Kalapara}, \bibinfo{person}{Edward Walczak}, \bibinfo{person}{Keenan Moore}, \bibinfo{person}{Heather Kaluzniak}, \bibinfo{person}{Joel Rosenberg}, \bibinfo{person}{Paul Blake}, \bibinfo{person}{Zachary Rengel}, \bibinfo{person}{Makinna Oestreich}, \bibinfo{person}{Joshua Dean}, \bibinfo{person}{Michael Tradewell}, \bibinfo{person}{Aneri Shah}, \bibinfo{person}{Resha Tejpaul}, \bibinfo{person}{Zachary Edgerton}, \bibinfo{person}{Matthew Peterson}, \bibinfo{person}{Shaneabbas Raza}, \bibinfo{person}{Subodh Regmi}, \bibinfo{person}{Nikolaos Papanikolopoulos}, {and} \bibinfo{person}{Christopher Weight}.} \bibinfo{year}{2020}\natexlab{}.
\newblock \bibinfo{title}{The KiTS19 Challenge Data: 300 Kidney Tumor Cases with Clinical Context, CT Semantic Segmentations, and Surgical Outcomes}.
\newblock
\newblock
\showeprint{1904.00445}
\urldef\tempurl%
\url{https://arxiv.org/abs/1904.00445}
\showURL{%
\tempurl}


\bibitem[Kim et~al\mbox{.}(2023)]%
        {kim2023fusionflow}
\bibfield{author}{\bibinfo{person}{Taeyoon Kim}, \bibinfo{person}{Chanho Park}, \bibinfo{person}{Mansur Mukimbekov}, \bibinfo{person}{Heelim Hong}, \bibinfo{person}{Minseok Kim}, \bibinfo{person}{Ze Jin}, \bibinfo{person}{Changdae Kim}, \bibinfo{person}{Ji{-}Yong Shin}, {and} \bibinfo{person}{Myeongjae Jeon}.} \bibinfo{year}{2023}\natexlab{}.
\newblock \showarticletitle{FusionFlow: Accelerating Data Preparation for Machine Learning with Hybrid {CPU-GPU} Processing}.
\newblock \bibinfo{journal}{\emph{Proceedings of the VLDB Endowment}} \bibinfo{volume}{17}, \bibinfo{number}{4} (\bibinfo{year}{2023}), \bibinfo{pages}{863--876}.
\newblock
\urldef\tempurl%
\url{https://doi.org/10.14778/3636218.3636238}
\showDOI{\tempurl}


\bibitem[Kotsiantis et~al\mbox{.}(2006)]%
        {10.5281/zenodo.1082415}
\bibfield{author}{\bibinfo{person}{Sotiris Kotsiantis}, \bibinfo{person}{Dimitris Kanellopoulos}, {and} \bibinfo{person}{P. Pintelas}.} \bibinfo{year}{2006}\natexlab{}.
\newblock \showarticletitle{Data Preprocessing for Supervised Learning}.
\newblock \bibinfo{journal}{\emph{International Journal of Computer Science}} (\bibinfo{year}{2006}).
\newblock


\bibitem[Kuchnik et~al\mbox{.}(2021)]%
        {kuchnik2019progressive}
\bibfield{author}{\bibinfo{person}{Michael Kuchnik}, \bibinfo{person}{George Amvrosiadis}, {and} \bibinfo{person}{Virginia Smith}.} \bibinfo{year}{2021}\natexlab{}.
\newblock \showarticletitle{Progressive Compressed Records: Taking a Byte out of Deep Learning Data}.
\newblock \bibinfo{journal}{\emph{Proceedings of the VLDB Endowment}} \bibinfo{volume}{14}, \bibinfo{number}{11} (\bibinfo{year}{2021}), \bibinfo{pages}{2627--2641}.
\newblock
\urldef\tempurl%
\url{https://doi.org/10.14778/3476249.3476308}
\showDOI{\tempurl}


\bibitem[Lee et~al\mbox{.}(2021)]%
        {lee2021refurbish}
\bibfield{author}{\bibinfo{person}{Gyewon Lee}, \bibinfo{person}{Irene Lee}, \bibinfo{person}{Hyeonmin Ha}, \bibinfo{person}{Kyung{-}Geun Lee}, \bibinfo{person}{Hwarim Hyun}, \bibinfo{person}{Ahnjae Shin}, {and} \bibinfo{person}{Byung{-}Gon Chun}.} \bibinfo{year}{2021}\natexlab{}.
\newblock \showarticletitle{Refurbish Your Training Data: Reusing Partially Augmented Samples for Faster Deep Neural Network Training}. In \bibinfo{booktitle}{\emph{2021 USENIX Annual Technical Conference}}. \bibinfo{publisher}{{USENIX} Association}, \bibinfo{pages}{537--550}.
\newblock
\urldef\tempurl%
\url{https://www.usenix.org/conference/atc21/presentation/lee}
\showURL{%
\tempurl}


\bibitem[Lee et~al\mbox{.}(2024)]%
        {lee2024presto}
\bibfield{author}{\bibinfo{person}{Yunjae Lee}, \bibinfo{person}{Hyeseong Kim}, {and} \bibinfo{person}{Minsoo Rhu}.} \bibinfo{year}{2024}\natexlab{}.
\newblock \showarticletitle{PreSto: An In-Storage Data Preprocessing System for Training Recommendation Models}. In \bibinfo{booktitle}{\emph{51st ACM/IEEE Annual International Symposium on Computer Architecture}}. \bibinfo{publisher}{{IEEE}}, \bibinfo{pages}{340--353}.
\newblock
\urldef\tempurl%
\url{https://doi.org/10.1109/ISCA59077.2024.00033}
\showDOI{\tempurl}


\bibitem[Liang et~al\mbox{.}(2024)]%
        {liang2024survey}
\bibfield{author}{\bibinfo{person}{Zijing Liang}, \bibinfo{person}{Yanjie Xu}, \bibinfo{person}{Yifan Hong}, \bibinfo{person}{Penghui Shang}, \bibinfo{person}{Qi Wang}, \bibinfo{person}{Qiang Fu}, {and} \bibinfo{person}{Ke Liu}.} \bibinfo{year}{2024}\natexlab{}.
\newblock \showarticletitle{A survey of multimodel large language models}. In \bibinfo{booktitle}{\emph{Proceedings of the 3rd International Conference on Computer, Artificial Intelligence and Control Engineering}}.
\newblock


\bibitem[Lin et~al\mbox{.}(2014)]%
        {cocodataset}
\bibfield{author}{\bibinfo{person}{Tsung{-}Yi Lin}, \bibinfo{person}{Michael Maire}, \bibinfo{person}{Serge~J. Belongie}, \bibinfo{person}{James Hays}, \bibinfo{person}{Pietro Perona}, \bibinfo{person}{Deva Ramanan}, \bibinfo{person}{Piotr Doll{\'{a}}r}, {and} \bibinfo{person}{C.~Lawrence Zitnick}.} \bibinfo{year}{2014}\natexlab{}.
\newblock \showarticletitle{Microsoft {COCO:} Common Objects in Context}.
\newblock   \bibinfo{volume}{8693} (\bibinfo{year}{2014}), \bibinfo{pages}{740--755}.
\newblock
\urldef\tempurl%
\url{https://doi.org/10.1007/978-3-319-10602-1\_48}
\showDOI{\tempurl}


\bibitem[Maetschke et~al\mbox{.}(2017)]%
        {nuts-flow/ml}
\bibfield{author}{\bibinfo{person}{Stefan Maetschke}, \bibinfo{person}{Ruwan~Bandara Tennakoon}, \bibinfo{person}{Christian Vecchiola}, {and} \bibinfo{person}{Rahil Garnavi}.} \bibinfo{year}{2017}\natexlab{}.
\newblock \showarticletitle{nuts-flow/ml: data pre-processing for deep learning}.
\newblock  (\bibinfo{year}{2017}).
\newblock
\showeprint[arXiv]{1708.06046}
\urldef\tempurl%
\url{http://arxiv.org/abs/1708.06046}
\showURL{%
\tempurl}


\bibitem[Makino et~al\mbox{.}(2019)]%
        {rnnt}
\bibfield{author}{\bibinfo{person}{Takaki Makino}, \bibinfo{person}{Hank Liao}, \bibinfo{person}{Yannis~M. Assael}, \bibinfo{person}{Brendan Shillingford}, \bibinfo{person}{Basilio Garcia}, \bibinfo{person}{Otavio Braga}, {and} \bibinfo{person}{Olivier Siohan}.} \bibinfo{year}{2019}\natexlab{}.
\newblock \showarticletitle{Recurrent Neural Network Transducer for Audio-Visual Speech Recognition}. In \bibinfo{booktitle}{\emph{{IEEE Automatic Speech Recognition and Understanding Workshop}}}. \bibinfo{publisher}{{IEEE}}, \bibinfo{pages}{905--912}.
\newblock
\urldef\tempurl%
\url{https://doi.org/10.1109/ASRU46091.2019.9004036}
\showDOI{\tempurl}


\bibitem[Massa and Girshick({[n.\,d.]})]%
        {massa2018mrcnn}
\bibfield{author}{\bibinfo{person}{Francisco Massa} {and} \bibinfo{person}{Ross Girshick}.} \bibinfo{year}{[n.\,d.]}\natexlab{}.
\newblock \bibinfo{title}{{maskrnn-benchmark: Fast, modular reference implementation of Instance Segmentation and Object Detection algorithms in PyTorch}}.
\newblock \bibinfo{howpublished}{\url{https://github.com/facebookresearch/maskrcnn-benchmark}}.
\newblock
\newblock
\shownote{Accessed: May 11, 2025}.


\bibitem[Mattson et~al\mbox{.}(2020)]%
        {mattson2020mlperf}
\bibfield{author}{\bibinfo{person}{Peter Mattson}, \bibinfo{person}{Christine Cheng}, \bibinfo{person}{Gregory Diamos}, \bibinfo{person}{Cody Coleman}, \bibinfo{person}{Paulius Micikevicius}, \bibinfo{person}{David Patterson}, \bibinfo{person}{Hanlin Tang}, \bibinfo{person}{Gu-Yeon Wei}, \bibinfo{person}{Peter Bailis}, \bibinfo{person}{Victor Bittorf}, {et~al\mbox{.}}} \bibinfo{year}{2020}\natexlab{}.
\newblock \showarticletitle{{MlPerf Training Benchmark}}.
\newblock \bibinfo{journal}{\emph{Proceedings of Machine Learning and Systems}} (\bibinfo{year}{2020}).
\newblock


\bibitem[Mazumder et~al\mbox{.}(2023)]%
        {mazumder2022dataperf}
\bibfield{author}{\bibinfo{person}{Mark Mazumder}, \bibinfo{person}{Colby Banbury}, \bibinfo{person}{Xiaozhe Yao}, \bibinfo{person}{Bojan Karla{\v{s}}}, \bibinfo{person}{William~Gaviria Rojas}, \bibinfo{person}{Sudnya Diamos}, \bibinfo{person}{Greg Diamos}, \bibinfo{person}{Lynn He}, \bibinfo{person}{Alicia Parrish}, \bibinfo{person}{Hannah~Rose Kirk}, {et~al\mbox{.}}} \bibinfo{year}{2023}\natexlab{}.
\newblock \showarticletitle{Dataperf: Benchmarks for data-centric AI development}. In \bibinfo{booktitle}{\emph{Advances in Neural Information Processing Systems 36: Annual Conference on Neural Information Processing Systems 2023}}.
\newblock


\bibitem[{MLCommons}({[n.\,d.]})]%
        {mlcommons-github}
\bibfield{author}{\bibinfo{person}{{MLCommons}}.} \bibinfo{year}{[n.\,d.]}\natexlab{}.
\newblock \bibinfo{title}{{MLPerf Benchmarking Suite - PyTorch implementation for image segmentation}}.
\newblock \bibinfo{howpublished}{\url{https://github.com/mlcommons/training/tree/master/image_segmentation/pytorch}}.
\newblock
\newblock
\shownote{Accessed: [May 5, 2025]}.


\bibitem[Mohan et~al\mbox{.}(2021)]%
        {mohan2020analyzing}
\bibfield{author}{\bibinfo{person}{Jayashree Mohan}, \bibinfo{person}{Amar Phanishayee}, \bibinfo{person}{Ashish Raniwala}, {and} \bibinfo{person}{Vijay Chidambaram}.} \bibinfo{year}{2021}\natexlab{}.
\newblock \showarticletitle{Analyzing and Mitigating Data Stalls in {DNN} Training}.
\newblock \bibinfo{journal}{\emph{Proceedings of the VLDB Endowment}} \bibinfo{volume}{14}, \bibinfo{number}{5} (\bibinfo{year}{2021}), \bibinfo{pages}{771--784}.
\newblock
\urldef\tempurl%
\url{https://doi.org/10.14778/3446095.3446100}
\showDOI{\tempurl}


\bibitem[Murray et~al\mbox{.}(2021)]%
        {tfdata}
\bibfield{author}{\bibinfo{person}{Derek~Gordon Murray}, \bibinfo{person}{Jiri Simsa}, \bibinfo{person}{Ana Klimovic}, {and} \bibinfo{person}{Ihor Indyk}.} \bibinfo{year}{2021}\natexlab{}.
\newblock \showarticletitle{tf.data: {A} Machine Learning Data Processing Framework}.
\newblock \bibinfo{journal}{\emph{Proceedings of the VLDB Endowment}} \bibinfo{volume}{14}, \bibinfo{number}{12} (\bibinfo{year}{2021}), \bibinfo{pages}{2945--2958}.
\newblock
\urldef\tempurl%
\url{https://doi.org/10.14778/3476311.3476374}
\showDOI{\tempurl}


\bibitem[Nouaji and Bitchebe(2025)]%
        {rahma_nouaji_2025}
\bibfield{author}{\bibinfo{person}{Rahma Nouaji} {and} \bibinfo{person}{Stella Bitchebe}.} \bibinfo{year}{2025}\natexlab{}.
\newblock \bibinfo{booktitle}{\emph{Rahm-no/MinatoLoader: MinatoLoader v1.0.1}}.
\newblock
\urldef\tempurl%
\url{https://doi.org/10.5281/zenodo.17201356}
\showDOI{\tempurl}


\bibitem[Nouaji et~al\mbox{.}(2024)]%
        {nouajispeedyloader}
\bibfield{author}{\bibinfo{person}{Rahma Nouaji}, \bibinfo{person}{Stella Bitchebe}, {and} \bibinfo{person}{Oana Balmau}.} \bibinfo{year}{2024}\natexlab{}.
\newblock \showarticletitle{{SpeedyLoader: Efficient Pipelining of Data Preprocessing and Machine Learning Training}}. In \bibinfo{booktitle}{\emph{{Proceedings of the 4th Workshop on Machine Learning and Systems}}}. \bibinfo{publisher}{{ACM}}, \bibinfo{pages}{65–72}.
\newblock
\urldef\tempurl%
\url{https://doi.org/10.1145/3642970.3655824}
\showDOI{\tempurl}


\bibitem[Panayotov et~al\mbox{.}(2015)]%
        {librispeech}
\bibfield{author}{\bibinfo{person}{Vassil Panayotov}, \bibinfo{person}{Guoguo Chen}, \bibinfo{person}{Daniel Povey}, {and} \bibinfo{person}{Sanjeev Khudanpur}.} \bibinfo{year}{2015}\natexlab{}.
\newblock \showarticletitle{Librispeech: An {ASR} corpus based on public domain audio books}. In \bibinfo{booktitle}{\emph{2015 IEEE International Conference on Acoustics, Speech and Signal Processing}}. \bibinfo{publisher}{{IEEE}}, \bibinfo{pages}{5206--5210}.
\newblock
\urldef\tempurl%
\url{https://doi.org/10.1109/ICASSP.2015.7178964}
\showDOI{\tempurl}


\bibitem[Park et~al\mbox{.}(2020)]%
        {park2020trainbox}
\bibfield{author}{\bibinfo{person}{Pyeongsu Park}, \bibinfo{person}{Heetaek Jeong}, {and} \bibinfo{person}{Jangwoo Kim}.} \bibinfo{year}{2020}\natexlab{}.
\newblock \showarticletitle{TrainBox: An Extreme-Scale Neural Network Training Server Architecture by Systematically Balancing Operations}. In \bibinfo{booktitle}{\emph{53rd Annual {IEEE/ACM} International Symposium on Microarchitecture}}. \bibinfo{publisher}{{IEEE}}, \bibinfo{pages}{825--838}.
\newblock
\urldef\tempurl%
\url{https://doi.org/10.1109/MICRO50266.2020.00072}
\showDOI{\tempurl}


\bibitem[Reddy et~al\mbox{.}(2020)]%
        {9036908}
\bibfield{author}{\bibinfo{person}{G.~Thippa Reddy}, \bibinfo{person}{M.~Praveen~Kumar Reddy}, \bibinfo{person}{Kuruva Lakshmanna}, \bibinfo{person}{Rajesh Kaluri}, \bibinfo{person}{Dharmendra~Singh Rajput}, \bibinfo{person}{Gautam Srivastava}, {and} \bibinfo{person}{Thar Baker}.} \bibinfo{year}{2020}\natexlab{}.
\newblock \showarticletitle{Analysis of Dimensionality Reduction Techniques on Big Data}.
\newblock \bibinfo{journal}{\emph{{IEEE} Access}}  \bibinfo{volume}{8} (\bibinfo{year}{2020}), \bibinfo{pages}{54776--54788}.
\newblock
\urldef\tempurl%
\url{https://doi.org/10.1109/ACCESS.2020.2980942}
\showDOI{\tempurl}


\bibitem[Sambasivan et~al\mbox{.}(2021)]%
        {sambasivan2021everyone}
\bibfield{author}{\bibinfo{person}{Nithya Sambasivan}, \bibinfo{person}{Shivani Kapania}, \bibinfo{person}{Hannah Highfill}, \bibinfo{person}{Diana Akrong}, \bibinfo{person}{Praveen~K. Paritosh}, {and} \bibinfo{person}{Lora Aroyo}.} \bibinfo{year}{2021}\natexlab{}.
\newblock \showarticletitle{"Everyone wants to do the model work, not the data work": Data Cascades in High-Stakes AI}. In \bibinfo{booktitle}{\emph{Conference on Human Factors in Computing Systems}}. \bibinfo{publisher}{{ACM}}, \bibinfo{pages}{39:1--39:15}.
\newblock
\urldef\tempurl%
\url{https://doi.org/10.1145/3411764.3445518}
\showDOI{\tempurl}


\bibitem[Shorten and Khoshgoftaar(2019)]%
        {shorten2019survey}
\bibfield{author}{\bibinfo{person}{Connor Shorten} {and} \bibinfo{person}{Taghi~M Khoshgoftaar}.} \bibinfo{year}{2019}\natexlab{}.
\newblock \showarticletitle{A survey on Image Data Augmentation for Deep Learning}.
\newblock \bibinfo{journal}{\emph{Journal of Big Data}}  \bibinfo{volume}{6} (\bibinfo{year}{2019}), \bibinfo{pages}{60}.
\newblock
\urldef\tempurl%
\url{https://doi.org/10.1186/S40537-019-0197-0}
\showDOI{\tempurl}


\bibitem[Soviany et~al\mbox{.}(2022)]%
        {soviany2022curriculum}
\bibfield{author}{\bibinfo{person}{Petru Soviany}, \bibinfo{person}{Radu~Tudor Ionescu}, \bibinfo{person}{Paolo Rota}, {and} \bibinfo{person}{Nicu Sebe}.} \bibinfo{year}{2022}\natexlab{}.
\newblock \showarticletitle{Curriculum Learning: A Survey}.
\newblock \bibinfo{journal}{\emph{International Journal of Computer Vision}} \bibinfo{volume}{130}, \bibinfo{number}{6} (\bibinfo{year}{2022}), \bibinfo{pages}{1526--1565}.
\newblock
\urldef\tempurl%
\url{https://doi.org/10.1007/S11263-022-01611-X}
\showDOI{\tempurl}


\bibitem[Spark(2019)]%
        {spark2019spark}
\bibfield{author}{\bibinfo{person}{Apache Spark}.} \bibinfo{year}{2019}\natexlab{}.
\newblock \bibinfo{title}{Spark}.
\newblock
\newblock


\bibitem[Um et~al\mbox{.}(2023)]%
        {um2023fastflow}
\bibfield{author}{\bibinfo{person}{Taegeon Um}, \bibinfo{person}{Byungsoo Oh}, \bibinfo{person}{Byeongchan Seo}, \bibinfo{person}{Minhyeok Kweun}, \bibinfo{person}{Goeun Kim}, {and} \bibinfo{person}{Woo{-}Yeon Lee}.} \bibinfo{year}{2023}\natexlab{}.
\newblock \showarticletitle{FastFlow: Accelerating Deep Learning Model Training with Smart Offloading of Input Data Pipeline}.
\newblock \bibinfo{journal}{\emph{Proceedings of the VLDB Endowment}} \bibinfo{volume}{16}, \bibinfo{number}{5} (\bibinfo{year}{2023}), \bibinfo{pages}{1086--1099}.
\newblock
\urldef\tempurl%
\url{https://doi.org/10.14778/3579075.3579083}
\showDOI{\tempurl}


\bibitem[Zhao et~al\mbox{.}(2022)]%
        {zhao2022understanding}
\bibfield{author}{\bibinfo{person}{Mark Zhao}, \bibinfo{person}{Niket Agarwal}, \bibinfo{person}{Aarti Basant}, \bibinfo{person}{Bugra Gedik}, \bibinfo{person}{Satadru Pan}, \bibinfo{person}{Mustafa Ozdal}, \bibinfo{person}{Rakesh Komuravelli}, \bibinfo{person}{Jerry Pan}, \bibinfo{person}{Tianshu Bao}, \bibinfo{person}{Haowei Lu}, \bibinfo{person}{Sundaram Narayanan}, \bibinfo{person}{Jack Langman}, \bibinfo{person}{Kevin Wilfong}, \bibinfo{person}{Harsha Rastogi}, \bibinfo{person}{Carole{-}Jean Wu}, \bibinfo{person}{Christos Kozyrakis}, {and} \bibinfo{person}{Parik Pol}.} \bibinfo{year}{2022}\natexlab{}.
\newblock \showarticletitle{Understanding data storage and ingestion for large-scale deep recommendation model training: industrial product}. In \bibinfo{booktitle}{\emph{The 49th Annual International Symposium on Computer Architecture}}. \bibinfo{publisher}{{ACM}}, \bibinfo{pages}{1042--1057}.
\newblock
\urldef\tempurl%
\url{https://doi.org/10.1145/3470496.3533044}
\showDOI{\tempurl}


\bibitem[Zhao et~al\mbox{.}(2023a)]%
        {recd}
\bibfield{author}{\bibinfo{person}{Mark Zhao}, \bibinfo{person}{Dhruv Choudhary}, \bibinfo{person}{Devashish Tyagi}, \bibinfo{person}{Ajay Somani}, \bibinfo{person}{Max Kaplan}, \bibinfo{person}{Sung-Han Lin}, \bibinfo{person}{Sarunya Pumma}, \bibinfo{person}{Jongsoo Park}, \bibinfo{person}{Aarti Basant}, \bibinfo{person}{Niket Agarwal}, \bibinfo{person}{Carole-Jean Wu}, {and} \bibinfo{person}{Christos Kozyrakis}.} \bibinfo{year}{2023}\natexlab{a}.
\newblock \bibinfo{title}{RecD: Deduplication for End-to-End Deep Learning Recommendation Model Training Infrastructure}.
\newblock
\newblock
\showeprint{2211.05239}
\urldef\tempurl%
\url{https://arxiv.org/abs/2211.05239}
\showURL{%
\tempurl}


\bibitem[Zhao et~al\mbox{.}(2023b)]%
        {zhao2023tectonic}
\bibfield{author}{\bibinfo{person}{Mark Zhao}, \bibinfo{person}{Satadru Pan}, \bibinfo{person}{Niket Agarwal}, \bibinfo{person}{Zhaoduo Wen}, \bibinfo{person}{David Xu}, \bibinfo{person}{Anand Natarajan}, \bibinfo{person}{Pavan Kumar}, \bibinfo{person}{Shiva~Shankar P.}, \bibinfo{person}{Ritesh Tijoriwala}, \bibinfo{person}{Karan Asher}, \bibinfo{person}{Hao Wu}, \bibinfo{person}{Aarti Basant}, \bibinfo{person}{Daniel Ford}, \bibinfo{person}{Delia David}, \bibinfo{person}{Nezih Yigitbasi}, \bibinfo{person}{Pratap Singh}, {and} \bibinfo{person}{Carole{-}Jean Wu}.} \bibinfo{year}{2023}\natexlab{b}.
\newblock \showarticletitle{{Tectonic-Shift: A Composite Storage Fabric for Large-Scale ML Training}}. In \bibinfo{booktitle}{\emph{2023 USENIX Annual Technical Conference}}. \bibinfo{publisher}{{USENIX} Association}, \bibinfo{pages}{433--449}.
\newblock


\end{thebibliography}
